\documentclass[aps,prb,reprint,
superscriptaddress,floatfix,longbibliography]{revtex4-1}
\usepackage{graphicx,amsmath,amsfonts,amssymb,amsthm,xr}
\usepackage{epsfig,amsmath,amssymb,color,dsfont,upgreek,physics}
\usepackage[utf8]{inputenc}
\usepackage{bm}
\usepackage{mathrsfs}
\usepackage{mathtools}
\usepackage{bbold}
\usepackage{comment}
\usepackage{soul}
\usepackage{float}

\newcommand{\bh}{{\hat{b}}}
\newcommand{\nh}{{\hat{n}}}
\newcommand{\thet}{{\hat{\theta}}}

\usepackage[bookmarks=true,colorlinks,linkcolor=OrangeRed,urlcolor=NavyBlue,citecolor=RoyalBlue]{hyperref}
\usepackage[dvipsnames]{xcolor}

\usepackage{ulem}

\begin{document}
\title{Exotic quantum liquids in Bose-Hubbard models with spatially-modulated symmetries}
\author{Pablo Sala}
\email{psala@caltech.edu}
\affiliation{Department of Physics and Institute for Quantum Information and Matter,
California Institute of Technology, Pasadena, California 91125, USA}
\affiliation{Walter Burke Institute for Theoretical Physics, California Institute of Technology, Pasadena, California 91125, USA}
\author{Yizhi You}
\affiliation{Department of Physics, Northeastern University, Boston, MA, 02115, USA}
\author{Johannes Hauschild}
\affiliation{
Technical University of Munich, 
TUM School of Natural Sciences, 
Physics Department,
James-Franck-Str.~1,
85748 Garching,
Germany
}
\affiliation{Munich Center for Quantum Science and Technology (MCQST), Schellingstr. 4, 80799 M{\"u}nchen, Germany}
\author{Olexei Motrunich}
\affiliation{Department of Physics and Institute for Quantum Information and Matter,
California Institute of Technology, Pasadena, California 91125, USA}

\begin{abstract}

We investigate the effect that spatially modulated continuous conserved quantities can have on quantum ground states. We do so by introducing a family of one-dimensional local quantum rotor and bosonic models which conserve finite Fourier momenta of the particle number, but not the particle number itself. These correspond to generalizations of the standard Bose-Hubbard model (BHM), and relate to the physics of Bose surfaces. First, we show that while having an infinite-dimensional local Hilbert space, such systems feature a non-trivial Hilbert space fragmentation  for momenta incommensurate with the lattice. This is linked to the nature of the conserved quantities having a dense spectrum and provides the first such example. We then characterize the zero-temperature phase diagram for both commensurate and incommensurate momenta. In both cases, analytical and numerical calculations predict a phase transition between a gapped (Mott insulating) and quasi-long range order phase; the latter is characterized by a two-species Luttinger liquid in the infrared, but dressed by oscillatory contributions when computing microscopic expectation values. Following a rigorous Villain formulation of the corresponding rotor model, we derive a dual description, from where we estimate the robustness of this phase using renormalization group arguments, where the driving perturbation has ultra-local correlations in space but power law correlations in time.
We support this conclusion using an equivalent representation of the system as a two-dimensional vortex gas with modulated Coulomb interactions within a fixed symmetry sector. 
We conjecture that a Berezinskii–Kosterlitz–Thouless-type transition is driven by the unbinding of vortices along the temporal direction. 
\end{abstract}

\date{\today}
\maketitle

%\tableofcontents

\section{Introduction}

Unconventional symmetries, including subsystem symmetries as well as dipole and higher-moment conservation laws have been extensively studied. These are the key ingredients to endow fractons with such exotic behavior at low energies (see review~\onlinecite{fractons_review_Nand}), and underlie the emergence of Bose surfaces~\cite{Paramekanti_2002,Tiamhock_2011, PhysRevB.100.024519,you2020fracton,Lake_2021} as well as the UV/IR-mixing observed in this type of systems~\cite{Seiberg_2020,gorantla2021lowenergy,you2021fractonic}. These symmetries have also been shown to play an important role out of equilibrium. For example, imposing dipole-moment (or higher-moment) conservation leads, among other rich phenomena, to a fragmentation of the Hilbert space into exponentially many sectors~\cite{Sala_PRX,khemani_localization_2020} (see also review~\onlinecite{Moudgalya_review}), which can be understood by the presence of extensively many non-local conserved quantities. When such fragmentation is not strong and can be ignored, these dipole (or higher) moments conservation laws give rise to universal subdiffusive behavior, which has been recently observed with ultra-cold atoms~\cite{Guardado_Sanchez_2020} and which can be completely characterized by the symmetries of the system~\cite{gromov2020_fractonhydro,morningstar2020_kinetic,Feldmeier_anomal,Zhang_2020,han2023scaling}. 

The motivation to consider such apparently artificial symmetries is two-fold. At the abstract level, these are analytically amenable models that provide new insights about the role and proper definition of symmetries for quantum many-body systems~\cite{Rakovszky_slioms,Moudgalya_2022,moudgalya2023symmetries}. Thus, this led to new insights on quantum thermalization and the eigenstate thermalization hypothesis, and motivated a formal mathematical framework to characterize symmetries in terms of commutant algebras~\cite{Moudgalya_2022}. 
On the experimental front, the flourishing development of new experimental platforms and quantum technologies, where engineered synthetic quantum matter has become a reality, is opening the door to realize and probe such less conventional systems in the lab~\cite{doi:10.1126/science.abi8378,doi:10.1126/science.abi8794}. Just in the last few years, dipole-conserving systems have been observed to approximately govern the dynamics of interacting cold atom systems in the presence of a linear tilted potential~\cite{Scherg_nature,PhysRevLett.130.010201}. These experimental observations are consistent with both the phenomenon of Hilbert space fragmentation, and the expected universal subdiffusive behavior~\cite{Guardado_Sanchez_2020}. Furthermore, one could also envision engineering such constrained systems with different platforms~\cite{Tortora_22}.

From this point of view, subsystem symmetries as well as dipole and higher-moment symmetries are just some of the many possible symmetries local quantum many-body Hamiltonians can realize. In fact, it is an open question to understand the possible conserved quantities local quantum many-body systems can have as well as their influence on the equilibrium and out-of-equilibrium behavior. A recent work~\cite{Sala_modsym} extended the notion of multipole and subsystem symmetries to more general spatially modulated symmetries, uncovering two novel instances with conserved quantities containing exponential and (quasi)-periodic spatial modulations, with dipole- and higher-moment conservation appearing as special cases. The latter case was shown to give rise to exotic forms of sub-diffusive behavior with a rich spatial structure influenced by lattice-scale features, while the former leads to infinitely
long-lived boundary correlations. Nonetheless, their effect at low-energies has remained an open question.

In this work we address this question by focusing on one-dimensional ($1$D) systems with quasi-periodic modulated symmetries, characterizing the zero-temperature (quantum ground state) phase diagram for a generalized Bose-Hubbard model (BHM).
For example, recent works~\cite{Lake_2022,Lake22_1DDBHM,Zechmann_22} examined (one-dimensional) bosonic dipole-conserving systems, finding a rich phase diagram including different types of unconventional phases that include dipole condensates and supersolids. In fact, some of these phases and their transitions were predicted in an earlier work using the fracton-elasticity duality~\cite{elast_dual_leo}; and in certain regimes, these also relate to earlier results for dipole-conserving spin$-1$ chains, where an exact mapping to an $XY$ chain with non-zero string order parameter was found~\cite{Rakovszky_slioms}. Moreover, higher-dimensional systems with either approximate or exact subsystem symmetries have been studied leading to a plethora of unexpected behavior including not only Bose metals~\cite{Paramekanti_2002} and UV-IR mixing~\cite{you2021fractonic,Gorantla_22,Seiberg_2020,gorantla2021lowenergy}, but also subsystem symmetry protected topological phases~\cite{You_2018}, fractal ``criticality''~\cite{zhou2021fractal} and fractal spin liquids~\cite{Yoshida_2013} among others.

The remainder of the paper is organized as follows. 
In Sec.~\ref{sec:bosmodels} we introduce the generalized BHM realizing different types of spatially modulated symmetries and analyze the spectrum of the associated conserved quantities. 
Sec.~\ref{sec:rotor} then introduces rotor Hamiltonians which will turn out to be useful to understand the quasi-long-range order phase analyzed in the later sections.
In this section, we also describe a generalized lattice duality for our models.
In Sec.~\ref{sec:frag} we show that both the bosonic and rotor systems are fragmented despite an infinite-dimensional local Hilbert space dimension, and identify extensively many non-local discrete conserved unitaries that can account for this behavior.  
In Sec.~\ref{sec:comm} we then discuss the phase diagram of systems with commensurate modulated symmetries, validating the analytical predictions against numerical tensor network calculations.
In Sec.~\ref{sec:incomm} we then discuss the case of incommensurate symmetries, which add new subtleties when dealing with the low-energy theory.
In Sec.~\ref{sec:vortex} we provide a complementary description of the system in terms of a two-dimensional Coulomb-like gas with rather unusual (and qualitatively important) microscopic details, that corroborates the existence of a quasi-long range order phase that can become disordered by the proliferation of topological defects.
We conclude in Sec.~\ref{sec:conclusion} by summarizing our main findings and discussing open questions.
Finally, we consign more technical aspects of our work to the appendices.

\begin{figure}[h!]
    \centering
    \includegraphics[width=\linewidth]{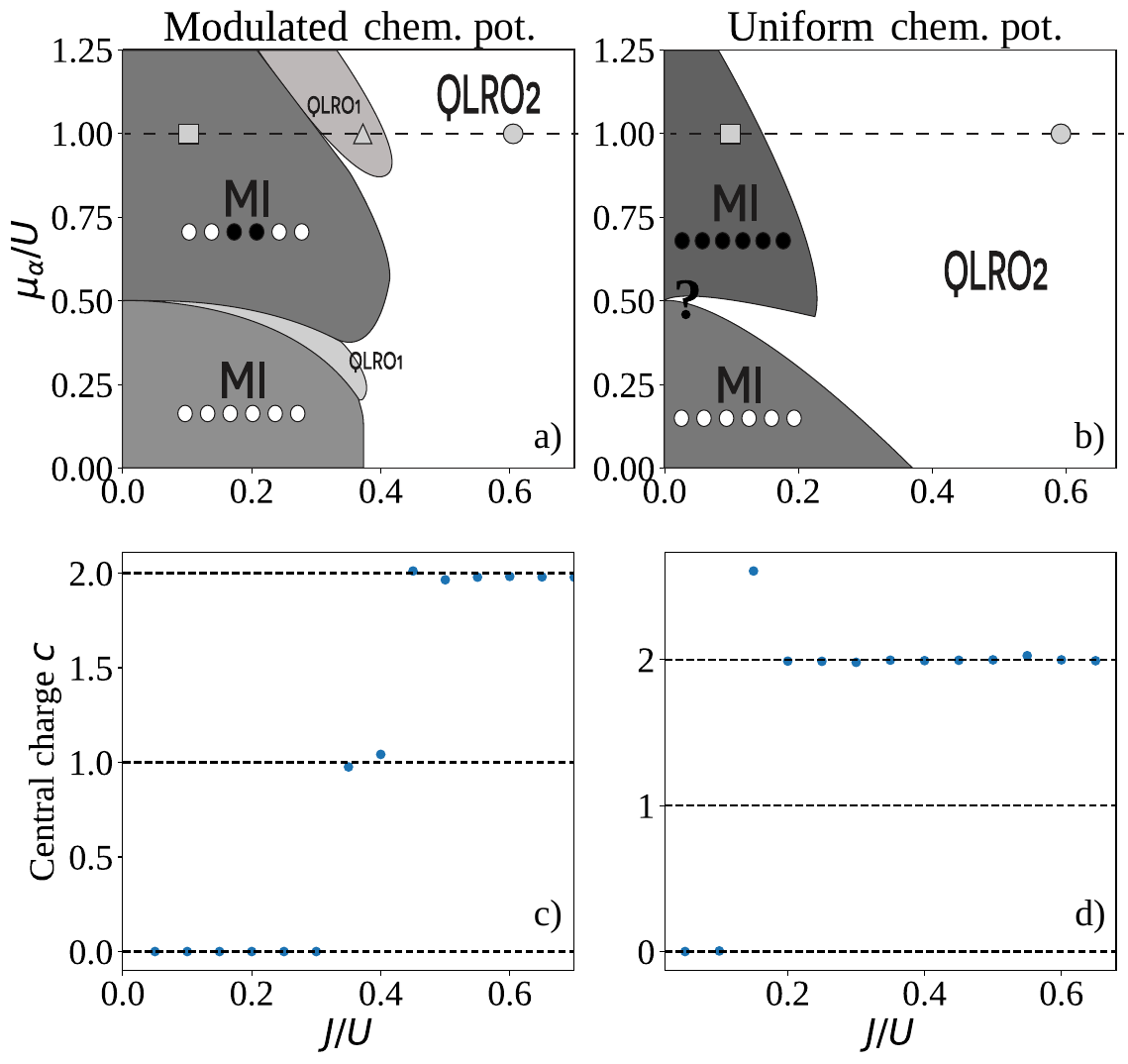}
    \caption{\textbf{Phase diagram for $\bf{q,p=1}$.} 
    Panels (a,b) show the schematic phase diagram for Hamiltonian $H_{1,1}$ in Eq.~\eqref{eq:H_11} in the presence of a modulated $\mu_A\sum_j \alpha^{(1,0)}_j \hat{n}_j$ (panel a) and uniform $-\mu \sum_j \nh_j$ (panel b) chemical potentials, as a function of $J/U$ and $\mu_\alpha/U$, with $\mu_\alpha=\mu_A, \mu$ respectively. In both cases, we find two different types of behaviors: Gapped phases for $J<J_c(\mu_{\alpha})$ corresponding to Mott insulating lobes, and a gapless quasi-long range order (QLRO$_2$) phase for $J>J_c(\mu_\alpha)$. The former acquires a particular particle density pattern in the case of modulated chemical potential, with the density patterns represented in the plot. One also finds intermediate lobes corresponding to gapless phases (QLRO$_1$) protected by bond inversion symmetry. Panels (c,d) show the effective central charge $c$ along a cut with $\mu_\alpha/U=1$ as obtained from infinite DMRG by scaling the bipartite entanglement entropy $S_\chi$ with the finite correlation length $\xi_\chi$ induced by a finite bond dimension $S_{\chi}=\frac{c}{6}\log(\xi_\chi)$~\cite{Pollmann_2009}. Both panels show a clear transition from $c=0$ to $c=2$, the latter corresponding to the QLRO phase. Gapless lobes with modulated chemical potential (QLRO$_1$) correspond to $c=1$ as seen over a narrow intermediate range in panel c, which we explain in detail in App.~\ref{app:QLRO_1}. 
    }
    \label{fig:Fig1}
\end{figure}

\section{Bosonic models} \label{sec:bosmodels}

We consider one-dimensional bosonic systems with annihilation/creation operators $\hat{b}_j, \, \hat{b}^\dagger_j$ satisfying canonical commutation relations $[\hat{b}_i,\bh_j^\dagger]=\delta_{ij}$. The minimal Hamiltonian models we consider take the form
\begin{equation} \label{eq:H_qp}
H_{q,p}=-\sum_{j=2}^{L-1} J_j\left(\hat{b}_{j-1}^q(\hat{b}^{\dagger}_{j})^p \hat{b}_{j+1}^q + \textrm{H.c.}\right) + \hat{V}(\{\hat{n}_j\}),
\end{equation}
on a chain of length $L$ for any choice of real couplings $J_j>0$.
Here $p\geq 0, q\geq 1$ are two integers. We refer to the first contribution as ``squeezing''~\footnote{Notice that for $p+2q$ odd, one can flip the sign of the squeezing term via the unitary transformation $e^{i\pi \sum_j \nh_j} \bh_j e^{-i\pi \sum_j \nh_j}=-\bh_j$.}, taking the form of a correlated hopping not necessarily conserving the total particle number. The second contribution $\hat{V}=\hat{V}_\text{interaction} + \hat{V}_\text{potential}$ is diagonal in the occupation basis, where the latter collects linear terms in $\nh_j$ on-site terms and the former includes powers of $\nh_j$.  Due to bosonic statistics, each such term in a situation with a large $O(N)$ number of particles on each of the three involved sites has energy scaling as $~-N^{\frac{2q+p}{2}}$.
Hence, to have a lower-bounded ground state energy one needs to include sufficiently high power of density-density interactions in $\hat{V}$ to stabilize it.
Thus, if $2q+p>4$, the usual on-site terms $\sim \hat{n}_j^2$ are not sufficient, but terms $\sim \hat{n}_j^m$ with any integer $m > \frac{2q+2}{2}$ will stabilize it.

All these Hamiltonians, regardless the choice of $J_j$'s and potential term $\hat{V}$,  are invariant under the unitary transformation $\bh_j\to e^{i\alpha_j }\bh_j$ (and correspondingly $\bh^\dagger_j\to e^{-i\alpha_j }\bh^\dagger_j$) where the real coefficients $\alpha_j$ satisfy the recurrence relation 
\begin{equation} \label{eq:LinRec}
    q\alpha_{j+1}-p\alpha_{j}+q\alpha_{j-1}=0,
\end{equation}
with either open boundaries (OBC) or compatible system sizes $L$ with periodic boundary conditions (PBC) for all sites $j$. 
When this is the case, two linearly-independent charges generating these symmetry transformations exist for any $J_j$'s, which we denote by 
\begin{equation} \label{eq:Q_AB}
\hat{\mathcal{Q}}^{A,B}=\sum_j \alpha^{A,B}_j \nh_j.
\end{equation}
Here, $\alpha^{A,B}_j\in\mathbb{R}$ are linearly-independent solutions of Eq.~\eqref{eq:LinRec}. As for the familiar Fibonacci sequence, these correspond to two linearly-independent choices of the initial conditions $\alpha_1, \alpha_2$. E.g., when $p=2q$ there are two U$(1)$ charges (for OBC), which correspond to particle number $(\alpha^A_j=1)$ and dipole conservation $(\alpha^B_j=j)$, whose zero-temperature phase diagram has been studied in Refs.~\onlinecite{Zechmann_22,Lake_2022,Lake22_1DDBHM} for translation invariant systems.
Nonetheless, dipole (and higher-moment) conservation is only one possible choice among the many spatially-modulated symmetries a local Hamiltonian can realize. A recent reference~\cite{Sala_modsym} showed that when $\frac{p}{2q}<1$, $\alpha^{A,B}_j$ are quasi-periodically modulated with wave-vectors $\pm k^*$ where $\cos(k^*)=\frac{p}
{2q}$, leading to two extensive conserved quantities $\hat{\mathcal{Q}}^{A,B}$~\footnote{ Notice that this fact has nothing to do with inversion symmetry, since the symmetry is present for any choice of $J_j$ couplings.}. Hence, instead of the total particle number, its finite Fourier components $\hat{\mathcal{Q}}_{c}\equiv\sum_j \cos(k^*j)\nh_j$, $\hat{\mathcal{Q}}_{s}\equiv\sum_j \sin(k^*j)\nh_j$ are conserved. This wave vector is commensurate with the lattice only when $\frac{p}{2q}=0,\frac{1}{2}$ as a consequence of Niven's theorem~\cite{niven_1956}. In these cases $k^*=\frac{\pi}{2}, \frac{\pi}{3}$ respectively, and one can choose $\alpha_1,\alpha_2$ such that the two linearly-independent solutions $\alpha^{A,B}_j$ are integer-valued and periodically repeat along the chain. The resulting $\hat{\mathcal{Q}}^{A,B}$ then correspond to two intertwined sub-lattice U$(1)$ symmetries. 
These are attained for the canonical choices $\hat{\mathcal{Q}}^{(1,0)}$, $\hat{\mathcal{Q}}^{(0,1)}$ with $\alpha_1=1, \alpha_2=0$ and $\alpha_1=0, \alpha_2=1$ respectively. For PBC, these are exactly conserved as long as one considers system sizes that are multiples of $4$ and $6$ respectively.  

On the other hand any other ratio $\frac{p}{2q} \neq 0,\frac{1}{2}$ leads to incommensurate $k^*$, in which case these are only exactly conserved for OBC. Moreover, $k^*$ being incommensurate translates into the extensive quantities $\hat{\mathcal{Q}}^{A,B}$ lacking an integer spectrum unless one redefines them as $\hat{\mathcal{Q}}^{A,B}\to \tilde{\hat{\mathcal{Q}}}^{A,B}=q^L\hat{\mathcal{Q}}^{A,B}$~\footnote{In general, $\alpha_j$ includes terms scaling as $(p/q)^j$.}. This implies that $\hat{\mathcal{Q}}^{A,B}$ have an exponentially small gap $\sim q^{-L}$ between consecutive different eigenvalues $Q_A,Q_B$.
Indeed, we numerically checked that this gap is uniform in the bulk of the spectrum leading to a continuous spectrum in the thermodynamic limit (TDL). Nonetheless, symmetry subspaces are still infinitely large (see additional details in App.~\ref{sec:spectrum}). 
Altogether, this means that it might be better to see $\hat{\mathcal{Q}}^{A,B}$ generating a unitary representation of the addition group $\mathbb{R}$ rather than of the U$(1)$ group. 
Moreover, $\hat{\mathcal{Q}}^{A,B}$ have an extensive operator norm which combined with their exponentially small gap leads to exponentially many $\sim q^L$ symmetry sectors labeled by the pair of eigenvalues $(Q_A, Q_B)$. Hence, in the language of Ref.~\onlinecite{Commutants}, all these systems have an exponentially large commutant and should be considered fragmented.
Note that this is
unlike previous studies
where, e.g., for dipole conserving models fragmentation always appears when the range of the terms and the dimension of the on-site Hilbert space are strictly bounded~\cite{Sala_PRX, khemani_localization_2020,Sanjay19}.
The described fragmentation is present in our case even when considering infinitely-large local Hilbert space dimension.
For completeness, we also note that when $\frac{p}{2q}>1$ one finds solutions $\alpha^{A,B}_j$ that are exponentially localized at the boundaries of the system (for OBC), leading to correspondingly localized conserved quantities $\hat{\mathcal{Q}}^{A,B}$. While here we focus on quasi-periodic modulated symmetries with $\frac{p}{2q}<1$, we leave the analysis of their zero-temperature physics for future work. 

Before closing this section it is important to mention that any other non-trivial local term conserving both $\hat{\mathcal{Q}}^{A,B}$, can be obtained by combining the local terms $\hat{b}_{j-1}^q(\hat{b}^{\dagger}_{j})^p \hat{b}_{j+1}^q$ (together with its hermitian conjugate) up to powers of the local densities $\nh_j$ (while it is obvious that all such generated terms have the same symmetry, the opposite direction is non-trivial; see proof in App.~\ref{app:longrang}).
For example, these include the $4$-local contribution $\hat{T}^4_{q,p} = \sum_j J_j^{(4)}\left(\bh^q_j \bh^{q-p}_{j+1} \bh^{q-p}_{j+2} \bh^q_{j+3} + \textrm{H.c.}\right)$ when $q>p$. 
Finally, we note that if $J_j = J$ for all sites (up to boundary conditions), these models have lattice translation symmetry and also inversion symmetry (assuming the potential terms $\hat{V}$ have the same symmetries).

\section{Rotor models and quasi-long range order phase} \label{sec:rotor} 
We also introduce the family of rotor Hamiltonians
\begin{equation} \label{eq:Hrotor_qp}
\begin{aligned}
H^{\textrm{rotor}}_{q,p}&=-\sum_j J_j \cos(\nabla^{q,p}_x \thet_j) +\frac{U}{2}\sum_j (\nh_j-\bar{n}_j)^{2},
\end{aligned}
\end{equation}
in terms of the canonical conjugate variables $[\thet_i, \nh_j]=i\delta_{ij}$ with $\thet_j\sim \thet_j + 2\pi$ having compact spectrum, and where $\bar{n}_j$ is the average on-site density.
Here we introduced the short-hand notation 
\begin{equation}
\nabla^{q,p}_x \thet_j\equiv -q\thet_{j-1} + p\thet_{j} -q\thet_{j+1}
\end{equation}
that will repeatedly appear in the following. E.g., with this notation the linear recurrence in Eq.~\eqref{eq:LinRec} reads $\nabla^{q,p}_x \alpha_j=0$. These Hamiltonians are invariant under the two continuous symmetry transformations 
\begin{equation}
\hat{\theta}_j \to \hat{\theta}_j + \alpha^{A,B}_j
\end{equation}
generated by $\hat{\mathcal{Q}}^{A,B}=\sum_j \alpha_j^{A,B} \nh_j$ with $\alpha^{A,B}_j$ satisfying $\nabla^{q,p}_x\alpha_j^{A,B}=0$.
One can intuitively understand the first contribution in Eq.~\eqref{eq:Hrotor_qp} as appearing in the regime $|J_j|/U \gg 1$ and large on-site particle numbers where the approximation $\bh_j\sim \sqrt{\bar{n}_j} e^{i\hat{\theta}_j}$ (and respectively $\bh^\dagger_j\sim \sqrt{\bar{n}_j} e^{-i\hat{\theta}_j}$) holds.
However, the main justification for using the rotor models is that they retain the same symmetries as the original boson models and are hence expected to show the same qualitative physics (phases and their long-distance properties) while allowing more direct connections to low-energy descriptions~\cite{Fisher_88, Fisher_89}.
Note that for the rotor models, the on-site terms $\sim (\hat{n}_j-\bar{n}_j)^2$ are already sufficient to obtain a mathematically well-defined (i.e., lower-bounded) spectrum.

Analogous to the familiar $XY$ model (see e.g., Ref.~\onlinecite{chaikin_lubensky_1995,dasgupta1981phase}), the rotor formulation $H^{\textrm{rotor}}_{q,p}$ admits a dual representation in terms of a new set of conjugate variables  $[N_i, \delta\phi_j]=i\delta_{ij}$, 
which relate to the original $\thet_j,\nh^r_j$ via the duality transformation
\begin{align}
\label{eq:dualmapN}
& \hat{N}_{j} \equiv -\nabla^{q,p}_x \thet_j,\\
\label{eq:dualmapphi} & \nh_{j} - \bar{n}_j= -\nabla^{q,p}_x \delta\phi_j.
\end{align}
Note, however, that the basic dual field definitions differ from the usual XY model and the dual variables here reside on the same lattice sites as the original lattice.
By ``solving" for $\delta\phi_j$ in terms of $\{ n_i, i<j \}$ as will be shown in Eq.~\eqref{eq:phi_sol}, i.e., number operators running from the left boundary, focusing on the bulk and ignoring boundaries (more precise treatment will be given in due time), we can indeed verify that the new variables satisfy the canonical commutation relations.

The Hamiltonian $H^{\textrm{rotor}}_{q,p}$ in terms of these becomes
\begin{equation} \label{modelrotordual}
\begin{aligned}
H^{\textrm{dual}}_{q,p}&= \frac{U}{2}\sum_j\left(\nabla^{q,p}_x\delta\phi_j\right)^{2}-\sum_jJ_j\cos(N_j).
\end{aligned}
\end{equation}
In the following sections, we will define and characterize the dual variables more precisely and will use this description to analyze the phase diagram of the bosonic Hamiltonian Eq.~\eqref{eq:H_qp} in the regime $|J_j|/U\gg 1$. In particular, we will find that this duality makes explicit that unlike for the standard Bose-Hubbard model~\footnote{For the standard Bose-Hubbard model the duality transformation reads $N_j=\theta_j - \theta_{j-1}$, $\hat{n}_j-\bar{n}_j=\delta \phi_j - \delta \phi_{j+1}$, which implies that for integer $\bar{n}_j$ the field $\delta \phi_j $ is integer-valued.}, the dual field $\delta \phi_j$ is not necessarily integer but rational-valued, translating into the fact that $e^{i2\pi \delta \phi_j}\neq \mathbb{1}$ is in general not trivial.

In this regime ---and taking $J_j>0$--- one can expand the cosine terms appearing in the rotor Hamiltonian \eqref{eq:Hrotor_qp} 
\begin{equation} \label{eq:Hrotor_qp_exap}
\begin{aligned}
H^{\textrm{rotor}}_{q,p}&\approx \sum_j \frac{J_j}{2} \left(\nabla^{q,p}_x \thet_j \right)^2 +\frac{U}{2}\sum_j (\nh_j-\bar{n}_j)^{2},
\end{aligned}
\end{equation}
or those in the dual description \eqref{modelrotordual}
\begin{equation} \label{modelrotordual_exp}
\begin{aligned}
H^{\textrm{dual}}_{q,p}&\approx \frac{U}{2}\sum_j\left(\nabla^{q,p}_x\delta\phi_j\right)^{2}+\sum_j\frac{J_j}{2}(N_j)^2,
\end{aligned}
\end{equation}
In the following, we will rigorously analyze the corresponding theory which will give a description of a gapless phase with two low-energy modes, and discuss the stability of this phase using field theoretic methods, justifying the above approximation in the appropriate regime.

\section{Hilbert space fragmentation for infinite-dimensional local Hilbert space} \label{sec:frag} 
%Before embarking into the phase diagram of these systems there is an additional consideration we need to address. 
When $q\neq 1$ ---namely whenever we consider incommensurate charges or commensurate ones with $q>1$--- and for OBC, one finds $L$ (with $L$ the system size) additional discrete symmetries as long as $\hat{\mathcal{Q}}^{A,B} $ are preserved.
These are generated by the following unitaries with support on sites $1$ to $j$:
\begin{equation} \label{eq:Uj}
U_j = \exp\left[i\frac{2\pi}{q}\sum_{1\leq i \leq j}\alpha^{(1,p/q)}_{j-i+1}\nh_i \right],
\end{equation}
for all $j\in\{1,\dots,L\}$, where $\alpha^{(1,p/q)}_{k}$ is a solution of Eq.~\eqref{eq:LinRec} with initial conditions $\alpha^{(1,p/q)}_1=1, \alpha^{(1,p/q)}_2=p/q$. These can be interpreted as a subgroup of the two continuous symmetry groups generated by $\hat{\mathcal{Q}}^A,\hat{\mathcal{Q}}^B$ on the spatial region $[1,\dots,j]$ for every $j$. 
The order of the group is not clear from the expression of $U_j$. To gain some intuition we can consider commensurate $k^*$ (i.e., $p/q = 0, 1$). Then $\alpha_j^{(1,p/q)}\in \{0,\pm 1\}$, which implies $(U_j)^q=1$ for all $j$. In fact, in this case the $\mathbb{Z}_q$ symmetry transformations can be locally implemented via $\tilde{U}_j= e^{i\frac{2\pi}{q}\hat{n}_j}$. However, for incommensurate $k^*$ only $(U_1)^q=1$ holds, and one can only show that $(U_j)^{q^j}=1$ for $j>1$, which suggests that each $U_j$ generates a discrete group $\mathbb{Z}_{q^j}$ whose order $q^j$ varies along the chain. Nonetheless, the different $U_j$'s are not independent of each other. In general, $(U_j)^q(U_{j-1}^{\dagger})^p(U_{j-2})^q=\mathbb{1}$, where we have defined $U_{-1}=U_{0}\equiv \mathbb{1}$, imposing strong restrictions on the independent eigenvalues that different $U_j$'s can take. Thence, once we fix the eigenvalues of $U_1$ and $U_2$, we can only choose among $q$ different choices for every new $U_j$ proceeding from $j=3$ to $j=L$. This means that the whole set $\{U_j\}_{j=1}^{L}$ generates a $(\mathbb{Z}_q)^L$ symmetry. Notice however that one cannot use this relation to define local $\tilde{U}_j$ transformations.
Altogether, this result implies the existence of exponentially many $q^L$ disconnected Krylov subspaces, i.e., a provable example of Hilbert space fragmentation for a (semi-)infinite-dimensional local Hilbert space, naively appearing together or as a part of the fragmentation associated with the dense spectrum of $\hat{\mathcal{Q}}^{A,B}$ in the incommensurate case. In Sec.~\ref{sec:funct_dep} and also in App.~\ref{app:Uj1}, we discuss the functional dependence of these additional symmetries on $\hat{\mathcal{Q}}^{A,B}$.
Moreover, any system with finite-range interactions locally preserving the $\hat{\mathcal{Q}}^{A,B}$  symmetries is equally fragmented. This follows from the fact that any such finite local interactions can be obtained from the $3$-local interactions $\bh_{j-1}^q(\bh^\dagger_j)^p\bh_{j+1}^q$ as shown in App.~\ref{app:longrang}, hence inheriting these non-local symmetries. 

This analysis directly extends to the family of rotor models introduced in Eq.~\eqref{eq:Hrotor_qp}. In fact, the expression for the non-local unitaries $U_j$ simplifies when writing $\delta \phi_j$ as a function of the density using Eq.~\eqref{eq:dualmapphi}. One then finds 
\begin{equation} \label{eq:phi_sol}
 \delta\phi_j =\frac{1}{q}\sum_{1\leq i <j}\alpha_{j-i}^{(1,p/q)}(\hat{n}_i-\bar{n}_i),
\end{equation}
implying that $U_j=e^{i2\pi \delta \phi_{j+1}}$ up to boundary contributions. This means that the duality transformation maps $U_j$ to a localized unitary while preserving the locality of the Hamiltonian. 
Moreover, it appears that both the bosonic and rotor models in Eqs.~\eqref{eq:H_qp} and \eqref{eq:Hrotor_qp} respectively, are invariant under any phase shift $\theta_j \to \theta_j + \alpha_j$ satisfying $\nabla_x^{q,p}\alpha_j=2\pi m_j$ with $m_j\in\mathbb{Z}$ for any site. While this corresponds to a trivial transformation in the standard Bose-Hubbard model, here it can be expressed in terms of the $U_j$'s as $\prod_j (U_j)^{m_j}$ and is hence not a new symmetry.

\section{Commensurate modulation and sublattice symmetries} \label{sec:comm} 
We first analyze systems with commensurate charges. As we found in Sec.~\ref{sec:bosmodels} this happens whenever $\frac{p}{q}=0,1$. In both cases, we can fix $q=1$ to avoid the presence of the extensively many additional $U_j$ symmetries. (Otherwise, one should restrict to a specific superselection sector of the $U_j$ symmetries, since by virtue of Elitzur's theorem, such local symmetries cannot be spontaneously broken.)
On the one hand when $p=0$ and since $\hat{V}$ only includes on-site contributions, $H_{q,0}$ exactly maps onto two decoupled chains defined on even and odd sites with local terms of the form $\bh_j^\dagger\bh_{j+1}^\dagger + \textrm{H.c.}$. The resulting systems still have a U$(1)$ symmetry generated by $\sum_j (-1)^j\nh_j$ within each sublattice. And although one cannot directly map these bosonic chains to the standard BHM via a unitary transformation, one finds a mathematically well-defined mapping at the level of rotor variables $\thet_j\to -\thet_j$, $\nh_j \to -\nh_j$ at every second site, hence recovering the same phase diagram as for the standard BHM.
Therefore, we focus on the choice $p=1$ and study the zero-temperature phase diagram of the following thermodynamically stable Hamiltonian
\begin{equation} \label{eq:H_11}
H_{1,1}=-J\sum_j\left(\bh_{j-1}\bh_{j}^\dagger \bh_{j+1}+\textrm{H.c.}\right) +\frac{U}{2}\sum_j \nh_j^2,
\end{equation}
where we have fixed a uniform coupling $J_j=J$ to simplify the upcoming analysis. This implies that $H_{1,1}$ is also invariant under translations by one lattice site $(T_1)$ when considering PBC and appropriate system sizes $(L\in 6\mathbb{N})$, as well as site-centered (or bond-centered) inversion symmetric.
This corresponds to a generalized formulation of the standard Bose-Hubbard model, resembling that considered for bosonic dipole-conserving systems~\cite{Lake_2022}. The two canonical modulated charges correspond to the subsystem symmetries with
$\alpha^{(1,0)}_{j\,\text{mod}\, 6} = (1,0,-1,-1,0,1)$ and $\alpha^{(0,1)}_{j\,\text{mod}\, 6} = (0,1,1,0,-1,-1)$, or equivalently $\alpha^{(0,1)}_{j}=\frac{2}{\sqrt{3}}\cos\left(\frac{2j-3}{6}\pi\right),
\alpha^{(1,0)}_{j}=  \frac{2}{\sqrt{3}}\cos\left(\frac{2j-1}{6}\pi\right)$.

\subsection{Overall phase diagram} \label{sec:comm_pd}

The ground state phase diagram of the standard Bose-Hubbard model with onsite interactions includes gaped Mott-insulating phases as well as gapless quasi-long range ordered phase depending on the relative sizes of the chemical potential and the tunneling rate $J$ with respect to the interaction strength $U$~\cite{PhysRevB.40.546,PhysRevLett.81.3108,PhysRevB.61.12474,Ejima_2012}. Moreover, a roughly similar phase diagram has been also found for dipole-conserving systems, which can now include a novel Bose-Einstein insulating phase (BEI), as well as a dipole condensate~\cite{Zechmann_22,Lake22_1DDBHM,Lake_2022}. Here, we combine the finite-size and infinite-size~\cite{iDMRG} density matrix renormalization group (DMRG) algorithm with analytical methods to characterize the zero-temperature phase diagram of the Hamiltonian in Eq.~\eqref{eq:H_11}.

Before discussing the results, let's list the several numerical challenges when simulating this type of system. First, as it happens for the standard BHM as well as the model with commensurate symmetries in Eq.~\eqref{eq:H_11}, the local infinite-dimensional Hilbert space needs to be truncated to a maximum finite occupation $n_{\textrm{max}}$~\cite{PhysRevB.61.12474}. This is an important parameter when dealing with quasi-long range order (or spontaneous symmetry breaking in higher dimensions) where particle fluctuations become large. Second, Hamiltonians in \eqref{eq:H_qp} include (at least) $3$-local terms which can lead to even larger local fluctuations and average particle number than a usual single-particle hopping process, hence requiring large values of $n_{\textrm{max}}$. Finally, the requirement of at least modifying the configuration on three sites to obtain a different configuration with the same quantum numbers, requires either the use of a mixer; or grouping two consecutive sites (see e.g.,~\cite{Tenpy}), which effectively increases the local dimension to $ n_{\textrm{max}}^2$, slowing down the simulations. While both approaches can be applied for commensurate symmetries, we find the latter to be better suited for incommensurate ones.  

Fig.~\ref{fig:Fig1} shows numerical results using infinite DMRG on a unit cell of size $\frac{2\pi}{k^*}=6$~\cite{iDMRG}. If $J/U\ll 1$, density fluctuations ---associated to $\hat{\mathcal{Q}}^{A,B}$--- are small and the system arranges itself in a gapped Mott insulator phase. 
These correspond to the lobes appearing for small $J/U$ in the upper panels of Fig.~\ref{fig:Fig1}. On the other hand, when $J/U\gg 1$ the system becomes gapless leading to large on-site particle numbers as well as large fluctuations (while phase fluctuations are small) and, as we show below, quasi-long range order. 
Results on different columns correspond to two possible choices of chemical potentials: (1) a modulated chemical potential of the form $\mu_A \hat{\mathcal{Q}}^{(1,0)}+\mu_B \hat{\mathcal{Q}}^{(0,1)}$ which explicitly breaks translation symmetry (panels a and c)~\footnote{We consider $\mu_A, \mu_B\geq 0$ as inverting their sign simply leads to exchanging the signs of $\alpha^A_j$ on the two non-trivial sublattices.}; and (2) a more standard uniform one $-\mu \sum_j \nh_j$, even though the Hamiltonian does not conserve particle number (panels b and d). 
To plot Fig.~\ref{fig:Fig1}(a) we obtained the ground state $\ket{\psi(J,\mu_A)}$ for several values of $J/U$ and $\mu_A/U$ and analytically identified the existence of Mott insulators along the $J=0$ line. We then located the lobes ---including the Mott insulating and the QLRO$_1$ phases---by calculating $\langle\psi(J,\mu_A)|\mathcal{Q}_{A,B}|\psi(J,\mu_A)\rangle$. For the former, both $\mathcal{Q}_{A}$ and $\mathcal{Q}_{B}$ take integer values per unit cell signifying incompressibility, while for the latter $\mathcal{Q}_{A}$ takes integer values per unit cell but $\mathcal{Q}_{B}$ varies continuously signifying the corresponding compressibility. Figure~\ref{fig:Fig1}(a) displays the boundaries among these regions.
On the other hand, in the setting of Fig.~\ref{fig:Fig1}(b) with $\mu_A = \mu_B = 0$ due to the choice of unit cell size in iDMRG ($L=6$ compatible with the symmetries) and translation invariance, all lobes have $\mathcal{Q}_A=\mathcal{Q}_B=0$, and hence cannot be distinguished by the conserved charges. However, we could analytically predict the existence of two different particle configurations when $J/U=0$ which---being gapped---are expected to survive to a finite value of $J/U$. To map the phase diagram we obtain the dependence of the correlation length (easily extracted in iDMRG from the second largest eigenvalue of the transfer matrix) on $J/U$ and $\mu/U$, which saturates in bond dimension to a finite value within the Mott insulators, while diverges in the QLRO phase. Figure~\ref{fig:Fig1}(b) just shows the boundaries between large and small correlation length.

In the following, we start characterizing the gaped phases by taking the strongly interacting regime $J/U \ll 1$. When $J/U = 0$, the Hamiltonian of interest becomes $\hat{H}_{\textrm{M.I.}}=\frac{U}{2}\sum_j (n_j)^2 -\mu \sum_j \nh_j + \mu_A \hat{\mathcal{Q}}^{(1,0)} + \mu_B \hat{\mathcal{Q}}^{(0,1)}$, that is diagonal in the occupation basis and has a finite gap except at special values of the parameters corresponding to the transitions between different lobes.
This allows us to perform a similar analysis to the standard BHM~\cite{PhysRevB.40.546}.

\subsubsection{Modulated chemical potential}
We start by fixing $\mu=0$ and simplifying our analysis by considering $\mu_B=0$. Then we are looking for the ground state of the gapped diagonal Hamiltonian $\hat{H}_{\textrm{M.I.}}=\frac{U}{2}\sum_j (\hat{n}_j)^2 + \mu_A\sum_j \alpha^A_j \hat{n}_j$. Since $\alpha^{(1,0)}_j=0,\pm 1$ is $6$-periodic, the ground state acquires a charge arrangement with period $6$ in the presence of on-site interactions, leading to a Mott-insulating state. Different lobes in Fig.~\ref{fig:Fig1}(a) are distinguished by the two quantum numbers $(Q_A, Q_B)$ and by different particle configurations that depend on the strength (and range) of interactions and the value of the chemical potential. To find the specific particle ordering, we require the energy of a given site to attain its minimum value. 
Focusing on the regime $\mu_A/U<1$, one finds~\footnote{The same analysis can be extended to larger values of $\mu_A/U$.}: The system orders in a charge configuration with an empty $6$-sites unit cell $|0\rangle =|\circ \circ \circ \circ \circ \circ  \rangle $ if $\mu_A/U<1/2$. On the other hand, for $\mu_A/U>1/2$ the unit cell configuration is given by $|2\rangle =|\circ \circ \bullet \bullet \circ \circ\rangle $ corresponding to the charges $(Q_A, Q_B)=(-2,1)$ per unit cell. Fig.~\ref{fig:Fig1}(a) shows that these lobes are robust to a finite value of $J/U \lesssim 0.3$, consistent with having a finite gap. At the transition point $\mu_A/U=1/2$, these two unit cell configurations are degenerate together with two more charge orderings $|L\rangle = |\circ \circ \bullet \circ \circ \circ  \rangle $ and $|R\rangle = |\circ \circ \circ \bullet \circ \circ  \rangle $, sharing the same $Q_A=-1$ but having different $Q_B=1$ or $0$ respectively.

For small but finite $J/U$, this degeneracy is partially broken at second order leading to the two different Mott insulators, but also to a quasi-long range order intermediate phase (QLRO$_1$ in Fig.~\ref{fig:Fig1}(a)) protected by bond-inversion symmetry $I_{\textrm{bond}}$ around a bond center between sites $6n+3$ and $6n+4$.
As shown in Fig.~\ref{fig:Fig1}(c), we find a central charge $c=1$ within this phase for $\mu_A/U=1$ and for a finite region of $J/U$. 
We numerically obtain it as the coefficient of the bipartite entanglement entropy $S_{\chi}$ with the finite correlation length $\xi_\chi$ induced by a finite bond dimension $\chi$ in infinite DMRG. This scales as $S_\chi=\frac{c}{6}\log(\frac{\xi_\chi}{a})$ for conformally invariant critical systems, as shown in Ref.~\onlinecite{Pollmann_2009}. The observed behavior can be perturbatively understood in the subspace spanned by $\ket{L},\ket{R}$ within each unit cell as we show in App.~\ref{app:QLRO_1}. 
Nonetheless, we found this phase disappears when considering a finite chemical potential $\mu_B$, which breaks bond inversion symmetry.

\subsubsection{Uniform chemical potential}
On the other hand when $\mu_A=0$ but for finite $\mu$, one finds the same charge ordering as for the standard Bose-Hubbard model if $J/U=0$. Being gapped, these phases are expected to be robust to small but finite values of $J/U$. However, an infinitesimal small $J/U$ now preserves a finite Fourier momentum of the particle number, while breaking particle number conservation. For $\mu/U<1.25$ we now observe two different charge orderings: A trivial Mott insulator with zero average density for $\mu/U<1/2$, while $\langle \nh_j\rangle\approx 1$ for $1/2<\mu/U<1.25$.
Both orderings correspond to global quantum numbers $Q_A=Q_B=0$, and hence one cannot directly say whether these two apparently different lobes correspond to the same phase of matter. Nonetheless, we notice that by considering a system size (or unit cell in the iDMRG simulations) that is not a multiple of $\frac{2\pi}{k^*}=6$, these two lobes acquire different quantum numbers $Q_A, Q_B$, then allowing us to distinguish them. Whether these correspond to different (crystalline) symmetry protected topological phases, or rather to symmetry protected trivial ones is left as an open question.

\begin{figure}[h!]
    \centering
    \includegraphics[width=1.02\linewidth]{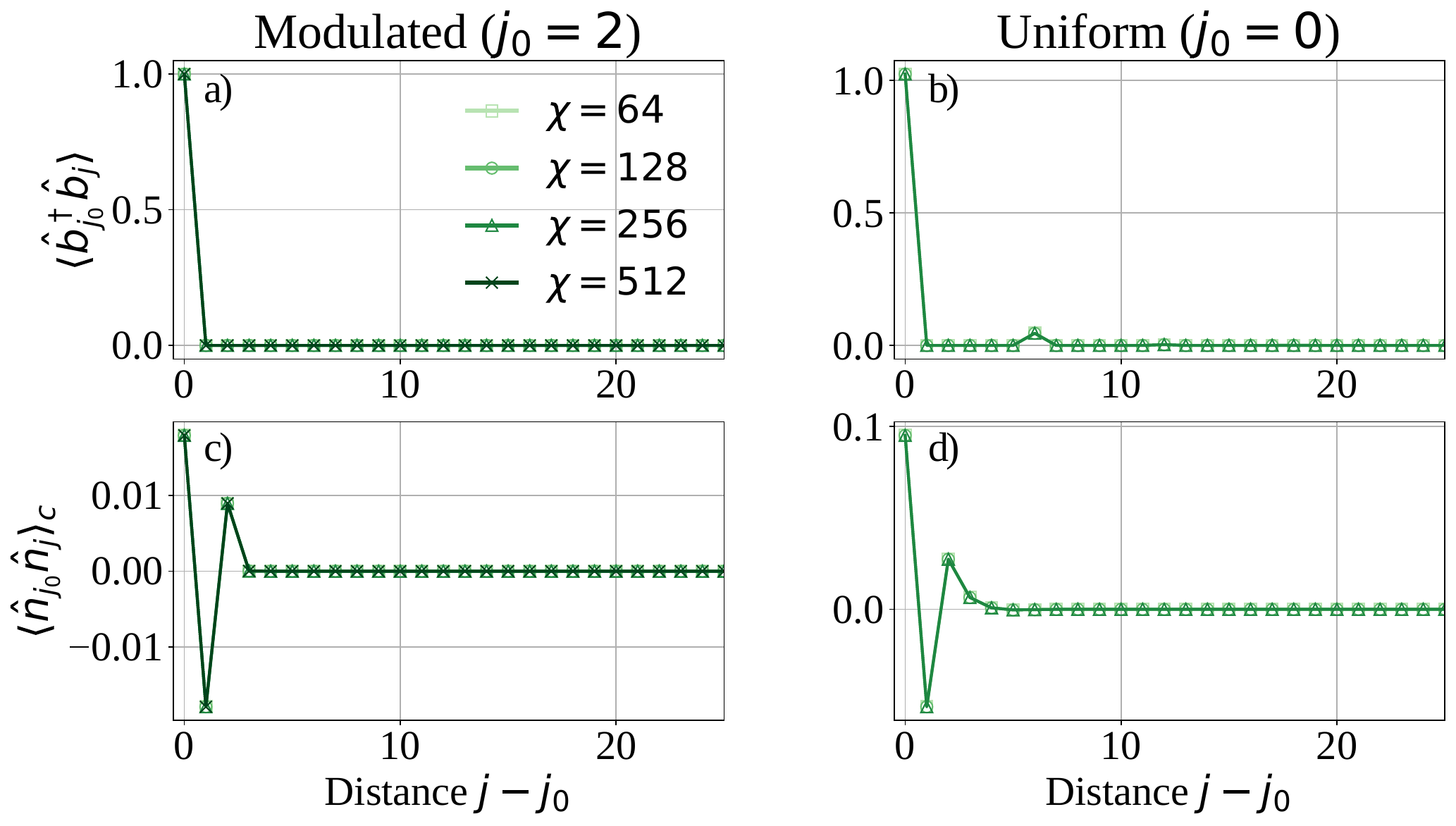}
    \caption{\textbf{Two-point correlations for Mott phases.} Decay of two-point correlations within the gapped phase for $J/U=0.1$ ($\blacksquare$ in Fig.~\ref{fig:Fig1} ) in the presence of modulated $\mu_A/U=1$ (panels a and c) and uniform $\mu/U=1$ (panels b and d) chemical potentials. First and second rows respectively show the decay of $\langle \bh_0^\dagger \bh_j\rangle$ and $\langle \nh_0 \nh_j\rangle$ with distance. }
    \label{fig:Fig2_Mott}
\end{figure}

Overall, for either modulated or uniform chemical potential a finite gap leads to exponentially decaying correlations. Moreover, due to the non-trivial spatial structure of $\hat{\mathcal{Q}}^{A,B}$, the operator $\bh_i \bh_j^\dagger$ is only charge neutral when $\alpha_i=\alpha_j$ for both $\alpha_j^A$ and $\alpha_j^B$, which for $p=q=1$ happens when $i-j\in 6\mathbb{Z}$, leading to vanishing correlations $\langle \bh^\dagger_i \bh_j\rangle$ everywhere else:
\begin{equation}
    \langle \bh^\dagger_i \bh_j\rangle = \begin{cases}
        e^{-|i-j|/\xi} \text{ if } |i-j|=6\times\text{integer}, \\[1ex]
        0 \text{ else,}
    \end{cases}
\end{equation}
where $\xi$ is the correlation length induced by the finite gap. 
This agrees with the numerical results we show in Fig.~\ref{fig:Fig2_Mott}(a,b) for modulated and uniform chemical potentials respectively. Density-density correlators $\langle \nh_i \nh_j\rangle$ also decay exponentially, and being charge neutral do not vanish for any particular distance. Nonetheless, these are dressed with periodic oscillations with wave vector $k^*$ for both modulated (Fig.~\ref{fig:Fig2_Mott}c) and uniform (Fig.~\ref{fig:Fig2_Mott}d) chemical potentials leading to $\langle \nh_i \nh_j\rangle\sim \cos(k^*(j-i))e^{-|i-j|/\xi_n}$. The oscillatory factor accompanying the exponential decay appears as a result of lowest gapped modes close $\pm k^*$.

\subsection{Quasi-long range order} \label{sec:QLRO_comm}
When the hopping amplitude becomes sufficiently large $J/U\gg 1$, the system becomes gapless and acquires QLRO (different from the one we found between two Mott lobes with $c=1$), with the quantum numbers $(Q_A, Q_B)$ varying within the phase. A clear characterization is shown in the lower panels of Fig.~\ref{fig:Fig1}, displaying the value of the central charge $c$ as a function of $J/U$. For large $J/U$ we find $c=2$, corresponding to two low-lying bosonic degrees of freedom at zero temperature and consistent with our preliminary analysis of the rotor model in Sec.~\ref{sec:rotor}. In this regime,
and for the purposes of a rough qualitative description rather than a precise mathematical mapping, the local particle number $\nh_j$ fluctuates around an average value  $\bar{n}_j$. To leading order we can substitute the bosonic degrees of freedom by a rotor variable $\bh_j\approx \sqrt{\bar{n}_j}e^{i\thet_j}$ (and $\bh_j^\dagger\approx \sqrt{\bar{n}_j}e^{-i\thet_j}$) such that fluctuations around this average value are captured by $\nh_j - \bar{n}_j$, leading to a natural description of the low-energy physics.
In the presence of the explicit translation symmetry-breaking terms $\mu_A \hat{\mathcal{Q}}^{(1,0)} + \mu_B \hat{\mathcal{Q}}^{(0,1)}$, the average density $\bar{n}_j$ is not uniform but just periodic modulo 6, which leads to space dependent rotor couplings $J_j$ in Eq.~\eqref{eq:Hrotor_qp}.

In the case of sublattice U$(1)$ symmetries ---as appearing for commensurate $k^*$---  the (two) relevant low-energy degrees of freedom can be identified decomposing the phase variable as $\thet_j=\alpha^{(1,0)}_j \hat{\vartheta}^A(x_j)+\alpha^{(0,1)}_j \hat{\vartheta}^B(x_j) $ where the equal sign is understood as retaining only low-energy modes. Then the action of each of the two continuous symmetry transformations is realized at low energies by a uniform shift of either $\hat{\vartheta}^A(x_j)$ or $\hat{\vartheta}^B(x_j)$, i.e., two conventional U$(1)$ symmetries. In App.~\ref{app:comm_approach} we construct an effective description utilizing this decomposition. In the following, we instead use a more general approach that will also prove to be useful for incommensurate charges.

\subsection{Villain action} \label{sec:Villain_11}
The main idea of the Villain formulation is to replace the cosine potential-term in the rotor Hamiltonian \eqref{eq:Hrotor_qp} without losing the $2\pi$-periodicity that is so relevant to characterize the QLRO to Mott insulator transition by the unbinding of vortices~\cite{Villain_75}. Proceeding in the standard way (see App.~\ref{app:derivation}) and neglecting boundary corrections expected to be irrelevant in the thermodynamic limit, the path integral partition function of the corresponding rotor Hamiltonian for any pair $q,p$ is given by $\mathcal{Z}(\beta)=\sum_{\{\mathcal{J}_\tau ,\mathcal{J}_x \in \mathbb{Z}\}} e^{-S[\mathcal{J}_\tau,\mathcal{J}_x]}$ where the action reads
\begin{align} \label{eq:S_11}
S[\mathcal{J}_\tau,\mathcal{J}_x]= \frac{1}{2}\sum_{j,\tau}\left(\frac{\mathcal{J}_x^2(j,\tau)}{\delta \tau J_{j}}+U\delta \tau\mathcal{J}_\tau^{2}(j, \bar{\tau})\right).
\end{align}
Here, $\mathcal{J}_\tau(j,\bar{\tau})=n_j(\bar{\tau})$ is the local density in the path integral defined on vertical links $\bar{\tau}\equiv\tau + \delta \tau/2$, and $\mathcal{J}_x(j,\tau)$ defined on sites of a $L\times L_\tau$ 2D lattice, where $\beta=L_\tau \delta \tau$ in the limit $\beta\to\infty$. 
Moreover, the configurations are constrained to satisfy 
\begin{equation} \label{eq:con_eq}
\nabla_\tau \mathcal{J}_\tau + \nabla^{q,p}_x \mathcal{J}_x =0
\end{equation}
around every site with coordinates $(j,\tau)$.
Considering a system with OBC, this constraint can be directly incorporated using the auxiliary height field $X_j( \bar{\tau})$ living on vertical links via
\begin{equation} \label{eq:cont_eq}
\mathcal{J}_x(j,\tau)=\nabla_\tau X_j( \bar{\tau}), \quad 
\mathcal{J}_\tau(j, \bar{\tau})=-\nabla^{q,p}_x X_{j}( \bar{\tau}) ~.
\end{equation}
Notice that in principle, one should consider the general solution $\mathcal{J}_x(j,\tau)=\nabla_\tau X_j( \bar{\tau}) + \bar{\mathcal{J}}_x(j,\tau)$, $\mathcal{J}_\tau(j, \bar{\tau})=-\nabla^{q,p}_x X_{j}( \bar{\tau})+ \bar{\mathcal{J}}_\tau(j,\tau)$ for not simply connected space-time manifolds, as it happens when taking periodic boundary conditions. These contributions fix the global symmetry sector since $Q_{A,B}=\sum_j \alpha^{A,B}_j n_j = \sum_j \alpha^{A,B}_j \bar{\mathcal{J}}_\tau(j,\tau)$. Hence, the background fields $\bar{\mathcal{J}}_x,\bar{\mathcal{J}}_\tau $ can be taken care of by fixing the global charges. A complementary description in terms of the winding number of a related stat-mech problem will be presented in Sec.~\ref{sec:stat_mech}. Assuming $\bar{\mathcal{J}}_x,\bar{\mathcal{J}}_\tau =0 $, i.e., $Q_A=Q_B=0$, one obtains
\begin{equation} \label{eq:X_j}
X_j( \bar{\tau})=\frac{1}{q}\sum_{i< j}\alpha_{j-i}^{(1,p/q)}n_i( \bar{\tau})
\end{equation} 
which explicitly shows that the field $X$ is integer valued when $q=p=1$. A more amenable description to pursue an analytical treatment follows by softening the field $X\in \mathbb{Z}\to \chi \in \mathbb{R}$ once including the potential term $-\lambda \sum_{j,\tau} \cos[2\pi\chi_j(\bar{\tau})]$. All together, plugging into Eq.~\eqref{eq:S_11} one finds a lattice action $S_{\textrm{lattice}}[\chi]=S_0[\chi_j(\bar{\tau})]-\lambda \sum_{j,\tau} \cos[2\pi\chi_j(\bar{\tau})]$ with the quadratic contribution $S_0$ given by
\begin{equation} \label{eq:S_11_soft}
\begin{aligned} 
S_{0}[\chi]&= \frac{1}{2}\sum_{j,\tau}\left(\frac{1}{\delta \tau J_{j}}(\nabla_\tau \chi_j(\bar{\tau}))^2+U\delta \tau(\nabla_x^{1,1}\chi_j(\bar{\tau}))^{2}\right).
\end{aligned}
\end{equation}
The term $\cos[2\pi\chi_j(\bar{\tau})]$ corresponds to a vortex creation/annihilation operator $e^{\pm i2\pi\chi_j(\bar{\tau})}$ expected to drive the QLRO to MI transition at fixed $Q_A, Q_B$ as for the standard BHM. However, unlike in that case, it does not insert a $2\pi$-phase shift in the boson phase $e^{i\thet_j}$ to the left of site $j$, but only at certain sites. 
A more detailed discussion of vortex degrees of freedom and their role to drive the transition between the two phases will be given in Sec.~\ref{sec:vortex}.
When $J_j=J$ is uniform ---as happens for a uniform chemical potential--- or when $J_j$ is approximately constant only varying over long wave-lengths, the quadratic action $S_0$ reads 
\begin{equation} \label{eq:S0UV_kw_main}
   S_0[\chi]=\frac{1}{2\beta L}\sum_{\omega_n,k}\left[  K_\tau\omega_n^2 + K_x(p-2q\cos(k))^2\right]|\chi_{(\omega_n,k)}|^2, 
\end{equation}
in (imaginary) frequency-momentum space, where we have introduced the parameters $K_\tau=\frac{1}{\delta \tau J}$, $K_x=U\delta \tau$.

\subsubsection{Continuum limit: Two species Luttinger liquid} \label{sec:conti_comm}
Eq.~\eqref{eq:S0UV_kw_main} shows that the energy dispersion has two zero-energy modes precisely at $\pm k^*$, corresponding to the wave-number of the modulated symmetries. Hence, using the low-energy expansion 
\begin{equation} \label{eq:chi_decomp}
\begin{aligned}
     \chi_j(\bar{\tau})  &=2\textrm{Re}[e^{ik^*j}(\varphi^1(x_j,{\tau}) + \varphi^2(x_j,{\tau}))],
\end{aligned}
\end{equation}
and keeping only non-oscillating contributions to derive an effective continuum description one finds
\begin{equation} \label{eq:S0_comm}
S_0[\varphi^a]= \frac{K}{2}\sum_{a=1,2} \int\,dxd\tau \left[\left(\partial_{\tau}\varphi^a\right)^2+ \left(\partial_{x}\varphi^a\right)^2 \right],
\end{equation}
after rescaling both space and time, with Luttinger parameter $K \propto \sqrt{K_\tau K_x} = \sqrt{U/J}$.
Namely, the QLRO phase corresponds to two (decoupled) Luttinger liquids, leading to the observed $c=2$ central charge shown in the lower panels of Fig.~\ref{fig:Fig1}~\footnote{We believe the data point indicating a central charge larger than $2$ is a numerical artifact appearing at the transition point. When extracting the central charge from fitting $S_\chi$ versus $\ln(\xi_\chi )$ for $J/U = 0.15$,
we obtain a $c > 2$, which decreases when considering all but the last data point. While we expect that the $c = 2$ of the QLRO also holds at the critical point, we leave the characterization of the phase transition between the Mott insulating and QLRO phases for future work.}. 

As for the lattice action in the previous section, this continuum quadratic description is valid as long as the compactness of the bosonic phase $\thet_j$ is irrelevant, which at the lattice level is controlled by the potential term $-\lambda \sum_{j,\tau} \cos[2\pi\chi_j(\bar{\tau})]$. Similar terms can appear in the continuum description driving the system unstable for sufficiently large $U/J$. To do so, we need to understand how the microscopic symmetries of the system are realized by the low-energy modes $\varphi^1, \varphi^2$ defined via Eq.~\eqref{eq:chi_decomp}, allowing for all those compatible cosine-contributions.
Let us start analyzing the case of commensurate $k^*$ with uniform chemical potential. With PBC, which is actually the case in the implementation of infinite DMRG, the system is invariant under translations by one lattice site $T_1$. Since $n_j= -\tilde{\nabla}^{q,p}_{x}\chi_j$ and $n_j\to n_{j+1}$ under this translations, we find that $T_1$ is realized at the infra-red via
\begin{equation}  \label{eq:bond_inv}
    \left(\begin{array}{c}
         \varphi^1\\
         \varphi^2
    \end{array} \right)\to  U_1 \left(\begin{array}{c}
         \varphi^1 \\
         \varphi^2
    \end{array} \right),
\end{equation}  
 where
\begin{equation} \label{eq:U_L}
   U_1\equiv \left(\begin{array}{cc}
         \cos(k^*) & -\sin(k^*) \\
         \sin(k^*)  & \cos(k^*) 
    \end{array} \right)
\end{equation}
is an orthogonal matrix. Defining $\bm{\varphi}\equiv (\varphi^1, \varphi^2)^T$, only the quadratic term $(\partial_x \varphi^1)^2+(\partial_x \varphi^2)^2$ is allowed, i.e., the Luttinger parameter $K$ is the same for both components $\varphi^1, \varphi^2$ which is consistent with a direct transition from a $c=2$ QLRO phase and a Mott insulator as shown in Fig.~\ref{fig:Fig1}(b). In fact, we can find cosine contributions whose arguments are linear in $\varphi^1, \varphi^2$, i.e., taking the form $\bm{\varphi}^T\cdot \bm{w}$, with $\bm{w}\in \mathbb{R}^2$. First, we can discard single cosine contributions of the form $\cos(2\pi\bm{\varphi}^T\cdot \bm{w} )$ since under  $T_1$ this term transforms as $\cos(2\pi\bm{\varphi}^T\cdot \bm{w} )\to \cos(2\pi\bm{\varphi}^T\cdot U_1^T\bm{w} )$, and $U_1$ has eigenvalues different from $\pm 1$. A natural next step is considering a finite linear combination of such cosine terms of the form $\sum_{n=1}^N\lambda_n \cos(2\pi\bm{\varphi}^T\cdot \bm{w}_n)$,
such that the sum remains invariant. This can only happen if $U_1^T\bm{w}_n=\pm \bm{w}_{n+1}$, and for finite $N$ this implies
$(U_1^T)^N\bm{w}_n=\pm \bm{w}_n$ for all $n$. Since $U_1^N$ has eigenvalues equal to $e^{\pm i Nk^*}$, one finds that for commensurate $k^*$ one just requires $N=\frac{\pi}{k^*}$ terms, i.e., $N=3$ terms for $k^*=\frac{\pi}{3}$. A possible choice of these terms is $\bm{w}_1=(1,0)^T, \bm{w}_2=(\cos(k^*), \sin(k^*))^T$ and $\bm{w}_3=(\cos(2k^*), \sin(2k^*))^T$. This analysis shows that a BKT-type transition is possible.

Let us now consider the case of modulated chemical potential $+\mu_A\hat{\mathcal{Q}}^{(1,0)}$ and a system size $L$ being a multiple of $\frac{2\pi}{k^*}$. In this case the coupling terms in the rotor Hamiltonian \eqref{eq:Hrotor_qp} are not uniform but rather $J_j\propto \sqrt{\bar{n}_{j-1} \bar{n}_j \bar{n}_{j+1}}$ acquire a similar modulation than that of the chemical potential. Hence, when expanding $\chi_j$ as in Eq.~\eqref{eq:chi_decomp}, the continuous action can pick up additional contributions  including cross terms of the form $\partial_{x,\tau}\varphi^1\partial_{x,\tau}\varphi^2$ between $\varphi^1,\varphi^2$. As before, we need to allow all those contributions that are compatible with the symmetries of the system. In particular, bond inversion $I_{\textrm{bond}}$ is the only relevant symmetry that acts non-trivially on the field $\chi_j$~\footnote{Notice that translations by $\frac{2\pi}{k^*}$ imposes no restrictions.}. Once again since $n_j= -\tilde{\nabla}^{q,p}_{x}\chi_j$ and $n_j\to n_{L-j}$ under this transformation, we find that $I_{\textrm{bond}}$ is realized at the infra-red via $\bm{\varphi}(x)
    \to  U_L  \bm{\varphi}(L-x)$ with
\begin{equation}  
   U_L\equiv \left(\begin{array}{cc}
         \cos(k^*L) & -\sin(k^*L) \\
         \sin(k^*L)  & \cos(k^*L) 
    \end{array} \right).
\end{equation}
However, for $L$ a multiple of $2\pi/k^*$, $U_L$ becomes the identity matrix, and the symmetry is realized at low energies as $\varphi^a\to \varphi^a$ with $a=1,2$. Hence, the quadratic contribution to the action takes the general form  
\begin{equation}
    S_0[\varphi^a]=\int d\tau dx\,\sum_{a,b}\left(g^\tau_{ab}\partial_\tau \varphi^a\partial_\tau\varphi^b+g^x_{ab}\partial_x \varphi^a\partial_x\varphi^b\right).
\end{equation}
Here $\underline{g}^\tau, \underline{g}^x$ are $2$ by $2$ real matrices, corresponding to the non-vanishing coupling between the $\varphi^1, \varphi^2$ modes, and constrained by the symmetries of the system. Simultaneously diagonalizing~\cite{Lai2010} $\underline{g}^\tau, \underline{g}^x$, the uncoupled fields can now have different Luttinger parameters. Since in the absence of translation symmetry cosine contributions of the form $\sum_{a=1,2} \lambda_a \cos(2\pi \varphi^a)$ as well as $-\cos(2\pi(\varphi_1\pm \varphi_2))$, are compatible with the symmetry, each mode can be independently gapped out, which could explain the different direct transitions between the $c=2$ QLRO phase to both the $c=1$ phase as well as to the short-range Mott insulating phase in Fig.~\ref{fig:Fig1}(a). In particular, when $Q_A=Q_B=0$ ---relevant to address the transition between the trivial Mott insulator and the $c=2$ phase---, the action becomes translation invariant and hence, both modes $\varphi^{1,2}$ have identical Luttinger parameter becoming simultaneously gapped out. On the other hand, the transition from QLRO$_2$ to QLRO$_1$ occurs at finite $Q_A,Q_B\neq 0$ where a single mode can acquire a gap. Everywhere else in the phase diagrams of Fig.~\eqref{fig:Fig1}(a-b), we expect a commensurate-incommensurate transition as in the standard BHM.

\subsubsection{Two point correlations and RG flow} \label{sec:corr_RG}

\begin{figure}[h!]
    \centering
    \includegraphics[width=0.55\linewidth]{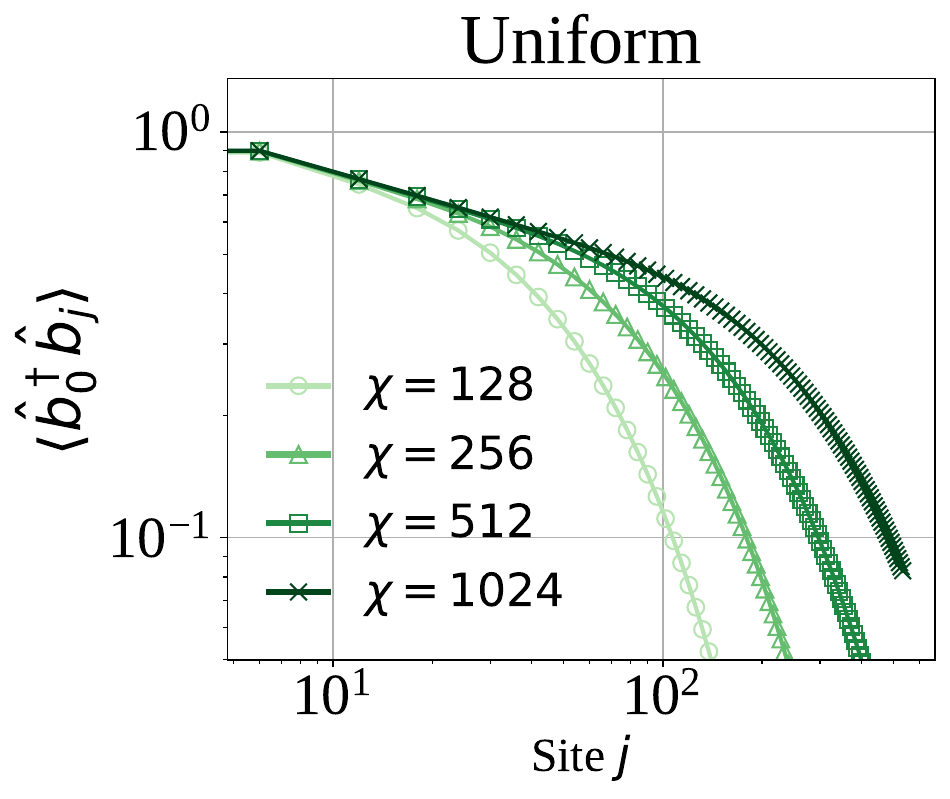}
    
    \caption{\textbf{Boson-boson correlator for QLRO phase with uniform chemical potential.} Decay of two-point correlations with the distance within the QLRO phase 
    in the presence of uniform chemical potential with $\mu/U=1$ and $J/U=0.6$. Convergence for increasing bond dimension $\chi$ tends to a power-law decay of $\langle \bh^\dagger_0 \bh_{j}\rangle$ when evaluating at sites $j=6\times \text{ integer}$. Data was obtained using iDMRG with fixed global charges $Q_A=Q_B=0.$
    }
    
\label{fig:Fig2_QLRO_bb}
\end{figure}

\begin{figure}[h!]
    \centering
    \includegraphics[width=1.0\linewidth]{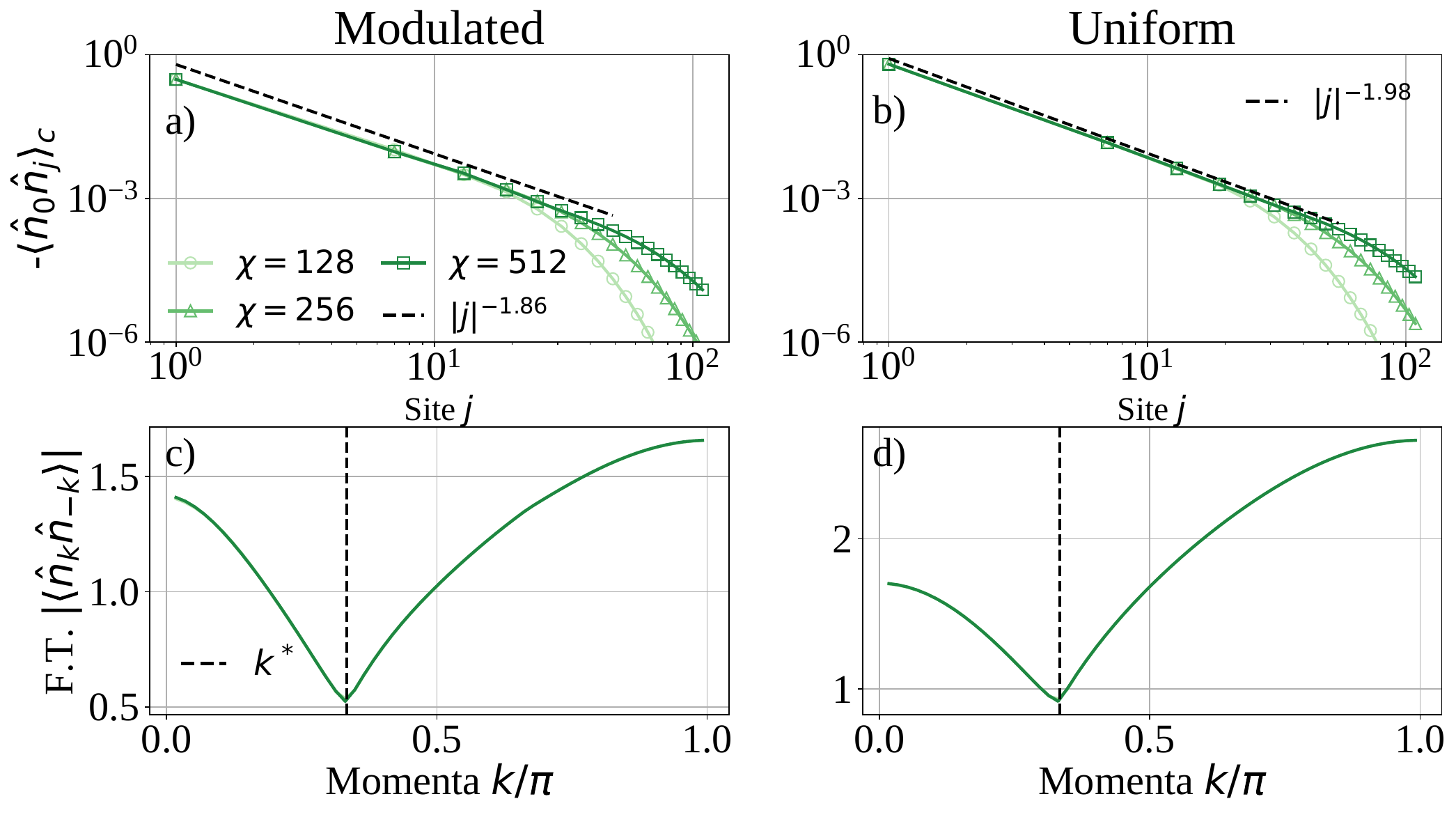}
    \caption{\textbf{Density-density correlator for QLRO phase.} Decay of two-point correlations with the distance for $J/U=0.6$ ($\bullet$ in Fig.~\ref{fig:Fig1} in the QLRO phase) in the presence of modulated $\mu_A/U=1$ (panels a and c) and uniform $\mu/U=1$ (panels b and d) chemical potentials. First row shows a power-law decay of $\langle \nh_0 \nh_{j}\rangle\sim -|j|^{-2}$ when evaluating at sites $j=6\times \text{ integer}$.
    Second-row panels show the amplitude of the Fourier transform of $\langle \nh_0 \nh_{j}\rangle$ as function of $k/\pi$. We find a V-shaped ($\sim|k\pm k^*|$) minimum at momenta $\pm k^*$, which is the expected momentum-space singularity corresponding to $1/r^2$ power law decay with oscillations with momentum $k^*$. Data was obtained using infinite DMRG without fixing global charges.}
\label{fig:Fig2_QLRO}
\end{figure}

The distinction between the Mott insulator and QLRO phases can also be detected by looking at the decay of symmetry invariant two-point correlations, which decay exponentially with the distance in the former case. To evaluate their dependence in the QLRO phase, we start by computing  $C_{\chi \chi}(j-j^\prime;\bar{\tau}-\bar{\tau}^\prime)=\langle e^{i\chi_j(\bar{\tau})}e^{-i\chi_{j^\prime}(\bar{\tau}^\prime)}\rangle$. Using that the action in Eq.~\eqref{eq:S_11_soft} is Gaussian, one obtains power-law decaying correlations (see details in App.~\ref{app:correlations})
\begin{equation} \label{eq:corr_chichi}
    C_{\chi \chi}(j-j^\prime, \bar{\tau} -\bar{\tau}^\prime)\sim  \begin{cases}
    |\bar{\tau}-\bar{\tau}^\prime|^{-C^\tau_{k^*}/K} \textrm{ if } j=j^\prime, \\[1ex]
     |j-j^\prime|^{-C^x_{k^*}/K} \textrm{ if } \tau=\tau^\prime,
    \end{cases}
\end{equation}
both in imaginary time and in space whenever $j-j^\prime$ is a multiple of $2\pi/k^*$ in the later case. Here, the power-law exponents depend on the Luttinger parameter $K$, and on the non-universal constants $C^{\tau}_{k^*}=(4\pi q|\sin(k^*)|)^{-1}$, $C^{x}_{k^*}=(2\pi q|\sin(k^*)|)^{-1}$. Similarly one finds~\footnote{This decays as correlators for $\chi_j$ but with $K\to 1/K$, after identifying $\bh_j\sim e^{i\thet_j}$. } that $\langle \bh^\dagger_0 \bh_j\rangle$ decays as a power-law when evaluated on $j=\frac{2\pi}{k^*}$, i.e., $j=6\times$integer for $q=p=1$. Results for a uniform chemical potential $\mu/U=1$ and $J/U=0.6$ are shown in Fig.~\ref{fig:Fig2_QLRO_bb}. These numerical results have been obtained within the global symmetry sector $Q_A=Q_B=0$.
 From here (or alternatively applying Wilson renormalization group approach as presented in App.~\ref{app:RG_wilson}) we conclude that the scaling dimension of the cosine contribution $\cos[2\pi p \chi ]$ is $\Delta_p\propto p^2/K$, and hence the cosine term in Eq.~\eqref{eq:S0_comm} becomes relevant when $U/J$ is sufficiently large. Therefore, we expect a quantum phase transition at a critical value $K_c$. From the duality mapping $n_j=-\nabla_x^{q,p} \chi_j$, one can then compute density-density correlations which decay as $\langle \nh_j \nh_{j^{\prime}}\rangle \sim -\cos(k^*|j-j^\prime|)/|j-j^\prime|^2$, similarly to as a Luttinger liquid~\cite{giamarchi2004quantum} times an oscillatory factor with momentum $k^*$, imprinted by the modulated symmetries. Its Fourier transform $\langle \hat{n}_k\hat{n}_{-k}\rangle$ then takes a V-shaped functional form $\langle \hat{n}_k\hat{n}_{-k}\rangle\sim |k\pm k^*|$ close to $\pm k^*$. Details on the computation can be found in App.~\ref{app:correlations}.
These dependencies are shown in Fig.~\ref{fig:Fig2_QLRO} for both modulated $\mu_A/U=1$ (panels a and c) and uniform $\mu/U=1$ (panels b and d) chemical potentials for $J/U=0.6$ (signaled with a $\bullet$ in the upper panels of Fig.~\ref{fig:Fig1}). Our numerical results are consistent with a power-law exponent close to $-2$, and oscillations with wave-vector $k^*=\frac{\pi}{3}$.

\subsection{Experimental realization: from many-body Aubry-Andr{\'e} to modulated symmetries}

Recent experimental works~\cite{Scherg_nature,PhysRevLett.130.010201}
have shown that one-dimensional dipole-conserving systems naturally emerge as approximate descriptions of interacting ultracold gases in the presence of a strong linear tilt potential, that is realized via a magnetic gradient~\cite{Scherg_nature,thomas_tilt}. In this section we show that using similar resources, one-dimensional models with quasi-periodic conserved quantities can be approximately realized considering modulated chemical potentials~\cite{MBAA_19}, although alternative approaches to engineer such systems can also exist, e.g., using superconducting circuits~\cite{Tortora_22}. 

Let us start considering systems with commensurate $k^*=\frac{\pi}{3}$ as discussed in the previous sections. Our goal is obtaining an effective Hamiltonian which commutes with both $\hat{\mathcal{Q}}^{(0,1)},\hat{\mathcal{Q}}^{(1,0)}$ 
properly normalized to have an integer spectrum. 
Similarly to the emergence of dipole conservation in tilted systems~\cite{Sala_PRX, Sanjay19,Scherg_nature}, the idea is coupling the system to a strong modulated chemical potential such that in the regime of this being large, the ensuing dynamics is governed by an effective Hamiltonian with quasi-periodic conserved quantities. In particular, we consider the bosonic Hamiltonian
\begin{equation} \label{eq:exp}
\begin{aligned}
	H_{\textrm{exp}} =& J \sum_j \left(\hat{a}_j^\dagger \hat{a}_{j+1}+\textrm{H.c.}\right) +\frac{U}{2}\sum_j (\nh_j)^2   \\
&+V\sum_j \nh_j \nh_{j+1}+ \Delta \left(\hat{\mathcal{Q}}^{(1,0)} + \beta \hat{\mathcal{Q}}^{(0,1)}\right),
\end{aligned}
\end{equation}
with $\beta$ a non-rational number $|\beta|<1$. The reason this is important, is to avoid any kind of resonance between  $\hat{\mathcal{Q}}^{(0,1)}$ and $\hat{\mathcal{Q}}^{(1,0)}$. This permits a direct application of the rigorous results obtained in Refs.~\onlinecite{de_roeck_very_2019,else_long-lived_2020}, which implies that in the limit $\Delta \gg J,U$, the dynamics (as well as the zero-temperature physics) is governed by an effective Hamiltonian $H_{\textrm{eff}}$ preserving both $\hat{\mathcal{Q}}^{(0,1)}$ and $\hat{\mathcal{Q}}^{(1,0)}$ for quasi-exponentially long times in $\Delta/J$. 

In fact, experimentally one does not need to tune two different (modulated) chemical potentials but just one~\footnote{We thank Monika Aidelsburger for this observation.} using that $\hat{\mathcal{Q}}^{(0,1)} + \beta \hat{\mathcal{Q}}^{(1,0)} =A\frac{2}{\sqrt{3}}\sum_j  \cos\left(\frac{\pi j}{3}+\varphi\right)\nh_j$
with $-A\sin(\varphi)=1+\frac{\beta}{2}$ and $A\cos(\varphi)=\frac{\sqrt{3}}{2}\beta$ for any random choice of $\varphi$ such that $\beta \not \in \mathbb{Q}$. The resulting Hamiltonian is known as many-body Aubry-Andr{\'e} model, extensively studied in the context of many-body localization~\cite{aubry1980analyticity,Iyer_2013,Schreiber_2015,MBAA_19}. 
For example, a spinful version with on-site interactions has been already experimentally realized with ultracold atoms, with a high level of control of the modulation of the chemical potential (here given by $k^*$) by tuning the ratio of primary
and detuning lattice wavelengths~\cite{MBAA_19}. Here, $\varphi$ corresponds to the relative phase difference between primary and detuning lattices.

Nonetheless, because the Hamiltonian in Eq.~\eqref{eq:exp} conserves the particle number, so does the effective one in the limit $\Delta \gg J,U$, apart from conserving $\hat{\mathcal{Q}}^{(0,1)}$ and $ \hat{\mathcal{Q}}^{(1,0)}$.
In fact, the most local off-diagonal terms include $a_j a_{j+6}^\dagger+ \textrm{H.c.}$, $\hat{a}_{j}\hat{a}_{j+1}^\dagger \hat{a}_{j+3}\hat{a}_{j+4}^\dagger + \textrm{H.c.}$ and $\hat{a}_{j}\hat{a}_{j+1}^\dagger \hat{a}_{j+2}\hat{a}_{j+3}^\dagger\hat{a}_{j+4}\hat{a}_{j+5}^\dagger + \textrm{H.c.}$ but not $\hat{a}_{j}\hat{a}_{j+1}^\dagger \hat{a}_{j+2} + \textrm{H.c.}$.
The resulting QLRO phase is hence different than the one studied in this work, which now includes an extra zero mode at $k=0$~\footnote{Recall also that by the proof in App.~\ref{app:longrang}, this effective Hamiltonian also conserves the $U_j$'s.}.
An alternative to get rid of this additional conservation is to start from an unperturbed Hamiltonian lacking any U$(1)$ symmetry. For example, one could consider a spin model with both longitudinal and transverse fields, with the latter being spatially modulated.
However, truncating to the first non-trivial off-diagonal contribution in perturbation theory, the resulting Hamiltonian will feature a much stronger fragmentation than the bosonic system.

While it is tempting to directly generalize our previous derivation to incommensurate $k^*$, neither the rigorous theory of prethermalization, nor that of local and non-local Schrieffer-Wolff perturbation theory~\cite{Bravyi_2011} appear to apply.
To invoke the latter results, which is less restrictive than the former as it focuses on particular low-energy physics, one  requires having a finite gap between the (degenerate) subspace $\Omega_0$ of interest (say the one with $Q_A=Q_B=0$) for which we want to find an effective description of, and the rest of the Hilbert space $\Omega_0^\perp$. However, as we found before this gap generically scales as $q^{-L}$ and hence it vanishes in the limit $L\to \infty$. While we cannot provide a rigorous formal derivation of the resulting ---if existent--- effective Hamiltonian, one can easily identify terms that commute with $\Delta\sum_j\cos(k^*j+\varphi)\nh_j$ as long as $\cos(k^*)=\frac{p}{2q}$ and which can be generated combining single-particle hoppings and interactions. Hence, this raises an interesting question: Can such terms be perturbatively generated in the regime of large $\Delta$, even though the interacting Aubry-Andre{\'e} model is expected to be in a many-body localized phase in that same regime? Can that results tell us something about the stability of the phase?
We leave addressing this question as future work.

\subsection{Close packed tiling problem on zig-zag stripe layers} \label{sec:stat_mech}
In this section, we provide an alternative view of the quantum rotor model in Eq.~\eqref{eq:Hrotor_qp} for $q=p=1$ by mapping it into a 2D statistical model.

\begin{figure}[h]
    \centering
    \includegraphics[width=0.4\textwidth]{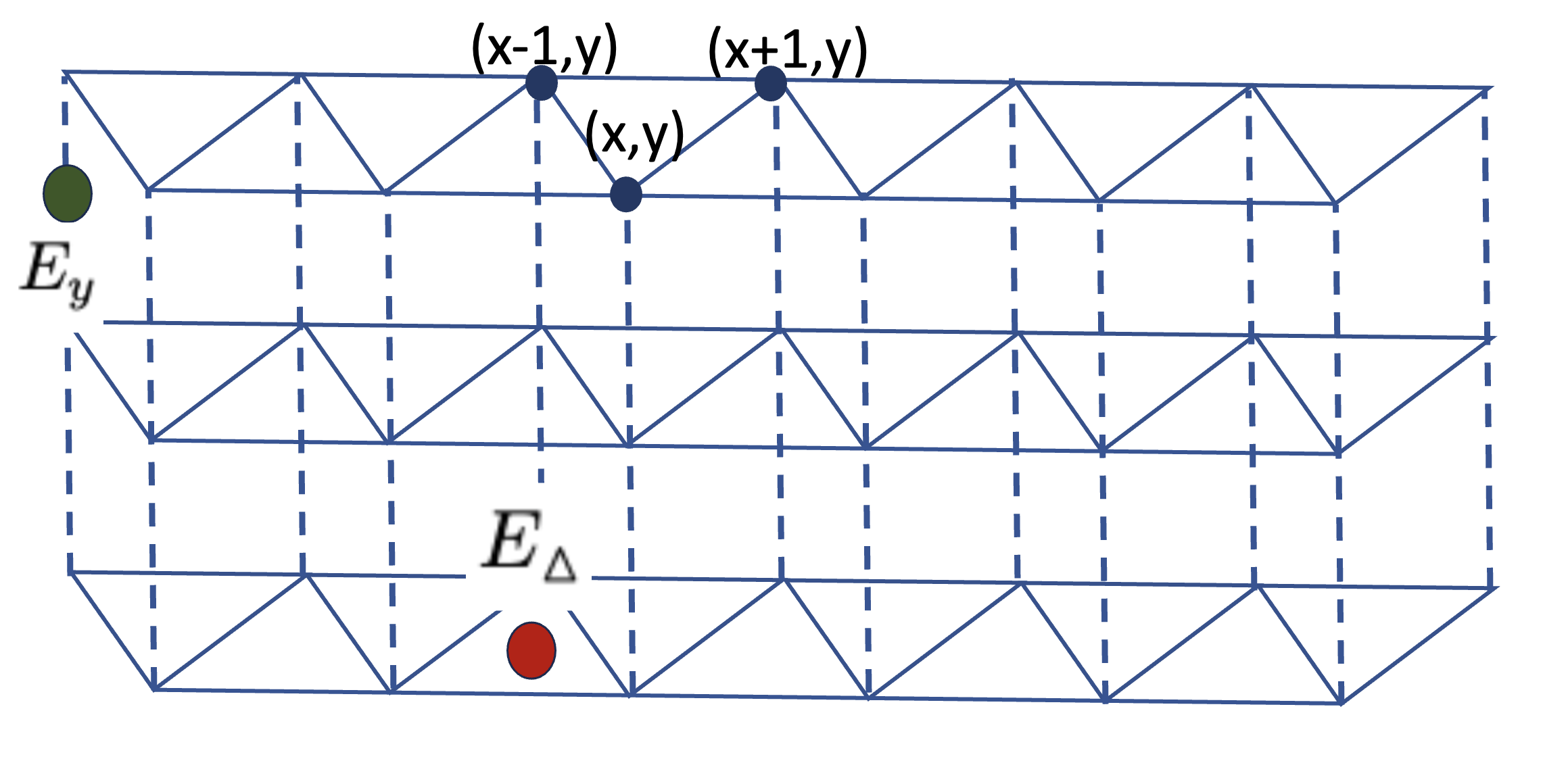}
    \caption{\textbf{2D statistical model on zig-zag stripe layers.}}
    \label{pack}
\end{figure}

\begin{figure}[h]
    \centering
    \includegraphics[width=0.45\textwidth]{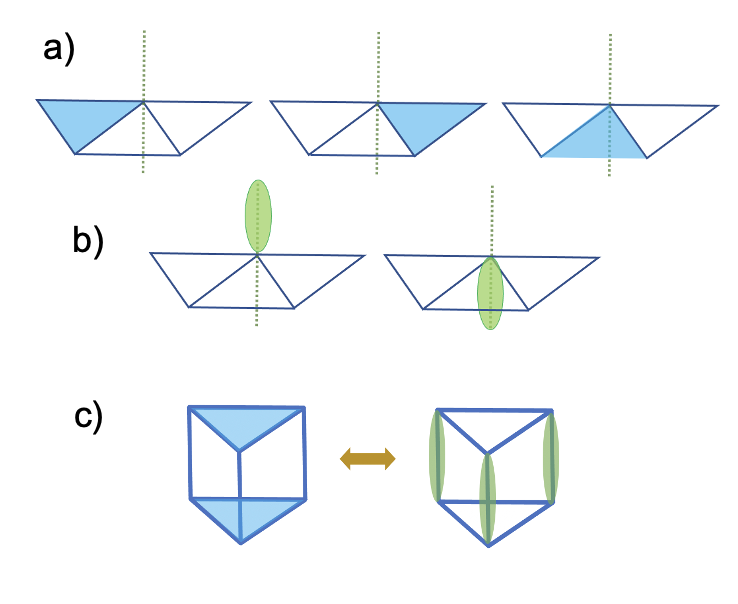}
    \caption{\textbf{Close-packed configurations.} At each site, the close-packed configuration restricts the site to be either connected (a) to one of the three trimers living on the plaquettes along the $x$-stripe (blue triangles); or (b) to be connected to the top/bottom layer via a dimer (green oval) lying along a $y$-link. (c) Local flipping between distinct close-packed patterns. }
    \label{gu}
\end{figure}

The system we are looking into is based on a close-packed trimer-dimer system defined on a zig-zag stripe layer shown in Fig.~\ref{pack}. At each site, the close-packed configuration restricts the site to be either connected to one of the three trimers (blue triangles) living on the plaquettes along the $x$-stripe (Fig.~\ref{gu}(a)); or to be connected to the top/bottom layer via a dimer (green ovals) lying along a $y$-link (Fig.~\ref{gu}(b)). 

\begin{figure}[h]
    \centering
    \includegraphics[width=0.45\textwidth]{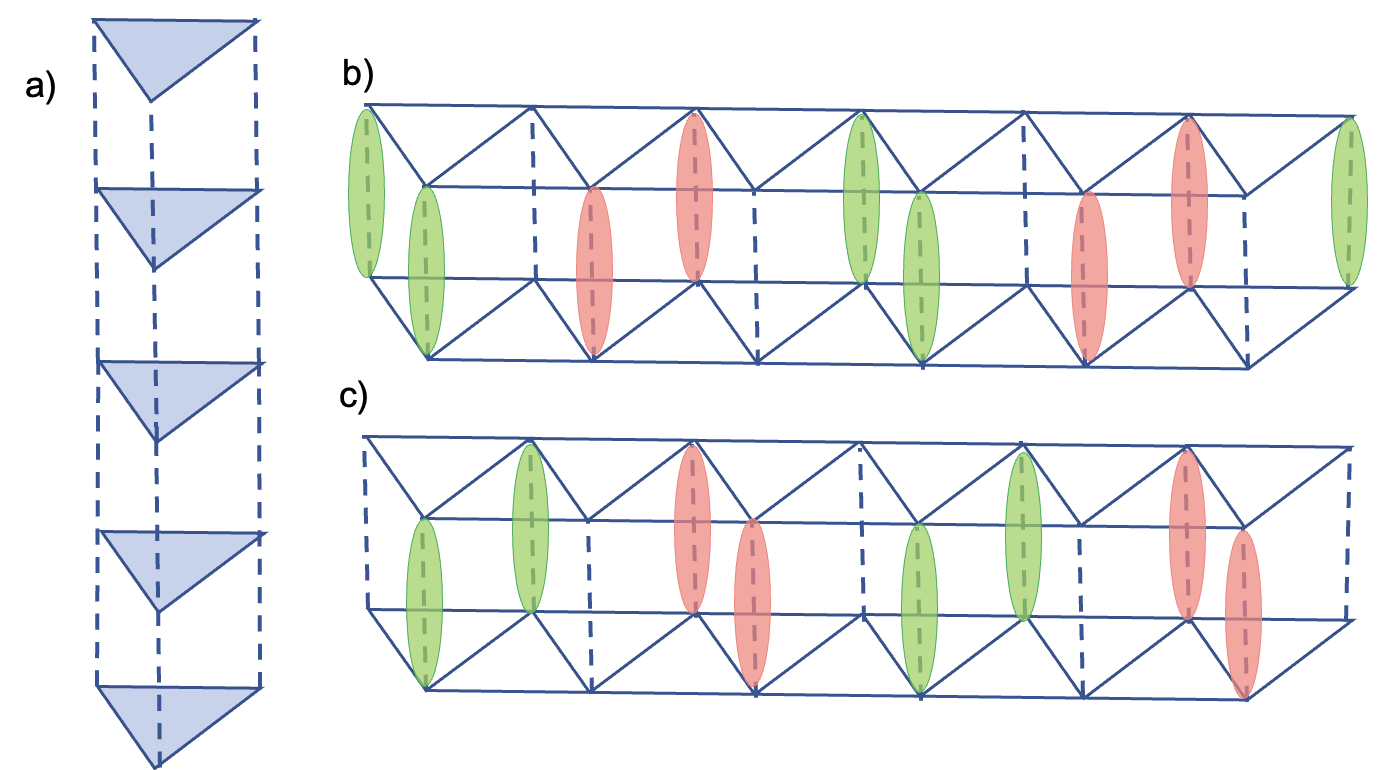}
    \caption{\textbf{Columnar phase and winding numbers.}(a) The winding number $m_y$ counts the number of trimers along each y-column. (b-c) The winding number $m_\vartriangle(y)=\sum \alpha(x) E_y(x,y)$ counts the number of dimers with a spatial modulation factor (illustrated as different colors, with $\alpha(x)=1$ for green, $\alpha(x)=-1$ for red and zero else wise) along each $x$-row. }
    \label{flux}
\end{figure}

To analyze the close-packed patterns, we account for the local dimer-trimer constraint by encoding the trimer and dimer coverage as a higher-rank electric field\cite{pretko2017subdimensionagl,bulmash2018higgs,vijay2016fracton,you2022fractonic},
\begin{align} 
E_{\vartriangle}=\eta T_{\vartriangle}, ~E_{y}=\eta D_{y}\ . \label{def}
\end{align}
$E_{\vartriangle}$ lives on the center of each triangle while $E_y$ lives on each y-link.
$T_{\vartriangle},D_y$ refer to the number of trimers and dimers living on the triangle and y-links, respectively. Here $\eta$ is the bipartite lattice factor with an alternating sign structure. Since we can uniquely associate each triangle  and $y$-link with a site of coordinates $\bm{r}=(x,y)$, we can define the bipartite lattice factor as $\eta=(-1)^{x+y}$. For example, we can associate to each triangle the coordinates of the left-most vertex, and the coordinates of the lowest site to each bond along the $y$ direction. Using this notation, the dimer-trimer constraint can be interpreted as the Gauss-law
\begin{align} \label{eq:cons}
&\nabla_x^{1,1} E_{\vartriangle}+\nabla_y E_{y}=\eta (1-Q)
\end{align}
where as before we define $\nabla_x^{1,1} E_{\vartriangle}(\bm{r})=-E_{\vartriangle}(\bm{r})+E_{\vartriangle}(\bm{r}+\hat{e}_x)+E_{\vartriangle}(\bm{r}-\hat{e}_x)$ along the $x$-stripe, and $\nabla_yE_y(\bm{r})=E_y(\bm{r})-E_y(\bm{r}-\hat{e}_y)$. This constraint resembles the continuity equation \eqref{eq:cont_eq} for the current fields in the Villain formulation of Sec.~\ref{sec:Villain_11} with $\mathcal{J}_x\to E_\vartriangle$, $\mathcal{J}_\tau\to E_y$. However, here a fixed pattern of background charge ($\eta$) has been introduced.
When considering close-packed configurations we choose $Q=0$, with the staggering background charge $\eta$ indicating each site is either connected with a dimer or trimer. To satisfy this local constraint, one can parameterize the electric field as
\begin{align} \label{hei}
E_{\vartriangle}=-\nabla_y X+\bar{E}_{\vartriangle},E_{y}=\nabla_x^{1,1} X+\bar{E}_{y},
\end{align}
Here $X$ is a discrete integer-valued field ---usually denoted height field--- living on the dual lattice at the center of each triangular prism, that characterizes the local fluctuations of the dimer-trimer pattern. This provides a solution of Eq.~\eqref{eq:cons} with vanishing right hand side. $\bar{E}_{y},\bar{E}_{\vartriangle}$ are background configurations that satisfy the inhomogeneous constraint Eq.~\eqref{eq:cons}, and can be chosen such that they are only a function of $y$ and $x$ coordinates respectively. For example, we can simply take the configuration $\bar{E}_{y}=0, \bar{E}_{\vartriangle}(x,y)=(2\cos(2\pi x/3)+1)/3$ for the trimer columnar phase, or $\bar{E}_{y}(x,y)=\eta(1+(-1)^y)/2, \bar{E}_{\vartriangle}=0$ for the dimer columnar phase. 
It is worth mentioning that a large number of distinct trimer-dimer configurations can be connected by changing the value of $X$ locally e.g., via the local process pictured in Fig.~\ref{gu}(c). In the meantime, the background patterns $\bar{E}_{y},\bar{E}_{\vartriangle}$ are responsible for the different global topological sectors (for a lattice with periodic boundary conditions) that cannot be connected by local $X$ fluctuations. We will return to this point shortly.

For the close packing problem $(Q=0)$, all allowed trimer/dimer configurations have equal Boltzmann weights. Since there are no energetic terms in the partition function, the free energy consists only of entropy. If we coarse-grained the $E_{\vartriangle},E_y$ field, flippable configurations with average $\bar{E}_{\vartriangle},\bar{E}_y=0$ correspond to a larger number of microscopic states, 
and hence to a larger coarse-grained entropy than any other non-flippable configuration with $\bar{E}_{\vartriangle},\bar{E}_y \neq 0$. 
Indeed, a flippable prism that can resonate between two trimers and three $y$-dimers (Fig.~\ref{gu}(c)) has zero average $E_{\vartriangle}, E_y$, so the coarse-grained energy should effectively favor such flippable patterns. 
This motivates the following ansatz for the height field partition function~\cite{Raghavan_1997}
\begin{equation} 
\begin{aligned} \label{th}
&\mathcal{Z}= \int \mathcal{D}E_y~ \mathcal{D}E_{\vartriangle}  e^{-\beta\sum_{\{\bm{r}\}}(E_y^2+E_{\vartriangle}^2)}\\
&=\int \mathcal{D} \bar{E}_y~ \mathcal{D}\bar{E}_{\vartriangle} e^{-\beta \sum_{\{\bm{r}\}}(\bar{E}_{\vartriangle}^2 + \bar{E}_{y}^2)} \\
&\times \int \mathcal{D}X~ e^{-\beta \sum_{\{\bm{r}\}}[(\nabla_y X)^2+(\nabla_x^{1,1} X)^2]-\lambda \sum_{\{\bm{r}\}}\cos(2\pi X)+...}\ ,
\end{aligned}
\end{equation}
where in the last equality we introduced the potential  term $\lambda \cos(2\pi X)$ that imposes the discreteness og $X$ energetically. 
By tuning $\beta$ (which is analogous to $K$ in Eq.~\eqref{eq:chi_decomp}), one changes the scaling dimension of $\cos(2\pi X)$ and drives a phase transition between a liquid phase with algebraic correlations for $X$, and an ordered phase with fixed values of $X$.
In fact, the effective action of the close pack trimer-dimer model exactly matches the dual theory of the hardcore boson model we found in Sec.~\ref{sec:Villain_11}.

\subsubsection{Topological sector and conserved quantities}

Now we scrutinize the \textit{topological sector} of close-packed configurations. While local fluctuations of the height field $X$ can change the local trimer-dimer pattern, these patterns inhabit distinct topological sectors\cite{moessner2010quantum} illustrated in Fig.~\ref{flux}, which can only be connected via non-local string-like updates, sometimes also known as large gauge transformations\cite{shirley2019fractional,xu2008resonating,you2022fractonic}.
We can define a topological winding number on a closed manifold of dimension $L_x\times L_y$ with periodic boundary conditions for $L_x$ a multiple of $6$ as
\begin{equation}
\begin{aligned} 
&m_y(x)=\sum_{y=1}^{L_y} E_{\vartriangle}(\bm{r}),\\
& m^A_\vartriangle(y)=\sum_{x=1}^{L_x}\alpha^A_xE_y(\bm{r}),\,\,m^B_\vartriangle(y)=\sum_{x=1}^{L_x}\alpha^B_xE_y(\bm{r}).
\end{aligned}
\end{equation}
with $\alpha^A_x, \alpha^B_x$ two linearly-independent solutions of Eq.~\eqref{eq:LinRec}.
Based on the sign structure defined in Eq.~\eqref{eq:cons}, 
$m_y$ encodes the total number of trimers (with a sign modulation given by $\eta$) on each column, while $m_{\vartriangle}^\alpha$ with $\alpha=A,B$; encode the number of dimers along each row with additional charge modulation (illustrated as different colors in Fig.~\ref{flux}).
These `winding numbers'' characterize the topological sectors, which cannot be changed by locally flipping among configurations. In fact, these quantities are independent of the height field $X$. For example, $m_y(x)=\sum_{y=1}^{L_y} \bar{E}_{\vartriangle}(\bm{r})$.
From the Gauss-law constraint in Eq.~\eqref{eq:cons}, it follows that
\begin{align} 
& \nabla^{1,1}_x m_y(x)=0 ,~\nabla_y m^\alpha_\vartriangle(y)=0,
\end{align}
when $Q=0$. 
This identity implies that the winding numbers of the electric field along each row and column are related. In particular, if we fix the value of $m_{y}(x)$ on two adjacent triangles on the $x$-stripe, all other topological sectors are fixed. This ``conserved quantity'' only appears due to the choice of boundary conditions in the $x$-direction, and does not relate to any physical symmetry of the original model \eqref{eq:H_11}. Likewise, the choices of $m^{A,B}_\vartriangle(y)$ are independent of coordinate $y$. In this case, these winding numbers correspond to the modulated conserved quantities $\hat{\mathcal{Q}}^{A,B}$ of the quantum model, with  $m^{A,B}_\vartriangle(y)$ counting the total charge of particles' worldlines crossing a given $y$-row.

To summarize, the close packed trimer-dimer models offer an alternative perspective on the commensurate $q=p=1$ modulated Bose Hubbard model in the imaginary time direction. This model provides insights into the transition from a liquid phase to a Mott phase, and this can be characterized by the disorder-order phase transition in the close-packed pattern.
One might consider whether it is possible to extend the close-packed formalism to other commensurate models or even to the incommensurate case. However, while close-packed patterns naturally hold when $p=1, q=1$, thereby allowing the effective ``Gauss law'' to manifest the commensurate modulated symmetry, this may not be easily extended to the general case where $p\neq 1$.

\section{Incommensurate charges} \label{sec:incomm} 
We will now discuss the case of incommensurate $k^*$ occurring whenever $\frac{p}{q}\notin \{0,\pm 1, \pm 2\}$. This implies that $q>1$ and as we found in Sec.~\ref{sec:frag}, it leads to the $L$ additional $\mathbb{Z}_q$ discrete symmetries $U_j$. In particular, we focus on the zero-temperature phase diagram of the bosonic Hamiltonian
\begin{equation} \label{eq:H_21}
	H_{q,p}=-J\sum_{j} \left(\hat{b}_{j-1}^q(\hat{b}^{\dagger}_{j})^p \hat{b}_{j+1}^q + \textrm{H.c.}\right) + U\sum_j \nh_j^{m_{q,p}},
\end{equation}
with $m_{q,p}=\lceil \frac{p+2q}{2}\rceil$, where $\lceil \cdot\rceil$ is defined to be the closest integer larger than the argument. E.g., $m_{2,1}=3$ for $q=2,p=1$. This unusual potential makes the system stable even in the regime of large $J/U$. Given the analysis in the previous sections, one anticipates finding short- and long-correlated phases by varying the ratio $J/U$. In the following, we provide numerical results supporting this expectation, preceded by an analytical characterization for a uniform chemical potential.
As we show in the following, the incommensurability  has consequences for the correlation functions as well as for more formal considerations when dealing with the Villain's action.

\subsection{Quasi-long range order phase}
Following a similar derivation as the one in Sec.~\ref{sec:comm}, one finds the action
\begin{equation} \label{eq:S_qp}
\begin{aligned} 
S_{0}[X]&= \frac{1}{2}\sum_{j,\tau}\left(K_\tau(\nabla_\tau X_j(\bar{\tau}))^2+K_x(\nabla_x^{q,p}X_j(\bar{\tau}))^{m_{q,p}}\right)
\end{aligned}
\end{equation}
in the regime where the approximation $\bh_j\sim \sqrt{\bar{n}_j}e^{i\thet_j}$ works.
Nonetheless, one cannot use Eq.~\eqref{eq:X_j} to show that $X_j(\bar{\tau})$ is integer-valued since in general $\alpha_j\notin \mathbb{Z}$. Hence, the compactness of $\theta_j$ does not appear to imply the discreteness of $X_j(\bar{\tau})$, naively preventing the appearance of a relevant cosine potential. 
This issue becomes explicit and can be addressed when recalling the existence of the discrete $U_j$ conserved quantities. Let us write $X_j(\bar{\tau})=I_j(\bar{\tau})+m_j$ with the condition $I_j(\bar{\tau})\in \mathbb{Z}$ and $|m_j|<1$. Then we can write $n_j(\bar{\tau})=-\nabla_x^{q,p}I_j(\bar{\tau})-\nabla_x^{q,p} m_j$, and using Eq.~\eqref{eq:X_j} we find $U_j\propto e^{i2\pi m_{j+1}}$ up to boundary conditions. Hence, the $m_j$'s are time-independent and fixed once the $U_j$'s conserved quantities are specified. In particular, if $U_j=1$ for all $j$, then $m_j=0$.
From here we conclude that one needs to fix the eigenvalues of all the $U_j$'s in order to write a well-defined action in the incommensurate case. 
Moreover, this also implies that the field that one needs to soften from integer to real-valued is $I_j(\bar{\tau})\in \mathbb{Z}\to \chi_j(\bar{\tau})\in \mathbb{R}$ by introducing the potential term $-\lambda \sum_{j,\tau} \cos(2\pi \chi_j(\bar{\tau}))$.
In the following section we provide a rigorous and exact derivation of the corresponding action for the height field $X_j(\bar{\tau})$ after resolving the $U_j$ conserved quantities, which in turn are functionally dependent on $\hat{\mathcal{Q}}^{A,B}$.

\subsubsection{Villain action in \texorpdfstring{$\hat{\mathcal{Q}}_A=\hat{\mathcal{Q}}_B=0$}{QA=QB=0}} \label{sec:funct_dep}
While the $U_j$ symmetries are linearly independent of $\hat{\mathcal{Q}}^{A,B}$ (when $1 \leq j \leq L-1$), it turns out they are not necessarily functionally independent. Let us consider the symmetry sector with quantum numbers $(Q_A, Q_B)=(0,0)$, which is one of the largest when truncating the local Hilbert space dimension to a maximum number of bosons per site $n_{max}$ on a finite system (see App.~\ref{sec:spectrum}). As we prove in App.~\ref{app:Uj1}, if $q$ and $p$ are coprime then $U_j=1$ for all $j$ within this symmetry sector. Moreover, while we currently miss a rigorous proof, we also numerically find that within any fixed $(Q_A, Q_B)$ sector, $U_j$'s take a fixed constant value. Hence, the $U_j$'s are completely determined by $\hat{\mathcal{Q}}^{A,B}$ when $q, p$ are coprime~\footnote{Nonetheless, two different sectors $(Q_A, Q_B), (Q_A^\prime, Q_B^\prime)$ can share the same set of eigenvalues for the $U_j$'s.}. In this section, we consider coprime $(q,p)$ tuples~\footnote{ An exhaustive and non-redundant list of co-prime pairs $(q.p)$ can be generated starting with the tuples $(q,p)=(2,1),\,(3,1)$ and generate all remaining ones iteratively following the three branches $(2q-p,q),\,(2q+p,q),\, (q+2p,p)$.} within the $(Q_A, Q_B)=(0,0)$ symmetry sector which exists for any system size, showing numerical results for $q=2, p=1$.

The continuity equation \eqref{eq:con_eq} can be solved within the $Q_A=Q_B=0$ sector by considering an integer-valued height field with boundary conditions $X_0=X_1=X_L=X_{L+1}=0$, and periodic boundary conditions in the time direction $X_j(0)=X_j(L_\tau)$ (see additional details in App.~\ref{app:Uj1}). This provides a rigorous $1$-to-$1$ mapping between height-field and number configurations $\{n_j(\bar{\tau})\}$ (rigorously) obtaining the partition function
\begin{equation} \label{eq:Z_00}
    \mathcal{Z}^{(0,0)}_{\textrm{height}}(\lambda)=\sum_{\{X_j(\bar{\tau})\in \mathbb{Z}\}}e^{-S_0[X_j(\bar{\tau})]}
\end{equation}
with $S_0[X_j(\bar{\tau})]$ given in Eq.~\eqref{eq:S_qp}. Then, we can replace $X_j(\bar{\tau})\in \mathbb{Z}\to \chi_j(\bar{\tau})\in \mathbb{R}$ by adding the potential term $-\lambda \cos(2\pi \chi_j(\bar{\tau}))$, and replacing $S_0[X_j(\bar{\tau})]$ by the action $S_0[\chi_j(\bar{\tau})]$,
\begin{equation} \label{eq:Z_Gauss}
    \mathcal{Z}^{(0,0)}(\lambda)=\int D\chi_j(\bar{\tau}) e^{-S_0[\chi_j(\bar{\tau})]-\lambda \cos(2\pi \chi_j(\bar{\tau}))}.
\end{equation}
Note that since $m_{q,p}>2$ for $q>1$, $S_0$ is not a quadratic action. However, quadratic terms like $(\nabla_x^{q,p}\chi_j(\bar{\tau}))^2$ compatible with both emergent U$(1)$ symmetries of $S_0$ given by $\chi_j\to \chi_j + \alpha_j^{A,B}$, will be generated at low energies being the leading relevant contributions. Hence, the fluctuations in this phase are expected to be controlled by the Gaussian action
\begin{equation} \label{eq:Squad_qp}
\begin{aligned} 
S_{0}[\chi]&= \frac{1}{2}\sum_{j,\tau}\left(K_\tau(\nabla_\tau \chi_j(\bar{\tau}))^2+K_x(\nabla_x^{q,p}\chi_j(\bar{\tau}))^2\right),
\end{aligned}
\end{equation}
coinciding with the path integral for the rotor model in Eq.~\eqref{eq:Hrotor_qp}. Hence, following Sec.~\ref{sec:conti_comm}, one finds that for any $(q,p)$ coprime, the ground state is described by two decoupled Luttinger liquids as in the continuum action \eqref{eq:S0_comm}, as long as the cosine contribution $\cos(2\pi \chi_j(\bar{\tau}))$ is not relevant.

\subsubsection{Imposing periodic boundary conditions}

Imposing PBC does not change the bulk physics, even though naively the specific conserved charges are present only for OBC systems.
In particular, all qualitative long-wavelength physics will be the same in the PBC and OBC systems in the thermodynamic limit.
Going into more details and taking spin-wave treatment as an example, imposing PBC on a system of size $L$ in the incommensurate case leads to a small size-dependent gap in the ``QLRO phase'' that vanishes in the thermodynamic limit.
This can be understood from the fact that only in that limit, the symmetry is exactly restored.
Hence, the choice of boundaries is not expected to affect the conclusions about the existence of the phase and its stability. The reason is the following: the incommensurate momentum $k^*$ can be decomposed as $k^*=2\pi n^*/L+\delta k$, i.e., a contribution $2\pi n^*/L$ lying in the Brillouin zone with $n^*=0,\dots, L-1$. Hence on a finite system the minimum energy is finite $\sim |k^*-2\pi n^*/L|$ (see e.g., the action in Eq.~\eqref{eq:S0UV_kw_main}), but vanishes in the thermodynamic limit as the lattice spacing for the momentum grid in the Brillouin zone decreases. On the other hand, it is even more simple that we do not expect the (gapped) Mott insulating phase to be affected by the boundary conditions.

\subsubsection{Continuum action and relevant contributions}
Analogous to our analysis in Sec.~\ref{sec:conti_comm} for commensurate $k^*$, one needs to add all possible contributions that are allowed by symmetry. In the infinite system-size limit, translation symmetry only allows the quadratic contribution in Eq.~\eqref{eq:S0_comm} diagonal in $\varphi^1,\varphi^2$. However, unlike for commensurate $k^*$, no finite number of cosine contributions $\sum_{n=1}^N\cos(2\pi\bm{\omega}_n\cdot \bm{\varphi})$ are invariant under $T_1$ which implies that no relevant term can be added to the infra-red action Eq.~\eqref{eq:S0_comm}!
However, this is in odds with the fact that such cosine contribution $\cos(2\pi \chi_j(\bar{\tau}))$ can become relevant at the lattice scale for sufficiently large interaction strength as we found in previous sections. It then appears that in order to account for the Mott insulating phase ---appearing as a result of vortex condensation--- one needs to remain at the lattice.

\subsubsection{Ultra-local correlations} \label{sec:ultra_local}

We can now use $S_0[\chi]$ to compute correlations on the lattice within the QLRO phase.
In the thermodynamic limit, one can neglect the boundary conditions for the height field and diagonalize the action \eqref{eq:Squad_qp} in momentum space (see App.~\ref{app:correlations}).
Similarly to Sec.~\ref{sec:corr_RG} one finds that density-density correlations decay as $\langle \nh_j \nh_{j^\prime} \rangle \sim \cos(k^*|j-j^\prime|)/|j-j^\prime|^2 $, i.e., they are imprinted by the wave vector associated to the conserved quantities. 
Given the small system sizes used in our numerical simulations, we do not provide numerical evidence of this scaling.
However, the slowly decaying texture in the local $\langle n_j \rangle$ induced by the boundaries in the OBC system provides indirect support to this prediction~\cite{EggertAffleck1995, RommerEggert2000, Kuhner2000, Lai2009}.

On the other hand, equal-time correlators $C_{\chi \chi}(j-j^\prime)=\langle e^{i\chi_j(0)}e^{-i\chi_{j^\prime}(0)} \rangle  \approx 0$ are ultra-local, vanishing for any $j\neq j^\prime$ as shown in App.~\ref{app:correlations}.
This happens as a result of the emergent invariance of the quadratic action \eqref{eq:Squad_qp} under
$\chi_j(\bar{\tau})\to \chi_j(\bar{\tau}) + \alpha_j^{A,B}$. 
This implies that $C_{\chi \chi}(j-j^\prime)$ vanishes unless $\alpha_j^{A,B}=\alpha_{j^\prime}^{A,B}$, which for incommensurate $k^*$ never holds. 
A similar reasoning in the spin-wave variables implies $\langle \bh_j^\dagger \bh_{j^\prime} \rangle=0 $ everywhere. 
Nonetheless, equal site but different times correlations decay as a power-law in $|\bar{\tau}-\bar{\tau}^\prime|$ with an exponent inversely proportional to the Luttinger parameter $K$. 
The fact that the spatial correlations of the vertex operator $e^{i\chi_j(0)}$ are ultra-local suggests that the cosine potential $\cos(2\pi\chi_j(\bar{\tau}))$ cannot become relevant. 
However, having power-law decaying correlations in the temporal direction is enough for this term to have a finite scaling dimension depending on the Luttinger parameter $K$ as shown at the end of App.~\ref{app:RG_wilson}.

\subsection{Numerical results}
In this section we provide numerical evidence for the presence of two different phases: a shortly-correlated Mott phase and one with quasi-long range order for the bosonic Hamiltonian $H_{2,1}$ in Eq.~\eqref{eq:H_21} with $q=2,p=1$. This is obtained using finite DMRG with open boundary conditions rather than infinite DMRG. The reason behind this choice is to be able to fix a $(Q_A, Q_B)$ symmetry sector in the numerical simulations; and avoid additional convergence issues in the size of the unit cell when dealing with infinite DMRG with incommensurate charges. Moreover, the large occupation and fluctuations of the local particle number make us fix $n_{\textrm{max}}=8,12$ to avoid strong truncation errors in the data. As shown in App.~\ref{app:conv_anal}, we required a sufficiently large $n_{\textrm{max}}$ for the numerical results to converge. Considering all these limitations, together with the fact that we group two consecutive sites when running DMRG, we restrict ourselves to system sizes $L\leq 40$. Moreover, we also include the term $ J^{(4)}\sum_j \left(\bh^2_j \bh_{j+1} \bh_{j+2} \bh^2_{j+3} + \textrm{H.c.}\right)$ commuting with $\hat{\mathcal{Q}}^{A,B}$ with small $J^{(4)}=0.1 J$ to improve convergence of the numerical algorithm.

\begin{figure}
    \centering
    \includegraphics[width=\linewidth]{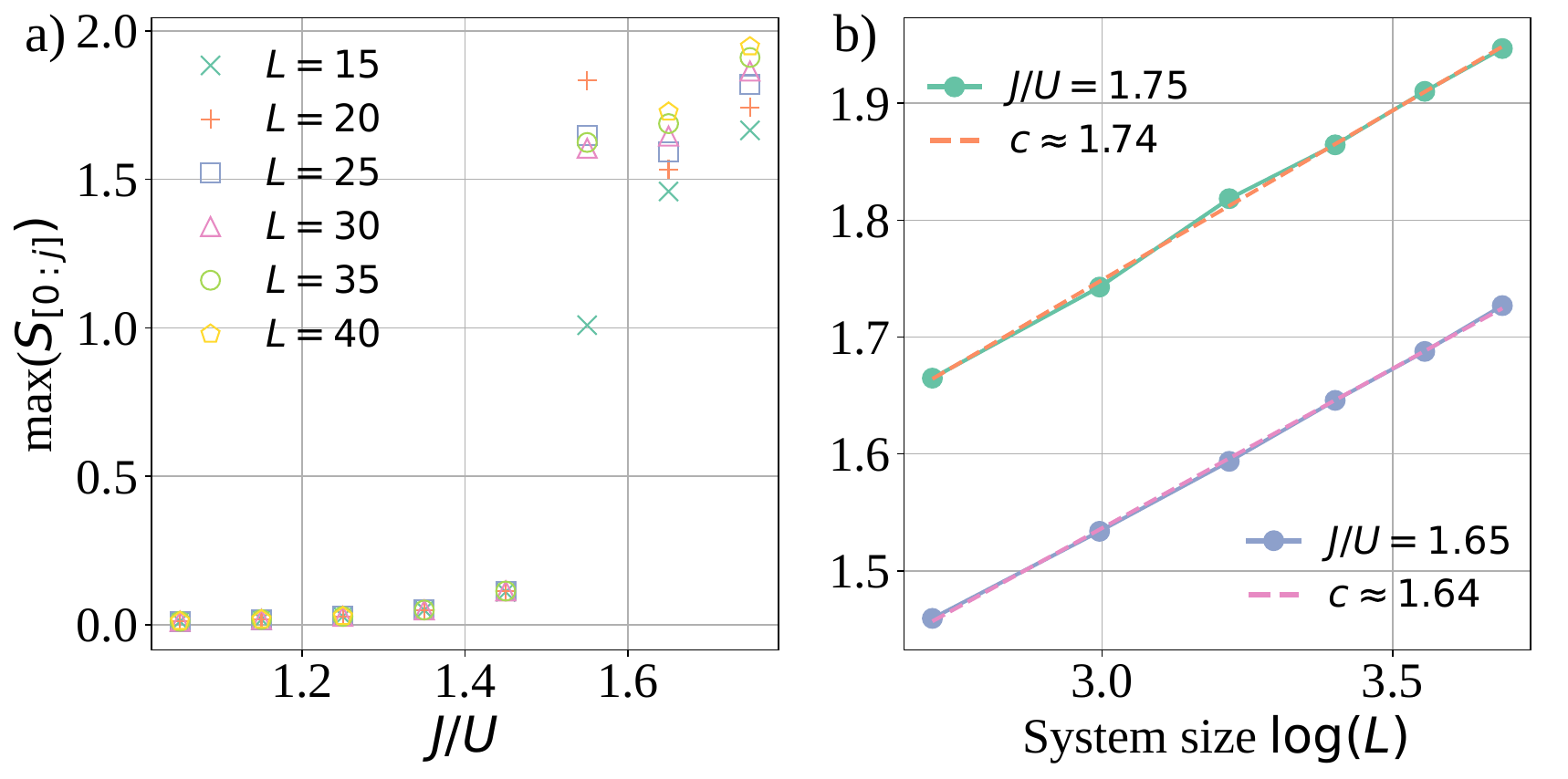}
    \caption{\textbf{Overall phase diagram for incommensurate symmetry $q=2,p=1$.} Scaling of the maximum entanglement entropy max($S_{[0:j]}$) with system size $L=15,20,25,30,35,40$ as a function of $J/U$ for $\mu/U=0.5$. Area-law scaling is found for $J<J_c$ with $J_c$ dependent on $\mu/U$, and logarithmic law for $J>J_c$. For $\mu/U=0.5$, $J_c/U\approx 1.5$.  (a) Dependence of max($S_{[0:j]}$) with $J/U$ for different system sizes. (b) Scaling of max($S_{[0:j]}$) with system size for $J/U=1.75$ (circles). This follows a scaling law $\max_j(S_{[0:j]})\approx \frac{c}{6}\log(L)$ with $c\approx 1.75$ for $J/U=1.75$ and $c\approx 1.64$ for $J/U=1.65$ obtained via linear fit. See complementary Fig.~\ref{fig:incomm_profile} showing profile $S_{[0:j]}$. The central charge obtained from a fit to the Cardy-Calabrese formula overestimates the value of the central charge in that case ($c\approx 2.84$ for $J/U=1.75$). Data has been obtained using finite DMRG with $n_{\text{max}}=8$, and bond dimensions $\chi=512$ finding convergence in both parameters (see main text for additional details). }
    \label{fig:incomm_pd}
\end{figure}

\begin{figure}[h!]
    \centering
    \includegraphics[width=\linewidth]{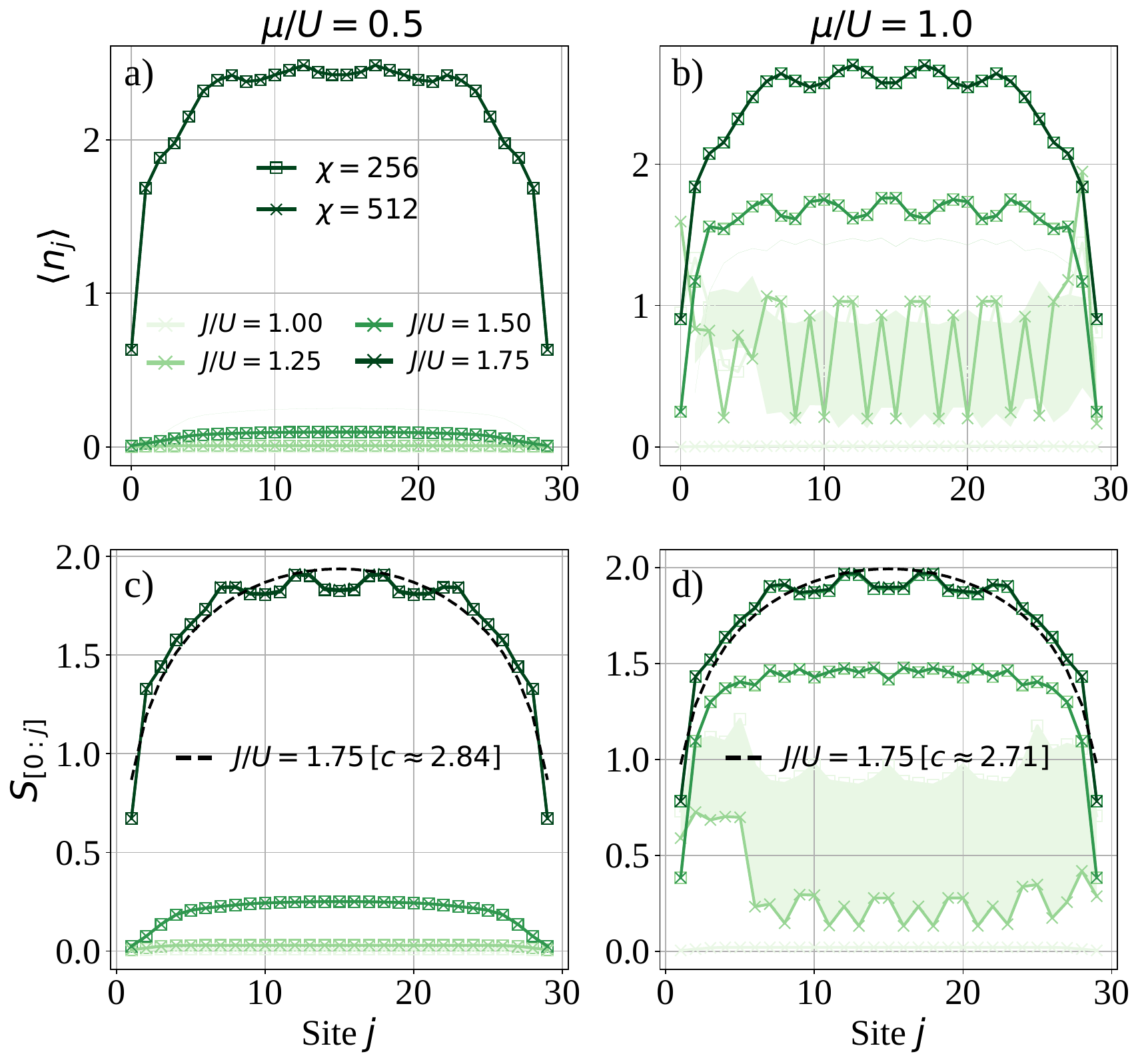}
    \caption{\textbf{Incommensurate $q=2, p=1$.} Dependence of the local density $\langle \nh_j\rangle$ and the bipartite entanglement entropy $S_{[0:j]}$ on $J/U\in [0.25,1.75]$ in steps of $0.25$ for $\mu/U=0.5$ (panels a and c) and $\mu/U=1$ (panels b and d).
    Both the density profile and $S_{[0:j]}$ show spatial oscillations whose wave vector is approximately given by $k^*=\arccos(\frac{p}{2q})$.
    Due to these superimposed oscillations, $S_{[0:j]}$ does not follow as nicely the Cardy-Calabrese formula~\cite{Calabrese_2009} as predicted for conformally invariant ground states, expected for $J>J_c$. Data has been obtained using finite DMRG with $n_{\text{max}}=8$, and bond dimensions $\chi=256$ (empty squares) and $512$ (crosses). We find convergence in both parameters (see main text for additional details), except manifest lack of convergence in bond dimension for $J/U=1.25, \mu/U=1$ which lies close to the transition point.}
    \label{fig:incomm_profile}
\end{figure}

Figure~\ref{fig:incomm_pd}(a) provides a rough estimate of the overall phase diagram as a function of $J/U$ for a representative uniform chemical potential $\mu/U=0.5$. The dependence of the half-chain entanglement entropy $\max_j(S_{[0:j]})$ with system size shows two qualitatively different scalings depending whether $J$ is smaller or larger than a critical value $J_c$. The scaling for $J>J_c$ illustrated in Fig.~\ref{fig:incomm_pd}(b) is consistent with the expected gapless QLRO phase with the half-chain entanglement entropy scaling logarithmically with system size. Assuming the scaling $\max_j(S_{[0:j]})\approx \frac{c}{6}\log(L) + \text{const.}$ with $c$ the central charge of the underlying conformal field theory, we find $c\approx 1.74$ for $J/U=1.75$, and $c\approx 1.64$ for $J/U=1.65$  lying within the critical regime. These two different regimes --- gapped versus gapless --- are further validated in Fig.~\ref{fig:incomm_profile} which shows the local density profile $\langle \nh_j \rangle$ (panels a and b), as well as the profile of $S_{[0:j]}$ for different bipartitions of the systems (panel c and d)  for two different values of the chemical potential $\mu/U=0.5, 1$. A different estimate of the central charge can be also obtained from the latter assuming that $S_{[0:j]}$ follows the Cardy-Calabrese formula~\cite{Calabrese_2009} derived for conformal field theories for open boundary conditions $S_{[0:j]}= \frac{c}{6} \log(\frac{2L}{\pi}\sin(\pi j/L))+\text{const.}$. This leads to $c\approx 2.8$ and $c\approx 2.7$ for $\mu/U=0.5, 1$ respectively. In these cases, this fitting overestimates the expected value $c=2$, compensating the under-estimate obtained in Fig.~\ref{fig:incomm_pd}(b). 
Fig.~\ref{fig:incomm_profile} shows characteristic slowly decaying modulations of both the local density profile $\langle \nh_j \rangle$ (similar to behavior in Luttinger liquids\cite{EggertAffleck1995, RommerEggert2000, Kuhner2000, Lai2009}) and $S_{[0:j]}$ with a wave-vector approximately equal to $k^*$.  
Such oscillations --- not appearing in the continuum conformal field theory describing the system in the QLRO phase --- lead to non-negligible deviations from the Cardy-Calabrese formula.
In fact, similar behavior has been found in the literature before~\cite{FAgotti}.

\section{Vortex formulation} \label{sec:vortex} 
We have found numerical evidence indicating that for both commensurate and incommensurate $k^*$, there exists a short-range correlated (Mott insulating) phase and a quasi-long-range ordered (gapless) phase even within a fixed $(Q_A,Q_B)$ symmetry sector.
We characterized the former by exponentially decaying correlations; and the latter by power-law density-density correlations imprinted with spatial oscillations at the lattice scale, and a finite central charge $c=2$.
Then, assuming the existence of the latter, we found that a relevant lattice-scale cosine term can drive a transition between the two. 
Nonetheless, we did not discuss the nature of this transition.
Similar phenomenology, although between a QLRO phase with $c=1$ and a Mott insulator, is found for the standard Bose-Hubbard model when tuning the chemical potential  and the hopping amplitude~\cite{PhysRevB.40.546}.
It is known that for fixed commensurate particle density, this cosine potential drives a Berezinskii–Kosterlitz–Thouless (BKT) transition where the unbinding of vortices disorders the system~\cite{chaikin_lubensky_1995}.
On the other hand, when the particle number can vary, this transition is in the commensurate-to-incommensurate (or Pokrovsky-Talapov) universality class (see e.g., Refs.~\onlinecite{Talapov,giamarchi2004quantum}).
To shed some light on the mechanism driving the transition for systems with modulated symmetries, we reformulate the theory in terms of topological defects loosely similar to vortices, focusing on the sector with $Q_A=Q_B=0$, which is similar to commensurate density case in the usual BHM in that one does not have non-trivial Berry phases.

We consider the partition function in Eq.~\eqref{eq:Z_00} for the integer-valued height field $X_j(\bar{\tau})$ with open boundary conditions $X_0=X_1=X_L=X_{L+1}=0$ in the spatial direction, and periodic in time $X_j(0)=X_j(L_\tau)$. Taking care of these we can rewrite 
\begin{equation}
S_0[X] = \frac{1}{2} \sum_{j,j^\prime=1}^L \sum_{\tau,\tau^\prime=1}^{L_\tau} X_j(\bar{\tau}) V^{-1}(j,j^\prime; \bar{\tau}^\prime-\bar{\tau}) X_{j^\prime}(\bar{\tau}^\prime)
\end{equation} 
with 
\begin{equation} \label{eq:Vm1}
V^{-1}(j,j^\prime; \bar{\tau}^\prime-\bar{\tau}) \equiv K_\tau(\nabla_\tau)^2 + K_x(\nabla_x^{q,p})^2.
\end{equation}
The precise meaning is that the R.H.S. is a quadratic form with the specific boundary conditions on $X$'s, which one can argue is invertible for finite $L$ and $L_\tau$ (indeed, the above height model with the specific boundary conditions is an exact rewriting of the OBC Villainized rotor model, which is a well-defined statmech model with a finite partition sum for any finite system size). 
We can now introduce ``topological defect'' (which we will loosely refer to as ``vortex'') degrees of freedom by using the exact rewriting $\sum_{X_j\in \mathbb{Z}}f(X_j)=\int_{-\infty}^{+\infty} d\chi_j \sum_{m_j\in \mathbb{Z}}e^{i2\pi m_j \chi_j}f(\chi_j)$ at each site and time slice to obtain $\mathcal{Z}_{\textrm{vortex}}=\sum_{\{m_j(\bar{\tau})\in\mathbb{Z}\}}e^{-S_{\textrm{vortex}}[m_j(\bar{\tau})]}$ with the action given by
\begin{equation} \label{eq:Z_vort}
    \begin{aligned}
        S_{\textrm{vor}}[m]=\frac{(2\pi)^2}{2}\sum_{j,j^\prime=2}^{L-1}\sum_{\bar{\tau}, \bar{\tau}^\prime=1}^{L_\tau} m_j(\bar{\tau})V(j,j^\prime; \bar{\tau}^\prime-\bar{\tau})m_{j^\prime}(\bar{\tau}^\prime),
    \end{aligned}
\end{equation}
where the vortex degrees of freedom $m_j(\bar{\tau})$ sit on vertical links, and we have explicitly used periodic boundary conditions in the time direction.
Since finding a set of elementary basis satisfying $X_0=X_1=X_L=X_{L+1}=0$ appears not to be feasible (unlike for boundary conditions with $X_{1/2}=X_{L+1/2}=0$ in the case of the usual HBM in the sector with zero total charge, with OBC basis spanned by sinusoidal functions), we numerically obtain $V_{(j,\bar{\tau}),(j^\prime, \bar{\tau}^\prime)}$ on a 2D lattice of size $L\times L_{\tau}$ by inverting the quadratic form on $X$'s.

\begin{figure*}[t!]
    \centering
    \includegraphics[width=\textwidth]{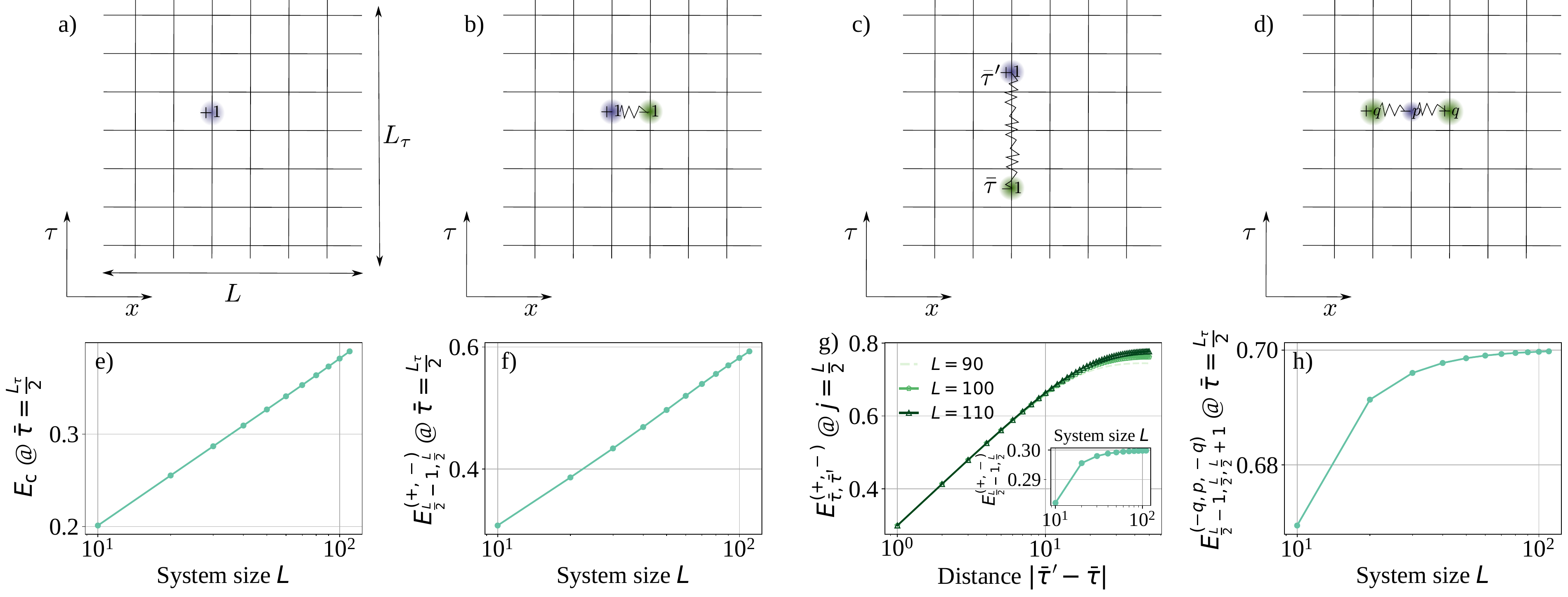}
    \caption{\textbf{Energetics of vortex configurations illustrated in the incommensurate $q=1,p=2$ case.} Space-time vortex configurations with $m_j(\bar{\tau})>0$ and $m_j(\bar{\tau})<0$ represented as blue and green circles respectively. All configurations are centered around the vertical link $(j,\bar{\tau})=(\frac{L}
    {2},\frac{L_\tau}
    {2})$: (a) Single vortex $m_{\frac{L}{2}}(\frac{L_\tau}
    {2})=+1$; (b) Two vortices on consecutive spatial coordinates $m_{\frac{L}{2}}(\frac{L_\tau}
    {2})=+1,m_{\frac{L}{2}+1}(\frac{L_\tau}
    {2})=-1$; (c) Two vortices on consecutive time coordinates with charges $m_{\frac{L}{2}}(\bar{\tau}')=+1, m_{\frac{L}{2}}(\bar{\tau})=-1$.
    (d) Three vortices on consecutive spatial coordinates with charges $m_{\frac{L}{2}-1}(\frac{L_\tau}
    {2})=-q,m_{\frac{L}{2}}(\frac{L_\tau}
    {2})=+p, m_{\frac{L}{2}+1}(\frac{L_\tau}
    {2})=-q$. 
    Panels (e), (f), (h) show the scaling with system size of the potential energy of the corresponding configuration for a square lattice of dimension $L\times L$. This diverges with system  size as $\sim \log(L)$ for configurations in panels (a) and (b), while it converges to a finite number for configurations (d). Panel (g) shows that the potential energy between two vortices of opposite charge and sitting at the same spatial coordinate as in panel (c) grows logarithmically with the distance. The inset shows that the energy of this configuration does not scale with system size for any finite $|\bar{\tau}'-\bar{\tau}|$. 
    While data is shown for $q=2, p=1$, all qualitative results (a), (c), (d) hold for any commensurate and incommensurate $k^*$, while the result (f) is modified for $j-j'=(2\pi/k^*) \times \text{integer}$, see text for details. Numerical results were obtained by numerically inverting the quadratic form in Eq.~\eqref{eq:Vm1}, setting $K_\tau=K_x=1$ for concretness, on a square space-time lattice of dimension $L \times L$, with spatial open boundary conditions as specified in the main text, and periodic temporal boundaries.}
    \label{fig:vortex_plot}
\end{figure*}

The resulting two-dimensional statistical model with Boltzmann weights $e^{-S_{\text{vortex}}[m_j(\bar{\tau})]}/\mathcal{Z}_{\text{vortex}}$ describes a gas of ``charged'' particles interacting via the potential $V(j,j^\prime; \bar{\tau}^\prime-\bar{\tau})$. In the standard XY model one can argue about the existence of a finite-temperature BKT transition by comparing the energy of a single-vortex configuration [Fig.~\ref{fig:vortex_plot}(a)] $E_C$ with its Boltzmann entropy, which is given by $k_B\log(L\times L_\tau)$. Such a transition from a confining-phase of topological defects, to a high-temperature disordered one where vortices proliferate will occur if $E_C$ follows the same scaling law. 
Figure~\ref{fig:vortex_plot}(e) shows that indeed $E_C\propto \log(L)$ when exactly computed on a square space-time lattice of size $L\times L$. 
While this energy is infinite in the thermodynamic limit, these defects are created in groups and are bound together in a configuration that only requires finite energy. For example the unbinding of vortex-antivortex pairs [see Figs.~\ref{fig:vortex_plot}(b,c)] drives the transition in the usual XY model, being confined by a two-dimensional Coulomb potential, which scales as the logarithm of their separation. 
However, we already noticed that the systems we are considering have important differences from the usual XY model, in particular they are not isotropic, and vortices might showcase different behavior along spatial and temporal directions. 

Let us start by considering a vortex-antivortex pair configuration occupying two spatial consecutive sites as shown in Fig.~\ref{fig:vortex_plot}(b).
As Fig.~\ref{fig:vortex_plot}(f) exhibits, its energy scales $E^{(+,-)}_{ \frac{L}{2}-1,\frac{L}{2}}\propto \log(L)$ and hence, these ``bound'' configurations will not appear as a result of thermal fluctuations, unlike the usual XY model where such configurations have finite energy in the thermodynamic limit (and this energy grows logarithmically with the separation).
We have verified that the above property of divergent energy with $L$ holds for any non-zero spatial separation between the vortex and antivortex.
On the other hand, if such a pair is oriented along the temporal direction [shown in Fig.~\ref{fig:vortex_plot}(c)], it only requires a finite energy in the thermodynamic limit for finite separation, as the inset of Fig.~\ref{fig:vortex_plot}(g) shows.
Moreover, the binding energy of this pair grows logarithmically with the distance along the temporal direction $E^{(+,-)}_{\bar{\tau},\bar{\tau}^\prime} \propto \log(|\bar{\tau}^\prime-\bar{\tau}|)$, until it reaches $|\bar{\tau}^\prime-\bar{\tau}| \sim L_\tau/2$ on a finite system [main panel in Fig.~\ref{fig:vortex_plot}(g)]. 
Thus, we conclude that the vortex-antivortex pair is energetically confined in the spatial direction, but can become unbound in the temporal direction leading to the destruction of the QLRO phase.
Since the binding potential in the temporal direction is logarithmic, the BKT energy-entropy balance arguments may still apply. 
While one might conclude that no possible bound configuration along the spatial direction can be created with finite energy, this conclusion turns out to be too quick as demonstrated in panels (d,h). 
Indeed, consider a configuration of the form $(m_{j-1}(\bar{\tau}),m_{j}(\bar{\tau}),m_{j+1}(\bar{\tau}))=(-q,p,-q)$ on any three consecutive sites $(j-1,j,j+1)$ with $m_{i}(\bar{\tau})=0$ everywhere else. 
As Fig.~\ref{fig:vortex_plot}(h) suggests, this only requires finite energy in the thermodynamic limit, and following the same discussion we just had, it can expand along the temporal direction analogous to the standard XY model. 
However, can we explain these observations given the knowledge we have about the system?

In the thermodynamic limit and neglecting boundary effects,
one finds that the field $\chi_j(\bar{\tau})$ mediates a long-range two-dimensional Coulomb interaction among vortex variables given by
\begin{equation} \label{eq:2d_Colum}
\begin{aligned}
    &V(j-j^\prime, \bar{\tau}-\bar{\tau}^\prime) = \langle \chi_j(\bar{\tau})\chi_{j^\prime}(\bar{\tau}^\prime) \rangle_{\text{Gauss}} \\
    & =  \int_{-\pi}^{\pi} \frac{dk}{2\pi} \int_{-\infty}^{\infty} \frac{d\omega}{2\pi}\frac{\cos(k(j-j^\prime)-\omega(\bar{\tau}-\bar{\tau}^\prime))  }{K_\tau\omega^2+K_x(p-2q\cos(k))^2}.
\end{aligned}
\end{equation}
However, this expression is size-dependent diverging as $\log(L)$ due to infra-red contributions.
The same is true for the standard XY model; what is finite is the energy of certain vortex configurations satisfying appropriate charge neutrality conditions. 
Hence, we need a regularized expression that is well-defined in the thermodynamic limit. To do so, we subtract (and add) the logarithmically-divergent term $\langle \chi^2_{\frac{L}{2}}(\frac{L_\tau}{2})\rangle_{\text{Gauss}} \propto \log(L)$ such that the resulting potential does not depend on system size
\begin{equation} \label{eq:V_reg}
\begin{aligned}
    &V^R(j-j^\prime, \bar{\tau}-\bar{\tau}^\prime)\\ &= \langle \chi_j(\bar{\tau})\chi_{j^\prime}(\bar{\tau}^\prime)\rangle_{\text{Gauss}} - \langle \chi^2_{\frac{L}{2}}(\frac{L_\tau}{2})\rangle_{\text{Gauss}} \cos(k^*(j-j^\prime)) \\
    & \simeq \cos(k^*(j-j')) V_{\text{2D Coul}}(j-j',\tau-\tau').
\end{aligned}
\end{equation}
Here $V_{\text{2D Coul}}(j-j',\tau-\tau')$ is now defined in the thermodynamic limit and is the familiar 2D Coulomb potential. 
Importantly, $V^R$ also includes an oscillatory factor at wave vector $k^*$ that takes positive and negative values. 
This naively suggests that even two equal-sign and highly charged vortices could indefinitely lower the energy of the system. However, as we have just discussed and will show in the following, the energy to create such a configuration grows logarithmically with system size.
Plugging Eq.~\eqref{eq:V_reg} into the vortex action \eqref{eq:Z_vort}, one finds
\begin{widetext}
\begin{equation} \label{eq:Z_vort_gen}
    \begin{aligned}
        S_{\textrm{vor}}[m]=\frac{(2\pi)^2}{2}\sum_{j,j^\prime}\sum_{\bar{\tau}, \bar{\tau}^\prime} m_j(\bar{\tau})V^R(j-j^\prime; \bar{\tau}-\bar{\tau}^\prime)m_{j^\prime}(\bar{\tau}^\prime) +
        \frac{(2\pi)^2}{2} \langle \chi^2_{\frac{L}{2}}(\frac{L_\tau}{2})\rangle_{\text{Gauss}}
        \left( Q_{A,\text{vor}}^2+Q_{B,\text{vor}}^2\right),
    \end{aligned}
\end{equation}
\end{widetext}
%\twocolumngrid
where $Q_{A,\text{vor}}=\sum_{j,\tau}\cos(k^*j)m_j(\bar{\tau})$, $ Q_{B,\text{vor}}=\sum_{j,\tau}\sin(k^*j)m_j(\bar{\tau})$ after using the trigonometric identity $\cos(k^*(j-j^\prime))=\cos(k^*j)\cos(k^*j^\prime)+\sin(k^*j)\sin(k^*j^\prime)$. 
The energetics thus requires that these $Q_{A/B,\text{vor}}$ are zero, which are the analog of the zero total vorticity condition $\sum_{j,\tau} m_j(\bar{\tau})=0$ in the standard XY model.

\begin{figure}[h!]
\centering
\includegraphics[width=\linewidth]{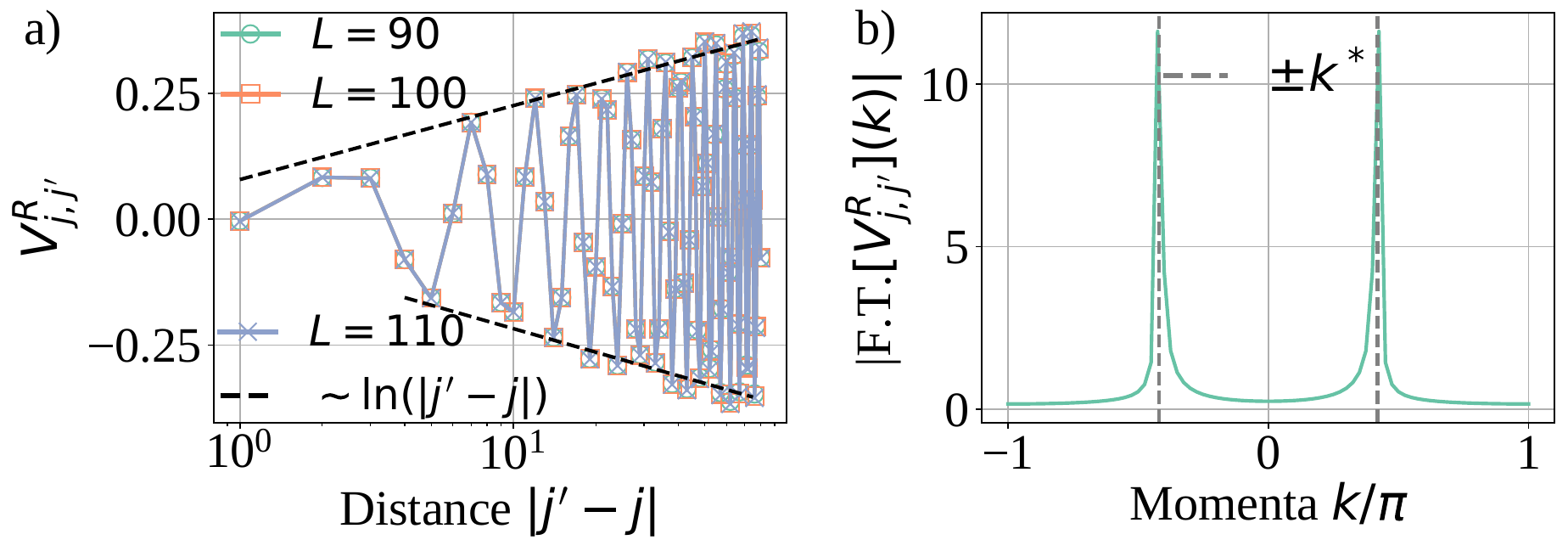}
\caption{\textbf{Regularized vortex potential in the $q=2, p=1$ case.} 
(a) Vortex potential $V^R(j-j^\prime, \bar{\tau}-\bar{\tau}^\prime)$ evaluated at $\bar{\tau}=\bar{\tau}^\prime=L_{\tau}/2$ regularized as in Eq.~\eqref{eq:V_reg} as a function of the distance $|j-j^\prime|$. 
(b) The amplitude of the Fourier components of $V^R_{j,j^\prime}$ shows a maximum at $\pm k^*$.
As explained in the text, the oscillatory $V^R_{j,j^\prime}$ does not mean any instability, since it is the total energy Eq.~(\ref{eq:Z_vort_gen}) that is always physically sensible. Numerical results were obtained with the same setting as in Fig.~\ref{fig:vortex_plot}.
}
\label{fig:2vortices_plot}
\end{figure}

The functional form of $V^R_{j,j^\prime}$ at $\bar{\tau}=\bar{\tau}^\prime=\frac{L_\tau}{2}$ is shown in Fig.~\ref{fig:2vortices_plot}(a). While OBC breaks spatial translation symmetry, we find that the potential only depends on the distance $j-j^\prime$ when sufficiently far from the boundaries. The envelope function approximately grows as $V^R_{j,j^\prime}\sim \log(|j-j^\prime|)$. 
Moreover, this features spatial oscillations with wave-vector $\pm k^*$ as shown in Fig.~\ref{fig:2vortices_plot}(b).
The absolute value of the Fourier amplitudes $|$F.T$[V_{j,j^\prime}](k)|$ is peaked at $\pm k^*$, matching with the modulation imprinted by the symmetry.

Recalling that $\langle \chi^2_{\frac{L}{2}}(\frac{L_\tau}{2})\rangle_{\text{Gauss}} \propto \log(L)$, the action  \eqref{eq:Z_vort_gen} leads to the results in the previous paragraphs regarding the energy of a single and multiple vortex configurations. 
First, a vortex-antivortex pair along the spatial direction $(m_{j}(\bar{\tau})=+1,m_{j+1}(\bar{\tau})=-1)$ as in Fig.~\ref{fig:vortex_plot}(b), has non-zero $Q_{A,\text{vor}}, Q_{B,\text{vor}} $ producing an infinite energy in the thermodynamic limit. 
On the other hand, such a pair along the temporal direction as in Fig.~\ref{fig:vortex_plot}(c) gives $Q_{A,\text{vor}}= Q_{B,\text{vor}}=0$ and then costs a finite energy. 
Finally, we note that the expression for $Q_{A,\text{vor}}, Q_{B,\text{vor}}$ is mathematically equivalent to that for the conserved charges $\hat{\mathcal{Q}}_{c,s}$ below Eq.~\eqref{eq:Q_AB}. 
Therefore, a configuration $(m_{j-1}(\bar{\tau}),m_{j}(\bar{\tau}),m_{j+1}(\bar{\tau}))=(-q,p,-q)$ on any three consecutive sites $(j-1,j,j+1)$ with $m_{i}(\bar{\tau})=0$ everywhere else, shown in Fig.~\ref{fig:vortex_plot}(d), satisfies the neutrality condition $Q_{A,\text{vor}}= Q_{B,\text{vor}}= 0$, and hence this configuration has a finite energy.
This appears then to be an analog of separating a vortex and an antivortex in the spatial direction in the usual XY model, which is clearly much more non-trivial in the modulated symmetry case (more below)!
The non-trivial contents of this $(-q,p,-q)$ object can then independently move along the temporal direction, e.g., $(m_{j-1}(\bar{\tau}_1),m_{j}(\bar{\tau}_2),m_{j+1}(\bar{\tau}_3))=(-q,p,-q)$ for any $\bar{\tau}_1,\bar{\tau}_2, \bar{\tau}_3$, while satisfying the neutrality conditions, where the energy is finite and grows logarithmically with their spreading in the temporal direction.
Nonetheless, as we will see, in the incommensurate case, spreading along the spatial direction appears not possible without changing the vortex ``charges" and requires additional energy. 
This is related to the fact that the energetically imposed charge neutrality conditions share the same structure as the original conservation laws.

To illustrate this, imagine we want to split the configuration $(m_{x}(\bar{\tau}),m_{y_{\text{ini}}=x+1}(\bar{\tau}),m_{y_{\text{ini}}+1}(\bar{\tau}))=(-q,p,-q)$ into two local lumps centered around sites $x$ and $y>x$ while satisfying the neutrality conditions. 
For commensurate $k^*$, e.g., $q=p=1$, and fixing the location of the defect originally at $x$, this can spread along the spatial direction with only logarithmic confinement as long as $y=x+3\times$integer or $y=x+1+3\times$integer without increasing the charges as $m_x, m_y, m_{y+1}$
[more specifically, in each case we need to consider two subcases: for $y=x+6n, n\in\mathbb{Z}$, the possible spreading is $(m_x,m_y,m_{y+1})=(-1,1,0)$ showing only non-zero charges, while for $y=x+3+6n$ it is $(m_x,m_y,m_{y+1})=(-1,-1,0)$; for $y=x+1+6n$ the spreading is $(m_x,m_y,m_{y+1})=(-1,1,-1)$, while for $y=x+4+6n$ it is $(m_x,m_y,m_{y+1})=(-1,-1,1)$]. 
This is consistent with the fact for commensurate $k^*$  the correlations $\langle e^{i(\chi_j(\bar{\tau})-\chi_{j^\prime}(\bar{\tau}))}\rangle$ decay as a power-law both along the temporal and spatial directions when $j-j' = 6 \times \text{integer}$, as found in Sec.~\ref{sec:corr_RG} (and $\langle e^{i(\chi_j(\bar{\tau})+\chi_{j^\prime}(\bar{\tau}))}\rangle$ is non-zero when $j-j' = 6 \times \text{integer} + 3$). 
On the other hand, for incommensurate $k^*$, the confinement along the spatial direction is stronger than logarithmic. 
While temporal correlations always decay as a power-law, the ultra-local spatial correlations $\langle e^{i(\chi_j(\bar{\tau})-\chi_{j^\prime}(\bar{\tau}))}\rangle$ found in Sec.~\ref{sec:ultra_local} are consistent with spatial confinement. 

To understand the difference between commensurate and incommensurate cases let us consider again the configuration $(m_{x}(\bar{\tau}),m_{y_{\text{ini}}=x+1}(\bar{\tau}),m_{y_{\text{ini}}+1}(\bar{\tau}))=(-q,p,-q)$, fixing the location at $x$, while shifting $y$ to the right. 
It turns out that in the incommensurate case, this requires to increase the charges $m_{x}(\bar{\tau}),m_{y}(\bar{\tau}),m_{y+1}(\bar{\tau})$ as $m_{x}(\bar{\tau}) = q^{ y-x}$, as well as $m_{y}(\bar{\tau}),m_{y+1}(\bar{\tau})\propto q^{y-x}$~\footnote{We find that $m_x=q^{y-x}$, and $m_y$ satisfies the linear recurrence $m_{y} = m_{y-1} - q^2 m_{y-2}$ when $y \geq 2$ with $m_0 = m_1 = 1$. Moreover, $m_{y+1} = -q m_{y-1}$.}.
This implies that the potential energy of this configuration from Eq.~\eqref{eq:Z_vort_gen} is given by $\sim V^R_{x,y}q^{2|y-x|}$ and grows exponentially with $y-x$, hence topological defects along the spatial direction are \textit{exponentially confined}.
Therefore, the unbinding of this bound vortex configurations is exponentially suppressed, leaving only the temporal direction as a way to drive the transition to the disordered phase.

We remark that this behavior is different from that of dipole-conserving systems.
Our exact height model derivation and the corresponding ``vortex gas'' model work in this case as well with $q=1,p=2$, but the resulting energetics of vortices is qualitatively different.
In the dipole case, while breaking up the bound configuration $(m_{j}(\bar{\tau}),m_{j+1}(\bar{\tau}) )=(+1,-1)$ into two separated lumps requires a polynomially large energy in the distance between the lumps, one finds a second type of configuration $(m_{j-1}(\bar{\tau}),m_{j}(\bar{\tau}),m_{j+1}(\bar{\tau}))=(-1,2,-1)$  which can be split as  $(m_{x}(\bar{\tau}), m_{x+1}(\bar{\tau}))=(-1,+1), (m_{y}(\bar{\tau}), m_{y+1}(\bar{\tau}))=(+1,-1)$ with the required energy remaining finite for any $y-x$.
Such $(m_{x}(\bar{\tau}), m_{x+1}(\bar{\tau}))=(-1,+1)$ ``particles", thus only experience a global constraint while having a finite energy, and hence we expect them to form a gas with a finite density, scrambling the naive ``order" that would be present by assuming the absence of the topological defects.
Allowing such $(m_x,m_{x+1})=(-1,+1)$ or $(+1,-1)$ objects is physically related to the presence of the relevant perturbation $\sim\cos(\partial_x \chi)$ in the continuum dual theory.
This physics can be interpreted as the reason why the quantum-Lifthshitz theory is not stable in the one-dimensional dipole-conserving systems~\cite{Zechmann_22,Lake22_1DDBHM}.

In conclusion, the order-to-disorder transition is driven, even for incommensurate $k^*$, by the proliferation and unbinding of vortices, but the unbinding happens by separating the vortices along the temporal direction only. 
Schematic BKT-like energy-entropy argument still works since the energy still grows logarithmically with the separation $|\tau - \tau'|$.
However, unlike the usual BHM where a single low-energy mode at $k=0$ gets gapped out, in the modulated symmetry case the two low-energy modes around $\pm k^*$ characterizing the QLRO as appearing in Eq.~\eqref{eq:S0_comm} are gapped out. 
This suggests that at long wavelengths two rather than one type of effective vortex excitations $m_j(\bar{\tau})\to m^{A,B}_j(\bar{\tau})$ exist associated to each of those modes and interacting via a standard 2D Coulomb potential. 
This is also the expectation from the presence of two phase and dual variables emerging at low energies.  
However, at this point we are not able to obtain such low-energy effective degrees of freedom in a rigorous manner starting from our lattice-scale topological defects $m_j(\hat{\tau})$.

\section{Conclusions and outlook} \label{sec:conclusion}

In this work, we considered different families of bosonic and rotor models which conserve finite Fourier momenta of the particle number $\hat{\mathcal{Q}}_{\pm}=\sum_j e^{\pm i k^*j}\nh_j$ but not the particle number itself.
We do so by considering generalizations of the standard Bose-Hubbard model where the standard single-particle hopping is replaced by a multi-boson correlated process commuting with $\hat{\mathcal{Q}}_{\pm}$, which is at least possible for $k^*$ satisfying $\cos(k^*) = p/(2q)$ with $p,q \in \mathbb{Z}$. 
These models can be understood as simple extensions of the previously introduced spin systems in Ref.~\onlinecite{Sala_modsym} that avoid Hilbert space fragmentation and other types of symmetries resulting from a finite-dimensional onsite Hilbert space and strictly finite range of interactions. 
Nonetheless, we proved that as long as $k^*$ is incommensurate, the models we consider showcase Hilbert space fragmentation that persists even when adding any longer range term compatible with the symmetry. 
This relates to the exponentially large number of distinct eigenvalues in the spectrum of the conserved quantities when $k^*$ is incommensurate, providing to the best of our knowledge the first (non-trivial~\footnote{We consider trivial the fragmentation appearing as a result of extensively-many local symmetries as is for example the case for a Hamiltonian with only pair hopping $(\bh_j^\dagger)^2(\bh_{j+1})^2+\textrm{H.c.}$.}) example of fragmentation with infinite-dimensional local Hilbert spaces.
It is still an open question how to extend commutant algebra framework~\cite{Moudgalya_2022} to study this scenario, even in the absence of particle number conservation. Along the way, we introduced a novel lattice-scale duality transformation which in the rotor-language localizes some of the underlying symmetries responsible for the fragmentation while keeping the Hamiltonian local.
While here we focused on bosonic statistics, analogous fermionic systems with the same symmetries can be constructed. However, these might be strongly fragmented unless sufficiently long range symmetry-preserving ``hopping'' terms are included~\cite{Sala_PRX,khemani_localization_2020}.

In the rest of this work, we characterized the zero-temperature phase diagram of the previously introduced bosonic models focusing on the quasi long-range order (QLRO) phases.
For commensurate $k^*$, these systems can be treated analogously to the standard Bose-Hubbard model~\cite{PhysRevB.40.546}, as the modulated symmetries correspond to two intertwined sublattice U$(1)$ symmetries. Here, we found a rich phase diagram which includes Mott insulating lobes (now labeled by eigenvalues of $\hat{\mathcal{Q}}_{\pm}$), as well as two different types of QLRO.
Even though the microscopic models we considered lack a quadratic kinetic energy term and are strongly interacting, the more generic QLRO phase can be captured by a quadratic effective action corresponding to two-species Luttinger liquid. 
These correspond to the gapless modes around $\pm k^*$, and can be understood as an exact (instead of emerging) Bose surface~\cite{Paramekanti_2002,Lake_2021,PhysRevB.100.024519,Tiamhock_2011,you2021fractonic}. 
In fact, when evaluating microscopic observables, one recovers the standard features associated with Luttinger liquids that are being modulated by oscillations with momentum $k^*$ as imprinted by the symmetry.
In the presence of a uniform chemical potential, we found two apparently disconnected Mott lobes which are nonetheless labeled by the same global quantum numbers associated to the modulated symmetries. We leave as an open question whether these can correspond to different (crystalline) SPT phases~\cite{Pollmann_12,Chen_2011,Schuch_2011} that could be distinguished by boundary degrees of freedom. Finally, while the model we studied can appear artificial at first glance, we showed that it actually emerges as an effective description when considering a standard interacting bosonic-model in the presence of a strong quasi-periodic chemical potential, the so-called generalized Aubry-Andr{\'e} model~\cite{aubry1980analyticity,Iyer_2013,Schreiber_2015,MBAA_19}. 
Moreover, we also provided a statistical model combining dimer and trimer degrees of freedom which captures the physics of the QLRO.

We then studied the phase diagram of systems with incommensurate charges $\hat{\mathcal{Q}}_{\pm}$. While the incommensurability complicates the characterization of the QLRO phase, we provide a rigorous derivation of the zero-temperature physics starting from the rotor model approximation and generalizing the standard duality to the corresponding Villain-action~\cite{Villain_75}. Unlike for commensurate $k^*$, we found that vortex-vortex correlators are ultra-local in space suggesting that the QLRO cannot be driven into a Mott insulating phase. Nonetheless, we showed that vortices can still become relevant and disorder the system into a Mott insulating phase. However, unlike in the standard XY model, vortices interact via a two-dimensional Coulomb potential in Euclidean space-time, which is dressed by an oscillatory factor with wave vector $k^*$. 
We find bound vortex configurations which require only a finite energy in the thermodynamic limit and hence can be produced by local thermal fluctuations.  
We argued that the unbinding of these objects along the temporal direction will eventually disorder the system into a Mott insulator for a fixed $(Q_A,Q_B)$ symmetry sector. Nonetheless, a complete characterization of the transition, as well as the possibility of a Pokrovsky-Talapov-like transition in the systems we studied is left as an open question. 

While vortex degrees of freedom play an important role in the characterization of the QLRO phase and its stability, there is not a clear connection between these and particular configurations of the bosonic phase.
For example, a vortex in the standard XY model can be associated with a singularity of the phase field at a point, the texture being smooth everywhere else. We notice that certain vortex configurations in dipole-conserving systems in $(2+1)$ dimensions have been discussed in Ref.~\onlinecite{dip_vortices}. An analogous formulation for systems with different types of unconventional symmetries (also in $(1+1)$-dimensional Euclidean space-time) is left for future work. Moreover, while the continuum effective description is given by two independent modes around momenta $\pm k^*$, we have not found a formal way to decompose the lattice-scale vortex field into two types of vortices, even for commensurate $k^*$. 
Our vortex formulation is completely exact for spatial OBC systems and applies also to the $1$D dipole-conserving systems (and possibly to higher moments once appropriate boundary conditions are fixed). 
This might provide an alternative explanation of the absence of so-called Bose-Einstein Insulating phase (a.k.a.\ Lifshitz theory) in this case, and its apparent stability in numerical simulations. Moreover, while we naively expect these systems to be unstable to the presence of single-hopping terms, a more careful analysis is left as an open question~\cite{Lesik_EBL, Lake_EBL}.

Reference~\onlinecite{Sala_modsym} also introduced exponentially-localized symmetries, as well as combinations between these with quasi-periodic symmetries, as well as dipole and higher-moment conservation. A naive analysis of the former class for the corresponding rotor models with exponential-localized symmetries, suggests that such systems are always gapped.
Hence, it would be interesting to extend our analysis not only to those but also general situations, as well as to different types of constrained models which can naturally appear in current experimental setups~\cite{Scherg_nature,PhysRevLett.130.010201}. 
Is there an overarching structure associated with the presence of such unconventional symmetries? Moreover, it would be interesting to extend the investigation on higher-dimensional systems~\cite{Paramekanti_2002,Tiamhock_2011,you2021fractonic,myersonjain2021pascals, gorantla2021lowenergy,gorantla2021modified,Seiberg_2020} beyond subsystem symmetries and a finite number of momenta modes being conserved, as well as many other types of symmetries potentially leading to novel phenomena, where the former can be understood as systems with Bose surfaces~\cite{Paramekanti_2002,Tiamhock_2011,gorantla2021lowenergy,gorantla2021modified,Seiberg_2020}. Understanding the relation (if any) to order-by-disorder transitions, as well as to systems with topological order is an intriguing open question (see recent Ref.~\onlinecite{delfino20232d} which studied some related questions for discrete exponentially-localized symmetries).

Finally, we note that while we were able to obtain effective models with commensurate symmetries by starting with the interacting Aubry-Andre{\'e} model in the large (commensurate) periodic potential, similar \textit{controlled} derivations do not appear to apply for a large incommensurate quasi-periodic potential. 
Still, one intuitively expects such commuting terms to arise in perturbation theory, providing many-body processes which connect exponentially many different configurations.
This regime naively lies within the extensively discussed many-body localized phase~\cite{Schreiber_2015,MBAA_19} induced by a quasi-periodic ``disorder"; however, our preceding discussion suggests that if such terms are indeed present, the localized phase will not exist at least for incommensurate modulations with wave vector satisfying $\cos(k^*)=\frac{p}
{2q}$.

\section*{Acknowledgments}
We thank Frank Pollmann for early collaboration at the beginning of this project. 
We thank Monika Aidelsburger, Nandagopal Manoj, Sara Murciano, Senthil Todadri, and Ruben Verresen for insightful comments and discussions. PS also acknowledges the help of Eleazer Johan Canda Orozco in making Fig.~\ref{fig:Fig1}(a,b). PS acknowledges support from the Caltech Institute for Quantum Information and Matter, an NSF Physics Frontiers Center (NSF Grant PHY-
1733907), and the Walter Burke Institute for Theoretical Physics at Caltech.
JH's research is part of the Munich Quantum Valley, which is supported by the Bavarian state government with funds from the Hightech Agenda Bayern Plus.
OIM acknowledges support by the National Science Foundation through grant DMR-2001186.
This work was completed in part at Aspen Center for Physics (PS, YY), which is supported by National Science Foundation grant PHY-2210452(PS, YY) and Durand Fund(YY).
Tensor network calculations were performed using the TeNPy Library~\cite{Tenpy}.

%\bibliography{biblio.bib}


\onecolumngrid
\begin{appendix}

\section{Spectrum of incommensurate charges} \label{sec:spectrum}
Let us consider the conventional particle number $\hat{N}=\sum_{j=1}^L \hat{n}_j$ for a system of size $L$.
It has non-negative integer spectrum $N\geq 0$ and contains exponentially large (in the system size $L$) invariant subspaces (sectors characterized by a fixed eigenvalue). 
Moreover, since we are dealing with bosonic systems, $N$ is unbounded from above. 
However, it is always the case that the minimum gap $\Delta$, i.e., the difference between two different consecutive eigenvalues, is one: $\Delta = \text{min}_{N_1\neq N_2\in \sigma(\hat{N})}(|N_2-N_1|) = 1$. 
While $\hat{\mathcal{Q}}^{A,B}$ are diagonal in the local occupation basis, these do not in general have integer spectrum due to the nontrivial $\alpha_j$.
In this section we take a closer look at their spectrum.

Quasi-periodic charges take the general form $\hat{\mathcal{Q}} = \sum_j \cos(k^*j + \phi) \nh_j$ with modulation $k^*$.
For commensurate $k^*$ with $\cos(k^*)=\frac{p}{2q}=0,1/2$, one can always choose two linearly-independent solutions $\alpha^{A,B}_j$ that periodically repeat along the chain taking values $\alpha^{A,B}_j = 0, \pm 1$. These modulated charges generate intertwined sublattice symmetry transformations such that $\hat{\mathcal{Q}}^{A,B}$ have integer spectrum with $\Delta_{q,p}=1$ and hence can be regard as microscopic U$(1)$ symmetries.
However, any other choice of $p/(2q)$ leads to non-integer spectrum.  
To analyze it we will study the spectrum as a function of system size, and as a function of $N_{\textrm{max}}$ corresponding to the maximum allowed number of bosons per site.

First of all, it is easy to convince oneself that the spectrum is extensive, i.e., the largest and smallest eigenvalues grow linearly with system size.  Recall that in this case $\alpha_j^{A,B}$ oscillate along the chain. Hence, one can distinguish between crests (where $\alpha_j>0$) and valleys ($\alpha_j<0$). The largest (smallest) eigenvalues of a given $Q_{\alpha_j}$ are attained by locating all bosons at the highest (lowest) points of the crests (valleys). As the number of periods grows linearly in system size we find that the spectrum is indeed extensive~\footnote{The growth is not strictly monotic since chains nearby in size can fit the same number of full periods.}. 

\begin{figure}[h!]
    \centering
    \includegraphics[width=0.95\textwidth]{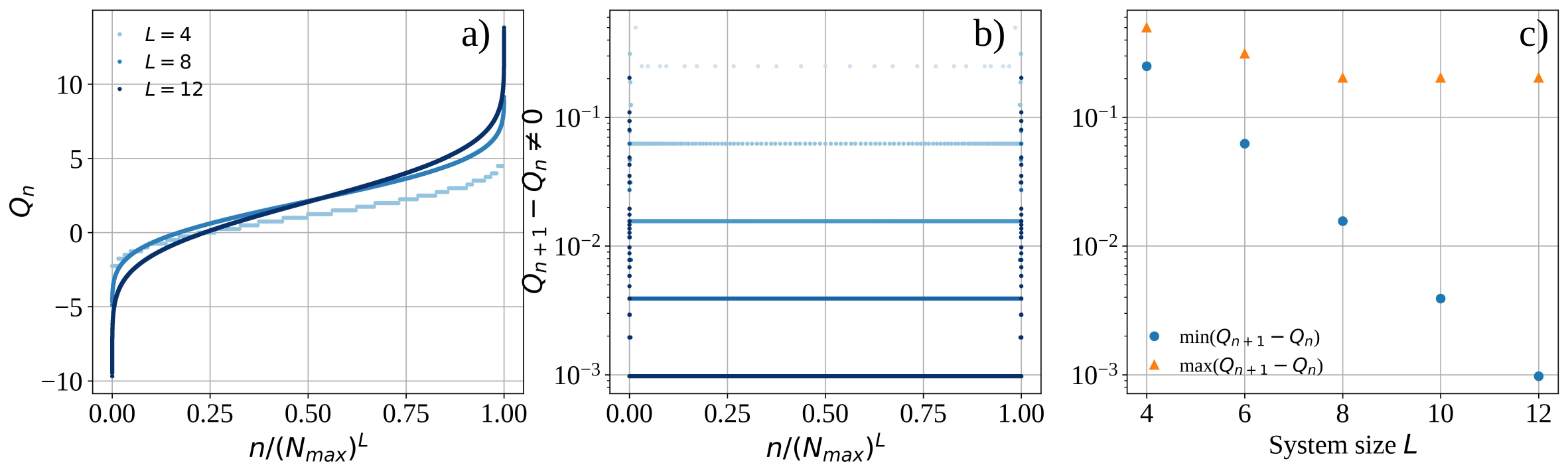}
    \caption{\textbf{Dependence on system size $L$ of the spectrum of incommensurate symmetries for $q=2,p=1$.} Data is shown for a finite maximum number of bosons per site $N_{\textrm{max}}=4$. (a) Eigenvalues $q_n$ in the spectrum of $\hat{\mathcal{Q}}_A$ for different system sizes. (b) Gap between two consecutive different eigenvalues $Q_A$. The gap is uniform within the bulk and scales as $q^{-L}$. (c) Scaling of the maximum and minimum gaps with system size.  }
    \label{fig:spect}
\end{figure}

\begin{figure}[h!]
    \centering
    \includegraphics[width=0.37\textwidth]{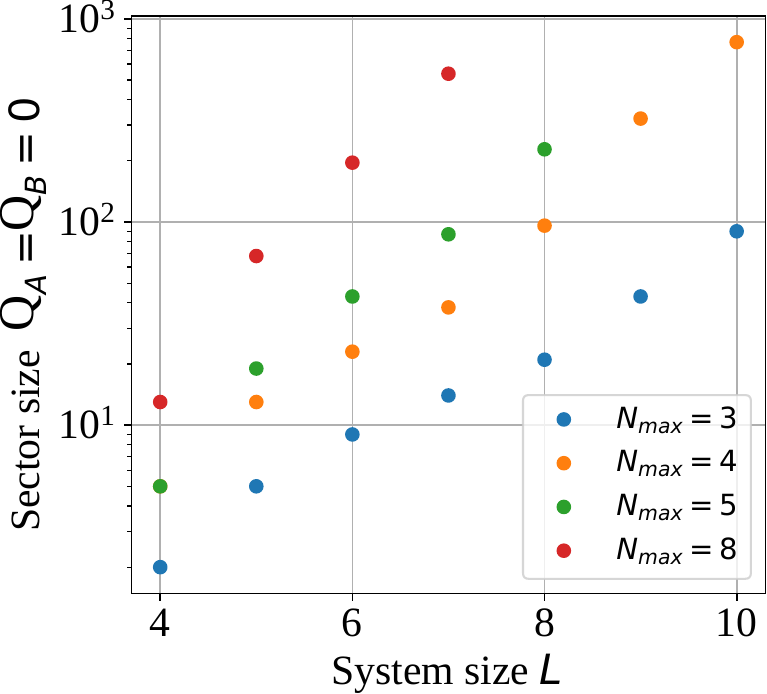}
    \caption{\textbf{Exponentially large symmetry sector $Q_A=Q_B=0$ for $q=2,p=1$.} (a) Scaling of the size of the symmetry sector $Q_A=Q_B=0$ with system size for different maximum number of bosons per site $N_{\textrm{max}}$. (b) Dimension of the Krylov space connected to the root state $\ket{2,1,1,2}\otimes \ket{0}^{\otimes (L-4)}$ relative to the full $(0,0)$ sector size. 
    }
    \label{fig:spect_nmax}
\end{figure}

For the remainder of this analysis and without loss of generality, we consider the case $p=1, q=2$ associated to incommensurate momentum $k^* = \arccos(1/4)$. In the following we summarize the main (numerical) findings shown in Fig.~\ref{fig:spect}.
First, we find that the spectrum (eigenvalues are denoted as $q_n$ with $Q_n<Q_{n+1}$) becomes denser with increasing system size as shown in panel (a). Indeed, we find that the minimum gap among non-equal eigenvalues $\Delta_{q,p}$  decreases exponentially with system size $L$ in the bulk $\Delta_{q,p}\sim q^{-L}$, and becomes uniform for sufficiently large $N_{\text{max}}$. The scaling with $q^{-L}$ can be easily predicted from the functional dependence of $\alpha_j$ that we prove in App.~\ref{app:Uj1}. 
Next, the boundary gaps lead to the largest gap among consecutive unequal eigenvalues and appear to saturate with increasing system size as shown in panel (c)
(orange triangles). 
All together, we find that the spectrum becomes dense (in the reals) in the thermodynamic limit. Namely, $Q^{A,B}$ have a continuous spectrum! Finally, we show that the global symmetry sector $Q_A=Q_B=0$ is exponentially large in system size in Fig.~\ref{fig:spect_nmax}, and plot the dependence of this degeneracy on $N_{\textrm{max}}$. 
Finally, we have numerically confirmed that even for finite $N_{\textrm{max}}$, the sector $Q_A=Q_B=0$ is almost fully connected by the action of the three-site terms introduced in the main text, with the exception of the empty state, when taking $\ket{\psi_0}=\ket{2,1,1,2}\otimes \ket{0}^{\otimes (L-4)}$ as the initial root configuration.
Nonetheless, the sector becomes fully connected when including four-sites terms $b_j^2 b_{j+1} b_{j+2} b_{j+3}^2+\textrm{H.c.}$ that also preserve $\hat{\mathcal Q}_{A,B}$.

\section{Systems with \texorpdfstring{$\hat{\mathcal{Q}}^{A,B}$}{QAB} conservation} \label{app:longrang}

In the following we show that any local Hermitian term $h_j$ acting on $k+1$ consecutive sites (i.e., geometrically $k$-local), commuting with $\hat{\mathcal{Q}}^{A,B}$ and which is not completely diagonal in the number basis, i.e., it cannot be completely written as a product of local densities $\hat{n}_j$, can be expressed as products of $(\bh_i)^q(\bh_{i+1}^\dagger)^p(\bh_{i+2})^q$ and its Hermitian conjugate, up to factors involving local densities $\hat{n}_j$.

Given the bosonic canonical commutation relations on $k+1$ sites, any local term $h_j$ can be written as 
\begin{equation}
h_j=\sum_{n=1}^{k}\sum_{\{q_i\in\mathbb{Z},n_i\in\mathbb{N}\}_{i=1}^n} J^{(n)}_{\{q_i,n_i\}}\bigotimes_{i=0}^n (\bh_{j+i}^{\textrm{sign}(q_{j+i})})^{|q_{j+i}|}(\hat{n}_{j+i})^{n_{j+i}} + \textrm{H.c.},
\end{equation}
with $\bh^{\textrm{sign}(q_{j+i})}=\bh, \bh^\dagger$ corresponding to $\#=-,+$ respectively, and $m_j\in \mathbb{Z}$ and for any choice of $n_j\in \mathbb{N}$.
Then
\begin{equation}
[h_j,\hat{\mathcal{Q}}^{A,B}]=-\sum_{n=1}^k J_n\underbrace{\left(\sum_{i=0}^nq_{j+i}\alpha^{A,B}_{j+i} \right)}_{\equiv R^j_n}\otimes_{i=0}^n (\bh_{j+i}^{\textrm{sign}(q_{j+i})})^{|q_{j+i}|}(\hat{n}_{j+i})^{n_{j+i}} - \textrm{H.c.},
\end{equation}
vanishes if and only if $R^j_n=0$ for all $n$ independently of the value of $j$. Since a general $\alpha_j$ solution of the linear recurrence \eqref{eq:LinRec} can be written as $\alpha_j=ae^{ik^*j}+be^{-ik^*j}$ with $a,b\in\mathbb{C}$, this is equivalent to the condition that the associated characteristic polynomial $P^j_n(x)=\sum_{i=0}^n q_{j+i} x^{j+i}=x^j\sum_{i=0}^n q_{j+i} x^{i}$ or simply $P_n(x)=\sum_{i=0}^n q_{j+i} x^{i}$, vanishes when evaluated at $x=e^{\pm i k^*}$, and hence $P_n(x)$ can be factorized as $P_n(x)=(qx^2-px+q)Q^j_n(x)$, with $Q^j_n(x)=\sum_{i=0}^{n-2}m_i(j)x^i$ where $m_i\in \mathbb{Z}$. In particular, this implies that $P_n(x)$ has a degree larger or equal than $2$ which translates into $k\geq 2$. Moreover, this also implies that $R^j_n=\sum_{i=0}^{n-2}m_i(j)(q\alpha^{A,B}_{j+i}-p\alpha^{A,B}_{j+i+1}+q\alpha^{A,B}_{j+i+2})$,
%, from where we deduce $q_{j}=qm_0, q_{j+1}=qm_1-pm_0$ and $q_{j+i}=qm_i-pm_{i-1}+qm_{i-2}$ for $j\geq 2$. 
i.e., $R^j_n$ is a linear combination of the linear recurrence $q\alpha^{A,B}_j-p\alpha^{A,B}_{j+1}+q\alpha^{A,B}_{j+2}=0$ (centered at different lattice sites) defining the symmetries $\hat{\mathcal{Q}}^{A,B}$. All together, the local term $h_j$ takes the form
\begin{equation}
h_j=\sum_{n=1}^{k}\sum_{\{m_i(j)\in\mathbb{Z}\}_{i=0}^{n-2}}\sum_{\{n_i\in\mathbb{N}\}_{i=0}^n} J_{\{m_i(j),n_i\}}\bigotimes_{i=0}^{n-2} \left((\bh_{j+i})^q(\bh_{j+i+1}^\dagger)^p(\bh_{j+i+2})^q\right)^{m_i(j)}\otimes_{i=0}^n (\hat{n}_{j+i})^{n_{j+i}} + \textrm{H.c.}.
\end{equation}
Hence, we conclude that every local term commuting with $\hat{\mathcal{Q}}^{A,B}$ can be written as a linear combination of products of $T^3_j\equiv (\bh_{j+i})^q(\bh_{j+i+1}^\dagger)^p(\bh_{j+i+2})^q$ acting around different sites $j$, where $\left(T^3_j\right)^{m_i(j)}$ is defined as $\left(T^{3,\dagger}_j\right)^{|m_i(j)|}$ for $m_i(j)<0$.

\section{\texorpdfstring{$Q_A=Q_B=0$}{QAB} implies \texorpdfstring{$U_j=1$}{Uj=1} for all \texorpdfstring{$j$}{j} with coprime \texorpdfstring{$q,p$}{qp} } \label{app:Uj1}

In this appendix we prove that if $q,p$ are coprime, then $U_j=1$ within the $(Q_A,Q_B)=(0,0)$ symmetry sector.
Moreover, we numerically show that for finite $N_{\textrm{max}}$ and our specific choices of coprime $q$ and $p$, the $U_j$ take constant value within any general symmetry sector $(Q_A,Q_B)$.

For the first proof, the goal is connecting any configuration $\ket{\{n_j\}}$ within the $(0,0)$ symmetry sector with the completely empty state $\ket{\{0\}}$ via symmetric moves. 
Let us denote by $\alpha_j$ the solution of the linear recurrence Eq.~\eqref{eq:LinRec} with initial conditions $\alpha_0=1, \alpha_1=p/q$ where for simplicity we are labeling sites with $j\in \{0, \dots, L-1\}$. Then, by induction one finds $q^{j-1}\alpha_j=z_j+p^j/q$ with $z_j\in \mathbb{Z}$. Indeed, assuming that is the case for $j\geq 2$ leads to $q^j\alpha_{j+1}=z_{j+1}+p^{j+1}/q$ with $z_{j+1}=pz_j-q^2z_{j-1}-qp^{j-1}\in \mathbb{Z}$.

Let us define the symmetry-preserving operator $G_j\equiv \bh_{j-1}^q(\bh_{j}^\dagger)^p \bh_{j+1}^q$, and consider the charge $\hat{\mathcal{Q}}=\sum_{j=0}^{L-1}\alpha_j \nh_j$, where without loss of generality we assume $\nh_j$ to correspond to the unbounded rotor density and $b_j^\dagger/b_j$ are the corresponding rotor raising/lowering operators (in particular, commuting with each other). 
We require this to be zero and consider the expression for $q^{L-2}\hat{\mathcal{Q}}$. Because of the above-proved property of $\alpha_j$, the contributions from $j\in \{0,1,2,\dots,L-2\}$ are all integer-valued, while the last one is $(z_{L-1}+p^{L-1}/q)n_{L-1}$. Since $q,p$ are coprime (and hence also $p$ and $q^j$) $\hat{\mathcal{Q}}=0$ implies that $n_{L-1}$ is a multiple of $q$, i.e., $n_{L-1}=qX_{L-2}$ with $X_{L-2}\in \mathbb{Z}$. Then, acting $X_{L-2}$ times with $G_{L-2}$ on the configuration $\ket{\{n_j\}}$, we obtain a new configuration $\ket{\{n_j^\prime\}}$  within the same symmetry sector but with $n^\prime_{L-1}=0.$ 
Iteratively continuing this procedure one finds that the filling of the right-most occupied site after each iteration $n^\prime_j=n_j+pX_{j}-qX_{j+1}$ is a multiple of $q$ given by $qX_{j-1}$. 
After $L-2$ such iterations iterations one then finds the configuration $\ket{\{n_j^\prime\}}\propto\prod_{j=1}^{L-2} (G_j)^{X_j}\ket{\{n_j\}}$ with $n_j^\prime=0$ for all $j\geq 2$, i.e., $\ket{\{n^\prime_j\}}=\ket{n_0^\prime, n_1^\prime, 0,\dots}$. 
Finally, imposing that a second (linearly-independent) charge $\hat{\mathcal{Q}}^{(0,1)}$ also vanishes leads to $n_0^\prime=n_1^\prime=0$. Therefore, any configuration $\ket{\{n_j\}}$ is connected to the empty state $\ket{\{0\}}$ via symmetric moves, which implies $U_j=1$ for all $j$ within the $(0,0)$ sector. Thus, we have proved that any two configurations within the sector $Q_A = Q_B = 0$ are connected by a finite number of moves when considering rotor degrees of freedom, and hence all such configurations have $U_j = 1$.
However, the previous argument also applies to bosons, as it is a claim about the property of a given configuration $\ket{\{n_j\}}$ regardless of its connectivity by the boson Hamiltonian moves. 
Regarding the connectivity by the latter which is also of interest, using numerical experiments, we have checked that with longer-range terms the bosonic models also satisfy the property that any two configurations within the sector $Q_A = Q_B = 0$ are connected by a finite number of boson Hamiltonian moves.

Moreover, we note that the previous argument provides a $1$-to-$1$ mapping between particle configurations $\{n_j\}$ on $L$ sites within the $(0,0)$ sector, and the $L-2$ integer variables $X_j$ given by
\begin{equation}
    n_j=qX_{j-1}-pX_j + qX_{j+1}
\end{equation}
after fixing $X_0=X_1=X_{L-1}=X_L=0$. As we find in the main text, this is equivalent to the duality mapping in Eq.~\eqref{eq:dualmapphi} and agrees with the solution to the continuity equation \eqref{eq:con_eq} once also taking $\mathcal{J}_x=\Delta_\tau X_j$. 
Moreover, we notice that this solution (while in a simplified version) also holds for the dipole-conserving case with $q=1, p=2$.

\section{Intermediate quasi-long range order with modulated chemical potential} \label{app:QLRO_1}
\begin{figure}[h!]
    \centering
    \includegraphics[width=0.7\textwidth]{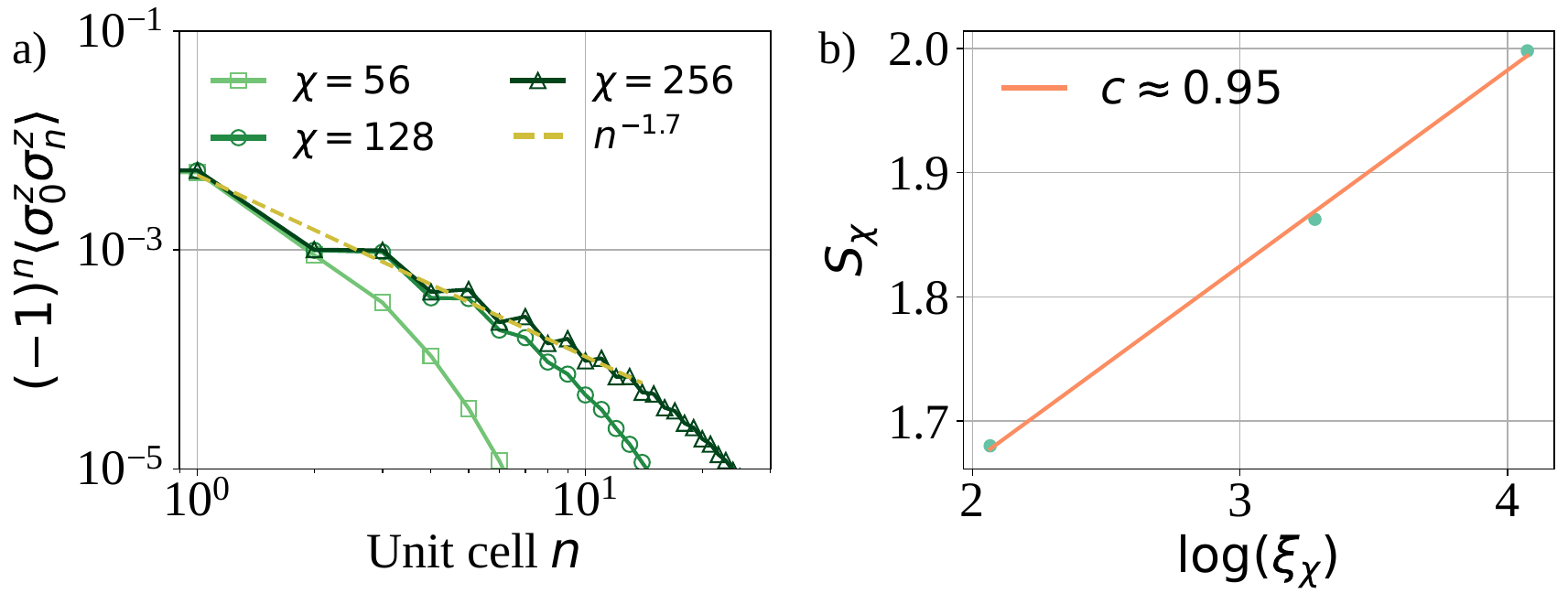}
    \caption{\textbf{Two-point correlations $\langle \sigma^z \sigma^z \rangle$ for $\bf{q,p=1}$ within the QLRO$_1$ phase.} 
    (a) Two-point spin correlations. (b) Scaling of half-chain entanglement entropy with correlation length $\xi_\chi$ induced by a finite bond dimension $\chi$. Data obtained using infinite DMRG with unit cell $L=6$.
    }
    \label{fig:QRLO1}
\end{figure}

In this section we fix $\mu=\mu_B=0$ and consider finite $\mu_A$. 
Since $\alpha^{(1,0)}_j=0,\pm 1$ is $6$-periodic, the ground state within a Mott lobe will have a charge arrangement with period $6$ in the presence of on-site interactions. To find the specific particle ordering, we require the energy of a given site to attain its minimum value. 
Focusing on the regime $\mu_A/U<1$, one finds that the system orders in a charge configuration with an empty $6$-sites unit cell $|{\mathbf 0}\rangle_n \equiv |\circ \circ \circ \circ \circ \circ \rangle_{6n+1,\dots,6n+6}$ ($n \in \mathbb{Z}$ labels unit cells) if $\mu_A/U<1/2$. 
On the other hand, for $\mu_A/U>1/2$, this unit cell configuration is given by $|\mathbf{2}\rangle_n \equiv |\circ \circ \bullet \bullet \circ \circ \rangle_{6n+1,\dots,6n+6}$
corresponding to the charges $Q_A=-2, Q_B=1$ per unit cell.
At the transition point $\mu_A/U=1/2$, these two unit cell configurations are degenerate together with two more charge orderings $|L\rangle_n \equiv |\circ \circ \bullet \circ \circ \circ \rangle_{6n+1,\dots,6n+6}$
and $|R\rangle_n \equiv |\circ \circ \circ \bullet \circ \circ \rangle_{6n+1,\dots,6n+6}$
, which have the same $Q_A=-1$ but different $Q_B$.
For small but finite $J/U$, this degeneracy is partially broken at second order in perturbation theory, with each of these configurations (except for $\ket{\bm{0}}$) shifted to lower energies
\begin{align}
    &\ket{\bm{2}}: \varepsilon_2^{(0)}\to\varepsilon_2=\varepsilon_2^{(0)}-\frac{8}{3}\frac{J^2}{U}, \text{ and }
    \ket{L},\,\ket{R}: \varepsilon_{L,R}^{(0)} \to \varepsilon_{L,R}=\varepsilon_{L,R}^{(0)}-2\frac{J^2}{U}.
\end{align}
Thus, while varying $\mu_A/U$ across $1/2$ at $J=0$ selects unique $\prod_n \ket{0}_n$ or $\prod_n \ket{2}_n$ Mott insulators, adding non-zero $J$ at $\mu_A/U=1/2$ selects $\ket{L}_n$ and $\ket{R}_n$ and keeps the degeneracy among $\ket{L}, \ket{R}$ within each unit cell configurations. 
While higher-order contributions can reduce this exponentially large degeneracy to two-fold, such as between $\otimes_n \ket{L}_n$ and $\otimes_n \ket{R}_n$, it is not sufficient to fully break it. 
In fact, this is protected by the inversion symmetry $I_{\textrm{bond}}$ around the bond center between sites $6n+3$ and $6n+4$ within each unit cell for any finite $\mu_A$ and vanishing $\mu,\mu_B$. 
This commutes with $\hat{\mathcal{Q}}^{A}$ and satisfies $\{I_{\textrm{bond}},\hat{\mathcal{Q}}^{B}\}=- \hat{\mathcal{Q}}^{A} I_{\textrm{bond}}$, implying that for any given energy eigenstate $\ket{\psi}$, $I_{\textrm{bond}}\ket{\psi}$ is a different eigenstate with the same energy as long as $2Q_B+Q_A\neq 0$. This suggests that the corresponding phase cannot be a `featureless' Mott insulator but the ground state is degenerate. Numerically we find that the corresponding phase (denoted as QLRO$_1$ in Fig.~\ref{fig:Fig1}(a)) is gapless with central charge $c=1$ as shown in Fig.~\ref{fig:Fig1}(c) or in Fig.~\ref{fig:QRLO1}(d). This behavior can be understood applying degenerate perturbation theory in the subspace spanned by $\ket{L},\ket{R}$ within each unit cell. Higher-order contributions in $J/U$ generate ``hopping'' processes among unit cells of the form $\ketbra{LR}{RL}_{n,n+1}+\textrm{H.c.}$ plus additional diagonal contributions allowed by symmetry. All together one finds an effective spin-$1/2$ Hamiltonian resembling an XXZ chain in its critical regime~\cite{giamarchi2004quantum}. Hence, one expects power-law decaying two-point correlations for the spin-spin correlation function $\langle \sigma^z_0 \sigma^z_n \rangle \sim |n|^{-2}$
with $\sigma^z_n\equiv \ketbra{L}_n - \ketbra{R}_n$, consistent with the numerical results shown in Fig.~\ref{fig:QRLO1}(c) for increasing bond dimension, as well as with the observed central charge $(c=1)$. Fig.~\ref{fig:QRLO1}(b) also shows the decay of density-density within this phase.
However, we note that this phase disappears when considering a finite chemical potential $\mu_B$.

\section{QLRO for commensurate \texorpdfstring{$q=p=1$}{q=p=1} model} \label{app:comm_approach}

Let us consider the rotor Hamiltonian 
\begin{align} \label{eq:Hrotor}
&H^{\textrm{rotor}}_{1,1}=-\sum_j J_j\cos(\thet_{j-1} - \thet_{j} +\thet_{j+1})+\frac{U}{2}\sum_j \left(\nh^r_j\right)^2,
\end{align}
where the hopping $J_j = {J}\sqrt{\bar{n}_{j-1} \bar{n}_{j} \bar{n}_{j+1}}$ is spatially modulated for a modulated chemical potential and hence satisfies $J_{j+6}=J_j$. Assuming that $J_j$ smoothly varies around a non-vanishing averaged value $J$ with $J_j=J + \delta_j$ such that $\abs{\delta_j}\ll J$ (this way $J_j$ does not change sign along the chain) or simply that $J_j=J$ as for uniform chemical potential, we can ignore the compactness of $\thet_j$ and expand the cosine around its minimum leading to 
\begin{equation} \label{eq:Hrotor_expansion}
	H_{\textrm{rotor}}\approx \sum_j\frac{J_j}{2}\left(\thet_{j-1} - \thet_{j} +\thet_{j+1}\right)^2+\frac{U}{2}\sum_j (\hat{n}_j^{\textrm{rotor}})^2 .
\end{equation}
The goal now is to derive a low-energy effective theory for the liquid phase. For example, in the standard Bose-Hubbard model the argument of the cosine contribution can be interpreted as a gradient of the field $\thet$ in the continuum limit, which is allowed by the conservation of the total particle number. Similarly, we want to identify the two low-energy degrees of freedom related to the two U$(1)$ (sublattice) symmetries generated by $\hat{\mathcal{Q}}_A$ and $\hat{\mathcal{Q}}_B$.
For the model with commensurate symmetries we are studying, we can accomplish this task decomposing the operator $\theta_j$ as %(pretty much the same as decomposing $\theta_j=e^{ik_* x}\Theta + e^{-ik_* x}\Theta^* $)
\begin{equation} \label{eq:decomp_11}
	\thet_j = \alpha^{(1,0)}_j \vartheta^A_j + \alpha^{(0,1)}_j \vartheta^B_j,
\end{equation}
where the equal sign is understood as retaining only low-energy modes, and where we assume $[\vartheta^A_i,\vartheta^B_j]=0$. Then the action of the two linearly independent symmetry transformations $\theta_j \to \theta_j + \gamma_{A/B}\alpha^{A,B}_j $ can be mapped to a uniform shift of only one of the fields
\begin{align}
	& \theta_j \to \theta_j + \gamma_{A}\alpha^{A}_j= \alpha^A_j \vartheta^A_j + \alpha^B_j \vartheta^B_j + \alpha^A_j\gamma_A =  \alpha^A_j (\vartheta^A_j + \gamma_A) + \alpha^B_j \vartheta^B_j \\
	& \theta_j \to \theta_j + \gamma_{B}\alpha^{B}_j= \alpha^A_j \vartheta^A_j + \alpha^B_j \vartheta^B_j + \alpha^B_j\gamma_B = \alpha^A_j \vartheta^A_j +  \alpha^B_j (\vartheta^B_j + \gamma_B).
\end{align}

In particular, for the symmetries of interest:
\begin{align} \label{eq:theta_j}
	\theta_j = \left\lbrace 
	 \begin{array}{l c}
		\vartheta^A_j & \textrm{if } j=1 \textrm{ mod}(6)\\
		\vartheta^B_j & \textrm{if } j=2 \textrm{ mod}(6)\\
		\vartheta^B_j- \vartheta^A_j & \textrm{if } j=3 \textrm{ mod}(6)\\
		-\vartheta^A_j & \textrm{if } j=4 \textrm{ mod}(6)\\
		-\vartheta^B_j & \textrm{if } j=5 \textrm{ mod}(6)\\
		\vartheta^A_j - \vartheta^B_j & \textrm{if } j=6 \textrm{ mod}(6)\\		
	\end{array} \right. ,
\end{align}
such that the action of the unitary transformations $U_A=e^{i\gamma_A \hat{\mathcal{Q}}_A},U_B=e^{i\gamma_B \hat{\mathcal{Q}}_B}$ on $\thet_j$ generated by $\hat{\mathcal{Q}}_A$ and $\hat{\mathcal{Q}}_B$ respectively leads to
\begin{align}
	\left\lbrace \begin{array}{l}
		\vartheta^A_j \to \vartheta^A_j + \gamma_A  \\
		\vartheta^B_j \to \vartheta^B_j
	\end{array} \right. ,
	\left\lbrace \begin{array}{l}
		\vartheta^A_j \to \vartheta^A_j  \\
		\vartheta^B_j \to \vartheta^B_j + \gamma_B
    \end{array} \right. 
\end{align} 
That is, the structure in these variables is like two uniform U$(1)$ symmetries acting on two different fields.
Now we can write $\theta_{j-1} - \theta_{j} +\theta_{j+1}$ in terms of $\vartheta^A$, $\vartheta^B$ as
\begin{align}
 \theta_{j-1} - \theta_{j} +\theta_{j+1}= \left\lbrace  \begin{array}{l c}
	\nabla_x \vartheta^B - 2\nabla_x\vartheta^A & \textrm{if } j=1 \textrm{ mod}(6)\\
	 -\nabla_x \vartheta^B - \nabla_x\vartheta^A & \textrm{if } j=2 \textrm{ mod}(6)\\
 -2\nabla_x \vartheta^B + \nabla_x\vartheta^A & \textrm{if } j=3 \textrm{ mod}(6)\\
	 -\nabla_x \vartheta^B + 2\nabla_x\vartheta^A& \textrm{if } j=4 \textrm{ mod}(6)\\
	\nabla_x \vartheta^B + \nabla_x\vartheta^A & \textrm{if } j=5 \textrm{ mod}(6)\\
	  2\nabla_x \vartheta^B - \nabla_x\vartheta^A& \textrm{if } j=6 \textrm{ mod}(6)\\		
\end{array} \right. 
\end{align}

To simplify the Hamiltonian, we can now assume that $\vartheta^{A/B}$ are smoothly varying fields which do not change at short distances and thus, we can average over a unit cell of six sites (we are interested in the long wavelength physics or rather, the physics for which $\vartheta^A$ and $\vartheta^B$ are the relevant degrees of freedom).
For finite $\mu_A \neq 0$ but $\mu_B = 0$, by inspection we have $J_{6n+1} = J_{6n+4}$, $J_{6n+2} = J_{6n+3}$, $J_{6n+5} = J_{6n+6}$.
Collecting terms over the six-site unit cell, one finds for the gradient terms:
\begin{equation}
    C [(\nabla_x \vartheta^A)^2 - \nabla_x \vartheta^A \nabla_x \vartheta^B] + C' (\nabla_x \vartheta^B)^2 = C_{AA} (\nabla_x \vartheta^A)^2 + 2 C_{AB} \nabla_x \vartheta^A \nabla_x \vartheta^B + C_{BB} (\nabla_x \vartheta^B)^2 ~, 
\end{equation}
i.e., a specific relation between the coefficients $C_{AA}$ and $C_{AB} \equiv C_{BA}$: $C_{AB} = -C_{AA}/2$, while the coefficient $C_{BB}$ is independent.
This can be traced to the inversion symmetry $I_\text{bond}$ in the bond center between $6n+3$ and $6n+4$, i.e., interchanging $6n+1 \leftrightarrow 6n+6$, $6n+2 \leftrightarrow 6n+5$, $6n+3 \leftrightarrow 6n+4$.
In the continuum, the inversion in bond center acts as (using Eq.~\eqref{eq:theta_j})
\begin{equation}
    I_\text{bond}: \vartheta^A(x) \to (\vartheta^A - \vartheta^B)(-x), \quad \vartheta^B(x) \to -\vartheta^B(-x) ~.
\end{equation}
Requiring this symmetry is enough to fix $C_{AB} = -C_{AA}/2$.

If $\mu_A = \mu_B = 0$, i.e., we have translation symmetry, $J_j = J = \text{const}$, then we also find $C_{BB} = C_{AA}$.
This can be traced to the action of the translation by one site:
\begin{equation}
    T_1: \vartheta^A \to \vartheta^B, \quad \vartheta^B\to -\vartheta^A + \vartheta^B ~.
\end{equation}
In fact, just requiring $T_1$ is enough to fix $C_{BB} = C_{AA}$, $C_{AB} = -C_{AA}/2$.

Analogously, we can also decompose $\hat{n}_j^{r}$ in terms of the integer-valued fields $n^{A,B}$ that are canonically conjugate to $\vartheta^{A,B}$ as follows
\begin{equation}
	\hat{n}_i^{\textrm{rotor}}=\frac{1}{(\alpha_i^A)^2 + (\alpha_i^B)^2}(\alpha^A_i n^A_i + \alpha^B_i n^B_i),
\end{equation}
which for the symmetries of interest reads
\begin{align}
	\hat{n}_j^{\textrm{r}} = \left\lbrace 
	 \begin{array}{l c}
		n^A_j & \textrm{if } j=1 \textrm{ mod}(6),\\
		n^B_j & \textrm{if } j=2 \textrm{ mod}(6),\\
		\frac{1}{2}(- n^A_j+b^B_j) & \textrm{if } j=3 \textrm{ mod}(6),\\
		-n^A_j & \textrm{if } j=4 \textrm{ mod}(6),\\
		-n^B_j & \textrm{if } j=5 \textrm{ mod}(6),\\
		\frac{1}{2}(n^A_j - n^B_j) & \textrm{if } j=6 \textrm{ mod}(6).\\		
	\end{array} \right. 
\end{align}

This choice is consistent with $n^A,n^B$ being the corresponding conjugate variables $[\vartheta^A,n^A]=[\vartheta^B,n^B]=i$ such that $[\vartheta^A,n^B]=[\vartheta^B,n^A]=0$. Analogous to the previous discussion, in the liquid phase we can then average the contribution over the $6$-sites unit cells 
\begin{equation}
    \sum_{j=1}^6 (\hat{n}_j^{\textrm{rotor}})^2= \Omega_{AA}(n^A)^2 + 2\Omega_{AB}n^An^B + \Omega_{BB}(n^B)^2
\end{equation}
and find $\Omega_{AA}=\Omega_{BB},\Omega_{AB}=-\Omega_{AA}/5$. All together we obtain the quadratic theory
\begin{align} \label{eq:rotor2TLL}
	H_{\textrm{rotor}}\approx \sum_j\frac{1}{2}\nabla_x \bm{\vartheta}_j^T\cdot \bm{K} \cdot \nabla_x \bm{\vartheta}_j +\frac{1}{2}\sum_j \bm{N}_j \cdot \bm{\Omega}  \cdot \bm{N}_j,
\end{align}
with $\bm{\vartheta}=(\vartheta^A,\vartheta^B)^T$, $\bm{N}=(n^A,n^B)^T$ and 
\begin{equation}
    \bm{K}= \left( \begin{matrix} C_{AA} & -\frac{C_{AA}}{2} \\ -\frac{C_{AA}}{2} & C_{BB} \end{matrix}\right), \hspace{15pt} \bm{\Omega}= \left( \begin{matrix} \Omega_{AA} & -\frac{\Omega_{AA}}{5} \\ -\frac{\Omega_{AA}}{5} & \Omega_{BB} \end{matrix}\right).
\end{equation}

The alternative decomposition $\thet_j =  2 \Re [e^{ik^* j} (\vartheta^{(1)}_j + i \vartheta^{(2)}_j) ]$ relates to Eq.~\eqref{eq:decomp_11} via the non-orthogonal transformation
\begin{equation}
    \vartheta^A = \vartheta^{(1)} - \sqrt{3} \vartheta^{(2)}, \quad
    \vartheta^B = -\vartheta^{(1)} - \sqrt{3} \vartheta^{(2)}.
\end{equation}
This diagonalizes the quadratic form $\bm{K}$ for uniform $J_j$ in the absence of the second term.
In fact, for uniform $J_j$ one finds 
\begin{equation} \label{eq:quadratic}
    (\nabla_x\vartheta^A)^2 - \nabla_x\vartheta^A \nabla_x\vartheta^B + (\nabla_x\vartheta^B)^2 = 
    3 [(\nabla_x\vartheta^{(1)})^2 + (\nabla_x\vartheta^{(2)})^2 ].
\end{equation}
In general one needs to simultaneously diagonalize $\bm{K}$ and $\bm{\Omega}$, e.g., following Ref.~\onlinecite{Lai2010}. From here one finds two decoupled harmonic oscillators consistent with the expected two-species Luttinger liquid that emerges in the large $J/U$ regime.

\section{Naive derivation of the Villain action for general \texorpdfstring{$q,p$}{qp}} \label{app:derivation}
The main idea of the Villain formulation is to replace the cosine potential-term in the rotor Hamiltonian \eqref{eq:Hrotor_qp} without losing the $2\pi$-periodicity that is so relevant to understand the role of vortices to disorder the system. 
The advantage is proceeding in a more systematic and controlled fashion when deriving the low-energy theory, which applies to all choices of $q$ and $p$, and not only to commensurate ones.
Proceeding in the standard way, the path integral of this system is given by
\begin{align} \label{eq:act_Villain}
 \nonumber	&\mathcal{Z}(\beta)\equiv \textrm{tr}\left[e^{-\beta H_{q,p}}\right]=\int \prod_{j,\tau}\frac{d\theta_j(\tau)}{2\pi} \sum_{n_j(\bar{\tau})} \exp\left\{\delta \tau \sum_{2\leq j\leq L-1,\tau}J_{j-1} \cos[q\theta_{j-1}(\tau)-p\theta_j(\tau) + q\theta_{j+1}(\tau)] \right. \\
	&\left. -\sum_{m=0}^{\lceil \frac{2q+p}{2}\rceil}\frac{U_m\delta \tau}{2}\sum_{j,\tau}(n_j(\bar{\tau}) - \bar{n}_j)^{m}+i\sum_{j,\tau} n_j(\bar{\tau})[\theta_j(\tau +1)-\theta_j(\tau)]\right\}
\end{align}
where $\bar{n}_j$ is the average density per site. Here, the variables $\theta_j(\tau)$ live on the nodes $(j,\tau)$ of the Euclidean (1+1)D lattice, while $n_j(\bar{\tau})$ live on the vertical links. Moreover, we are considering spatial open boundary conditions such that $\hat{\mathcal{Q}}^{A,B}$ are conserved quantities regardless the choice of $q,p$, and in the end assuming an infinitely long chain. While a detailed analysis can be carried for finite systems, boundary effects do not change the outcome. 
%Notice that the time-independent contribution $\bar{n}_j$ vanishes when integrating over the time(-periodic) component
%\begin{equation}
%    \sum_{j}\bar{n}_j\sum_{\tau}\left(\theta_j(\tau+1)-\theta_j(\tau)\right)=0.
%\end{equation}

We then Villainize the action by replacing the cosine potential as
\begin{align} \nonumber
	&e^{\delta \tau \sum_{j,\tau} J_{j-1}\cos(q\theta_{j-1}(\tau)-p\theta_j(\tau) + q\theta_{j+1}(\tau))}\approx \sum_{\{\ell(j,\tau)\in \mathbb{Z}\}}e^{-\sum_{j,\tau}\frac{\delta \tau J_{j-1}}{2}[q\theta_{j-1}(\tau)-p\theta_j(\tau) + q\theta_{j+1}(\tau)-2\pi \ell(j,\tau)]^2}
\end{align}
where $J_j$ on the right hand side is proportional to that on the left at low temperatures. Notice that the resulting theory is still $2\pi$-periodic in $\theta_j$. Using Poisson resummation formula
\begin{equation}
    \sum_{\ell\in\mathbb{Z}}g(\ell)= \sum_{m\in \mathbb{Z}}\int_{-\infty}^{+\infty}dx\, g(x)e^{-i2\pi m x},
\end{equation}
we finally obtain
\begin{align} \nonumber
&\sum_{\{\ell(j,\tau)\in \mathbb{Z}\}}e^{-\sum_{j,\tau}\frac{\delta \tau J_{j-1}}{2}[q\theta_{j-1}(\tau)-p\theta_j(\tau) + q\theta_{j+1}(\tau)-2\pi \ell(j,\tau)]^2} \\ &=\prod_{j,\tau}\frac{1}{2\pi \sqrt{\delta \tau J_j}}\sum_{\{R_j(\tau)\in \mathbb{Z}\}} e^{-\sum_{j,\tau}\frac{R_j^2(\tau)}{2\delta \tau J_{j-1}} + iR_j(\tau)[q\theta_{j-1}(\tau)-p\theta_j(\tau) + q\theta_{j+1}(\tau)]}.
\end{align}

We can now integrate out $\theta_j$ and write a theory in terms of $n$ and $R$ variables, the latter living on sites $(j,\tau)$. To do so we collect all terms linear in $\theta_j$
\begin{align}
&\sum_{j,\tau} in_j(\bar{\tau})[\theta_j(\tau +1)-\theta_j(\tau)] + \sum_{2\leq j\leq L-1,\tau}iR_j(\tau)[q\theta_{j-1}(\tau)-p\theta_j(\tau) + q\theta_{j+1}(\tau)] \\
&= \sum_{j,\tau} -i \theta_j(\tau)[n_j(\bar{\tau})-n_j(\bar{\tau}-1)+pR_j(\tau) - qR_{j+1}(\tau) - qR_{j-1}(\tau)],
\end{align}
such that integrating over $\theta_j(\tau)$ 
\begin{align}
\mathcal{Z}(\beta)=\prod_{j,\tau} \int_0^{2\pi} \frac{d\theta_j(\tau)}{2\pi}\exp\left({ -i \theta_j(\tau)[n_j(\bar{\tau})-n_j(\bar{\tau}-1)+pR_j(\tau) - qR_{j+1}(\tau) - qR_{j-1}(\tau)]}\right)\\
\times \sum_{n_j(\bar{\tau})}  \sum_{R_j=-\infty}^{\infty} \exp\left[-\sum_{j,\tau}
\frac{R_j^2(\tau)}{2\delta \tau J_{j-1}}
-\sum_{m=2}\frac{U_m\delta \tau}{2}\sum_{j,\tau}(n_j(\bar{\tau}) - \bar{n}_j)^{m}\right],
\end{align}
leads to the constrained action
\begin{align}
\mathcal{Z}(\beta)&=\sum_{\{n_j(\bar{\tau})\}}  \sum_{\{R_j(\tau)\}} \exp\left[-\sum_{j,\tau}\frac{R_j^2(\tau)}{2\delta \tau J_{j-1}}-\sum_{m=2}\frac{U_m\delta \tau}{2}\sum_{j,\tau}(n_j(\bar{\tau}) - \bar{n}_j)^{m}\right]\\
&\times \prod_{j,\tau}\delta\left[n_j(\bar{\tau})-n_j(\bar{\tau}-1)+pR_j(\tau) - qR_{j+1}(\tau) - qR_{j-1}(\tau)\right],
\end{align}
where $\delta[x]$ denotes Kronecker delta $\delta_{x,0}$.

Interpreting $J_{\tau}=n$, $J_x=R$ as temporal and spatial components of an appropriate ``current'', the constraint
\begin{align}\label{eq:cons_nm}
	&n_j(\bar{\tau})-n_j(\bar{\tau}-1)+pR_j(\tau) - qR_{j+1}(\tau) - qR_{j-1}(\tau)=0, 
\end{align}
at every site $(j,\tau)$ corresponds to the conservation of total charge around site $(j,\tau)$
\begin{equation}
	\nabla_{\tau} J_{\tau} + \tilde{\nabla}^{q,p}_x J_{x} = 0,
\end{equation}
with the modified spatial derivative $\tilde{\nabla}^{q,p}_x$ defined as $\tilde{\nabla}^{q,p}_x R_j(\tau)\equiv pR_j(\tau) - qR_{j+1}(\tau) - qR_{j-1}(\tau)$. 
This constraint can be directly incorporated writing $n$ and $R$ in terms of a ``height field'' $X$ living on vertical links $(\tau, \tau+1)$ \footnote{Notice that the second expression can be rewritten as
\begin{equation}
    n_j(\bar{\tau})=\tilde{\nabla}^{q,p}_{x}\phi=-q(\nabla_x)^2\phi + (p-2q)\phi,
\end{equation}
where the second term vanishes for dipole-conserving systems ($p=2q$). In this case the theory developed in Ref.~\onlinecite{Lake22_1DDBHM} applies.} via 
\begin{align} \label{eq:currentdualitymj}
&R_j(\tau)=\nabla_{\tau}X_j=X_j(\bar{\tau}) - X_j(\bar{\tau}-1)\\
&\label{eq:currentdualitynj} n_j(\bar{\tau})=-\tilde{\nabla}^{q,p}_{x}X= qX_{j-1}(\bar{\tau}) -pX_j(\bar{\tau})  + qX_{j+1}(\bar{\tau}).
\end{align}
As previously stated, while here we are not taking explicit care of boundary contributions for finite systems ---where the previous equations are slightly modified--- these do not modify the results.  

In the standard Bose-Hubbard model one finds similar relations but with $\tilde{\nabla}^{q,p}_{x}$ replaced by the first-order finite difference $\nabla_x$. This leads to the conclusion that integer-valuedness of $n_j, R_j$ implies the same for $X_j(\bar{\tau})$. Hence, to proceed one usually softens this constraint and consider a real-valued field $\chi_j(\bar{\tau})$ via a potential contribution of the form $-\lambda \cos(2\pi \chi_j(\bar{\tau}))$ at every spacetime site $(j,\tau)$. However, this is not the case here for generic $q,p$. 
Equation~\eqref{eq:currentdualitymj} only implies that the difference $X_j(\bar{\tau})-X_j(\bar{\tau}-1)$ between two consecutive times at spatial site $j$ is an integer. Let us fix a reference time slice $\tau_{\textrm{ref}}$ and find the value of $X_j$ on that slice. 
Equation~\eqref{eq:currentdualitynj} (and \eqref{eq:currentdualitymj}) implies that 
\begin{equation}
    X_j(\bar{\tau})=\frac{1}{q}\sum_{i< j}\alpha_{j-i}^{1,p/q}n_i(\bar{\tau}) ,
\end{equation}
with $\alpha_0^{1,p/q}=1, \alpha_1^{1,p/q}=p/q$ and $ X_{L-1}= \frac{1}{p}(n_{L-1}+qX_{L-2})$.
However, for general $q,p$ (unlike for commensurate choices) $\alpha_{j-i}^{1,p/q}$ is not integer-valued but only rational! Hence, $X_j(\tau_{\textrm{ref}}+1/2)$ is only constrained to be $\mathbb{Q}$-valued such that at any other time $X_j(\bar{\tau})-X_j(\bar{\tau}_{\textrm{ref}})\in \mathbb{Z}$. 

To proceed further we decompose $X_j(\bar{\tau})=I_j(\bar{\tau})+m_j(\bar{\tau})$ with the condition that $I_j\in \mathbb{Z}$ and $| m_j |<1$. Then we find
\begin{align} \label{eq:I_m_decomp}
&R_j(\tau)=\nabla_{\tau}X=I_j(\bar{\tau}) - I_j(\bar{\tau}-1)+m_j(\bar{\tau}) - m_j(\bar{\tau}-1), \\
&n_j(\bar{\tau})=-\tilde{\nabla}^{q,p}_{x}X_j= -\tilde{\nabla}^{q,p}_{x}I_j -\tilde{\nabla}^{q,p}_{x}m_j.
\end{align}
We now recall the existence of the discrete symmetries $U_j$ satisfying the constraints $(U_j)^q=(U_{j-1})^p(U^\dagger_{j-2})^q$ for all $j$ with $U_0=U_{-1}=1$. Moreover, from the duality mapping in Section~\ref{sec:frag}  we know that we can write $U_j=e^{i2\pi X_{j+1}}=e^{i2\pi m_{j+1}}$.  Hence, the $m_j(\bar{\tau})$'s are fixed after fixing the $(\mathbb{Z}_q)^L$ symmetry. For example, fixing $U_1$ fixes the value of $m_1$, since $(U_1)^q=e^{i2\pi qm_1}=1$, which implies $m_1=n_1/q$ with $n_1\in\{0,1,\dots, q-1\}$. However, recall that the $\mathbb{Z}_q$ are not locally realized, meaning, each $U_j$ does not generate a $\mathbb{Z}_q$ factor but one rather needs all of them to find $\mathbb{Z}_q^L$ after imposing the condition relating the $U_j$'s at different sites. In fact, this condition is equivalent to imposing $Z_j\equiv\tilde{\nabla}^{q,p}_{x}m_j\in \mathbb{Z}$, which is precisely the second term in the second line of Eq.~\eqref{eq:I_m_decomp}: $(U_j)^q(U_{j-1})^{-p}(U_{j-2})^q=e^{i2\pi \tilde{\nabla}^{q,p}_{x}m_j}=e^{i2\pi Z_j}=1$.  Hence, $n_j(\bar{\tau})=-\nabla^{q,p}_x I_j(\bar{\tau})-Z_j$, with $Z_j\in \mathbb{Z}$ fixed by the $\mathbb{Z}_q$ symmetries.  Therefore, the field that we should relax from integer- to real-valued is $I_j$: $I_j\to \chi_j\in \mathbb{R}$ via introducing the softening potential $\cos(2\pi \chi_j)$. To get rid of $\bar{n}_j$ in the Hamiltonian, we split $\chi=\delta \chi + \bar{\chi}$ such that 
\begin{equation} \label{eq:dens_cond}
n_j-\bar{n}_j = -\nabla^{q,p}_x  \delta \chi_j, \text{ and } \,\,\bar{n}_j+Z_j=-\nabla^{q,p}_x \bar{\chi}_j,
\end{equation}
since $Z_j$ are fixed by the symmetries. 

Overall, this leads to the partition sum $\mathcal{Z}(\beta) \propto \sum_{\{\delta \chi_j(\bar{\tau})\}} e^{-S[\delta \chi] }$ with action
\begin{equation} \label{eq:inc_action}
	S[\delta \chi]=\frac{1}{2}\sum_{j,\tau}\left[\frac{1}{\tilde{J}_j}(\nabla_{\tau} \delta \chi_{j})^2 + \sum_{m=2}^{\lceil \frac{p+2q}{2} \rceil} \tilde{U}_m(\tilde{\nabla}^{q,p}_x \delta \chi_j)^{m}   \right]-\lambda \sum_{j,\tau}\cos(2\pi (\delta \chi_j +\bar{\chi}_j)),
\end{equation}
and $\tilde{J}_j\equiv \delta \tau J_j$, $\tilde{U}_{{m}} \equiv \delta\tau U_{{m}}$. For example, for a uniform potential we need to solve $-\tilde{\nabla}^{q,p}_{x}\bar{\chi}_j=\bar{n}+Z_j$ which can be achieved via $\bar{\chi}_j= \frac{\bar{n}}{2q-p} + \frac{1}{q}\sum_{i\leq j}\alpha_{j-i}^{1,p/q}Z_i+\alpha_j$ with $\alpha_j$ any solution of the recurrence equation $q\alpha_j -p\alpha_{j+1} +q\alpha_{j+2}=0$ (which appears in the finite difference operator $\tilde{\nabla}_{x}^{q,p}$) and $p\neq 2q$.  Compare this with the solution for the standard Bose-Hubbard model $\bar{\chi}_j=\bar{n}j + a$ which is fixed up to a constant value $a$~\footnote{This corresponds to a uniform shift $2\pi p a$ in the argument of the cosine.}. Similarly to the Bose-Hubbard model, we fix the trivial solution with $\alpha_j=0$. We can also consider a modulated chemical potential in which case $\bar{\chi}_j$ is given by
\begin{equation}
	\bar{\chi}_{j+1}=\frac{1}{q}\sum_{i\leq j} \alpha^{(1,p/q)}_{j-i} (\bar{n}_i+Z_i).
\end{equation}

Moreover, since $U_m$-terms with $m>2$ becomes less and less relevant with increasing $m$, we can just keep the $m=2$ contribution, leading to the quadratic theory
\begin{equation} \label{eq:S0UV}
	S_0[\delta \chi]=\frac{1}{2}\sum_{j,\tau}\left[\frac{1}{\tilde{J}_j}(\nabla_{\tau} \delta \chi_{j})^2 + \tilde{U}(\tilde{\nabla}^{q,p}_x \delta \chi_j)^{2}   \right],
\end{equation}
or in frequency-momentum space (taking the time-continuum limit)
\begin{equation} \label{eq:S0UV_kw}
   S_0[\delta \chi]=\frac{K}{2\beta L}\sum_{k,\omega_n}\left[  \frac{\omega_n^2}{c} + c(p-2q\cos(k))^2\right]|\chi_{k,\omega_n}|^2, 
\end{equation}
where we have defined Fourier components by 
\begin{equation} 
\delta\chi_j(\tau) = \frac{1}{\beta L}\sum_{k, i\omega_n}e^{i(kj-\omega_n\tau)}\chi_{k,\omega_n},
\end{equation}
and $k=\frac{2\pi}{L}n\in [-\pi,\pi]$, $\omega_n=\frac{2\pi}{\beta}n\in (-\infty,+\infty)$.

As we have shown in the main text, this derivation simplifies when fixing the global symmetry sector $Q_A=Q_B=0$, where the field $X_j$ is directly integer-valued,  and the mapping to the integer-valued height model with appropriate boundary conditions on $X$ is exact on a finite system with OBC.

\section{Correlation functions and Wilson RG} \label{app:correlations}

\subsection{Vertex-vertex correlations}
Here we compute two-point correlations of the form 
\begin{equation} \label{eq:Cchichi}
C_{\chi \chi}(j-j^\prime, \bar{\tau}-\bar{\tau}^\prime)\equiv\langle e^{i\chi_j(\bar{\tau})}e^{-i\chi_{j^\prime}(\bar{\tau}^\prime)}\rangle_{0} = e^{-\frac{1}{2}\langle\left(\chi_j(\bar{\tau})-\chi_{j^\prime}(\bar{\tau}^\prime)\right)^2\rangle_{{0}}},
\end{equation}
for a system with incommensurate $k^*$, where in the second equality we have used that the corresponding action \eqref{eq:S0UV_kw} defining $\langle \dots \rangle_0$ is quadratic. 
Assuming a uniform coupling $J_j=J$ one finds
\begin{equation}
    \langle \chi_{k,\omega_n}\chi_{k^\prime,\omega_n^\prime}\rangle = \frac{\beta L\delta_{k,-k^{\prime}}\delta_{\omega_n,-\omega_n^{\prime}}}{\frac{1}{J}\omega_n^2+U(2q\cos(k)-p)^2}.
\end{equation}
From here we obtain the general expression
\begin{equation}
    -2\log C_{\chi \chi}(j-j^\prime, \bar{\tau}-\bar{\tau}^\prime)=\int_{-\infty}^{+\infty}\frac{d\omega}{2\pi}\int_{-\pi}^{+\pi}\frac{dk}{2\pi} \frac{2\left[1-\cos\left[k(j-j^\prime)-\omega(\tau-\tau^\prime)\right]\right]}{\frac{1}{J}\omega^2+U(2q\cos(k)-p)^2}.
\end{equation}

$\bullet$ When $j=j^\prime$ one finds
\begin{align}
    &-2\log C_{\chi \chi}(0, \bar{\tau}-\bar{\tau}^\prime)=\int_{-\infty}^{+\infty}\frac{d\omega}{2\pi}\int_{-\pi}^{+\pi}\frac{dk}{2\pi} \frac{2\left[1-\cos\left[\omega(\bar{\tau}-\bar{\tau}^\prime)\right]\right]}{\frac{1}{J}\omega^2+U(2q\cos(k)-p)^2} \\
   & = \frac{1}{2K}\int_{-\pi}^{+\pi}\frac{dk}{2\pi} \frac{1-e^{-\sqrt{JU}|\bar{\tau}-\bar{\tau}^\prime||2q\cos(k)-p|}}{|2q\cos(k)-p|}\approx \frac{4}{2K}\int_{0}^{+\epsilon}\frac{d\delta k}{2\pi} \frac{1-e^{-\sqrt{JU}|2q\sin(k^*)(\bar{\tau}-\bar{\tau}^\prime)\delta k|}}{2q|\sin(k^*)|\delta k}
\end{align}
after expanding $2q\cos(k)-p$ around $k^*$. Then considering sufficiently distanced times $|(\bar{\tau}-\bar{\tau}^\prime)\delta k|> 1$ one finally gets
\begin{equation}
    -2\log C_{\chi \chi}(0, \bar{\tau}-\bar{\tau}^\prime) \approx \frac{4}{4qK|\sin(k^*)|}\int_{1/|\bar{\tau}-\bar{\tau}^\prime|}^{+\epsilon}\frac{d\delta k}{2\pi} \frac{1}{\delta k}\sim \frac{1}{2\pi qK|\sin(k^*)|}\log(|\bar{\tau}-\bar{\tau}^\prime|),
\end{equation}
i.e., $ C_{\chi \chi}(0, \bar{\tau}-\bar{\tau}^\prime)\sim |\bar{\tau}-\bar{\tau}^\prime|^{-1/(4\pi qK |\sin(k^*)|)}$.

$\bullet$ However, spatial correlations are ultra-local for incommensurate $k^*$: Indeed, take $\tau=\tau^\prime$
\begin{align}
    &-2\log C_{\chi \chi}(j-j^\prime, 0)=\int_{-\infty}^{+\infty}\frac{d\omega}{2\pi}\int_{-\pi}^{+\pi}\frac{dk}{2\pi} \frac{2\left[1-\cos\left[k(j-j^\prime)\right]\right]}{\frac{1}{J}\omega^2+U(2q\cos(k)-p)^2}\\
    &= \frac{1}{K}\int_{-\pi}^{+\pi}\frac{dk}{2\pi} \frac{1-\cos k(j-j^\prime)}{|2q\cos(k)-p|},
\end{align}
the resulting integral can diverge $k=\pm k^*$. Expanding around those point gives 
\begin{align}
    &-2\log C_{\chi \chi}(j-j^\prime, 0)=\frac{1}{K}\int_{-\epsilon}^{+\epsilon}\frac{d\delta k}{2\pi} \frac{2-\cos ( k^* + \delta k)(j-j^\prime)-\cos (- k^* + \delta k)(j-j^\prime)}{2q|\sin(k^*)||\delta k|} \\
    &= \frac{4}{K}\int_{0}^{+\epsilon}\frac{d\delta k}{2\pi} \frac{1-\cos  k^* (j-j^\prime)\cos  \delta k(j-j^\prime)}{2q|\sin(k^*)||\delta k|} \to -\log(0^+)=\infty,
\end{align}
and then, $-2\log C_{\chi \chi}(j-j^\prime, 0)\sim 0.$ All together for incommensurate symmetries
\begin{equation} \label{eq:corrincom}
    C_{\chi \chi}(j-j^\prime, \tau -\tau^\prime)= \begin{cases}
    1/|\tau-\tau^\prime|^{\frac{1}{4\pi qK|\sin(k^*)|}} \textrm{ if } j=j\prime, \\[1ex]
    0 \textrm{ if } j\neq j^\prime.
    \end{cases}
\end{equation}
For commensurate $k^*$ and for $j-j^\prime=\frac{2\pi}{k^*}N$ with $N\in \mathbb{Z}$, spatial correlations instead decay as a power-law
\begin{equation}
    C_{\chi \chi}(j-j^\prime, 0) \sim |j-j^\prime|^{-\#/K}.
\end{equation}

\subsection{Density-density correlations}

Another signature of the quasi-long range order in the liquid phase is the polynomial decay of density-density correlators $\langle \hat{n}_j\hat{n}_{j^\prime} \rangle_c=\langle (\hat{n}_j-\bar{n})(\hat{n}_{j^\prime}-\bar{n}) \rangle$ with the distance $x=j-j^\prime \in \mathbb{Z}$. In the following we present the result of this computation. To start we recall that $n_j(\bar{\tau})=-\nabla^{q,p}_x\chi_j(\bar{\tau})$, hence
\begin{align}
    \langle \hat{n}_j\hat{n}_{j^\prime} \rangle_c=\frac{1}{\beta L}\sum_{k,\omega_n}e^{i\left(kx-\omega(\tau-\tau^\prime)\right)}\frac{(2q\cos(k)-p)^2}{\frac{1}{J}\omega^2+U(2q\cos(k)-p)^2}
\end{align}
and evaluating at $\bar{\tau}=\bar{\tau}^\prime$ we find
\begin{align}
    \langle \hat{n}_j\hat{n}_{j^\prime} \rangle_c &= \frac{1}{2K}\int_{-\pi}^{\pi}\frac{dk}{2\pi}e^{ikx}|2q\cos(k)-p|= \frac{2q}{2K}\int_{0}^{\pi}\frac{dk}{2\pi}\cos\left(kx\right)|\cos(k)-\cos(k^*)|\\
    &=\frac{2q}{2K}\left(2f(k^*)-f(\pi)-f(0)\right)
\end{align}
with
\begin{equation}
    f(k)=\frac{1}{x^2-1}\left(x\cos(k)\sin(kx)-\sin(k)\cos(kx) \right)-\frac{1}{x}\cos(k^*)\sin(k^*x).
\end{equation}
Therefore, in the limit $|x|\to\infty$ and using that $x\in \mathbb{Z}$ for the lattice calculation we find that
\begin{align}
    \langle \hat{n}_j\hat{n}_{j^\prime} \rangle_c \approx \frac{-2q\sin(k^*)}{K}\cos(k^*(j-j^\prime))\frac{1}{|j-j^\prime|^2} + O\left(\frac{1}{|j-j^\prime|^3}\right).
\end{align}
Recall that for the standard Luttinger liquid theory one obtains a polynomial decay with the same power-law exponent but without the spatial modulation. Moreover, one can then also conclude that the static structure factor $\langle \hat{n}_k\hat{n}_{-k} \rangle_c $ becomes $\langle \hat{n}_k\hat{n}_{-k} \rangle_c \sim |k\pm k^*|$ close to $k\approx \pm k^*$ as shown in Fig.~\ref{fig:Fig2_QLRO}. 

\subsection{Wilson renormalization group for incommensurate symmetries}
\label{app:RG_wilson}

In this appendix we calculate the renormalization group (RG) eigenvalue of the cosine term $\delta S\equiv -\lambda \int d\tau \int dx \cos(2\pi \chi_j)$ for incommensurate $k^*$ with respect to the quadratic action
\begin{equation}
    S=\int_0^{L_\tau}d\tau \left\{\sum_{j=1}^{L}\frac{1}{2J}(\partial_\tau \chi_j)^2+\frac{U}{2}(q\chi_{j-1}-p\chi_j +q\chi_{j+1})^2\right\}.
\end{equation}
First, we write
\begin{equation}
    \chi(x,\tau)=\int_{-\infty}^{+\infty} \frac{d\omega}{2\pi}\int \frac{dk}{2\pi}\chi(k,\omega)e^{i(kx-\omega\tau)}
\end{equation}
where we have only kept low-energy momenta close to $\pm k^*$, i.e., $|k\pm k^*|\leq \Lambda$. We now integrate out fast spatial-modes $|k\pm k^*|\in (\frac{\Lambda}{b},\Lambda)$ denoted as $\chi_>$ with $b>1$, while keeping $\omega \in (-\infty,+\infty)$. We obtain
\begin{align}
   & \langle \delta S \rangle_{>,\text{conn}}=-\lambda \int d\tau \int dx \langle \frac{1}{2}e^{i2\pi (\chi_>+\chi_<)}+\textrm{c.c.}\rangle_{>,\text{conn}}\\
   & =-\lambda \int d\tau \int dx \cos(2\pi \chi_<)e^{-\frac{1}{2}(2\pi)^2\langle \chi_>^2(x,\tau)\rangle_{>}},
\end{align}
where 
\begin{align}
    \nonumber &\langle \chi_>^2(x,\tau)\rangle_{>}=\int_{-\infty}^{+\infty} \frac{d\omega}{2\pi}\int_{|k\pm k^*|\in (\frac{\Lambda}{b},\Lambda)} \frac{dk}{2\pi}\frac{2}{\frac{\omega^2}{J}+U(2q\cos(k)-p)^2}\\
    &=4\times \sqrt{\frac{J}{U}}\int_{\Lambda/b}^{\Lambda}\frac{d\delta k}{2\pi}\frac{1}{2q\sin(k^*)\delta k}=\sqrt{\frac{J}{U}}\frac{4}{4\pi q\sin(k^*)}\ln(b).
\end{align}
All together we find
\begin{equation}
     \langle \delta S \rangle_{>,\text{conn}}-\lambda \int d\tau \int dx \cos(2\pi \chi_<)b^{-\sqrt{2\pi}{q\sin(k^*)K}},
\end{equation}
which implies that $\delta S$ has a scaling dimension $\Delta[\delta S]\propto \frac{1}{K}$, and hence it can be become relevant for a finite values of the Luttinger parameter $K_c$.

\section{Numerical analysis and convergence} \label{app:conv_anal}

\begin{figure}
    \centering
    \includegraphics[width=0.5\linewidth]{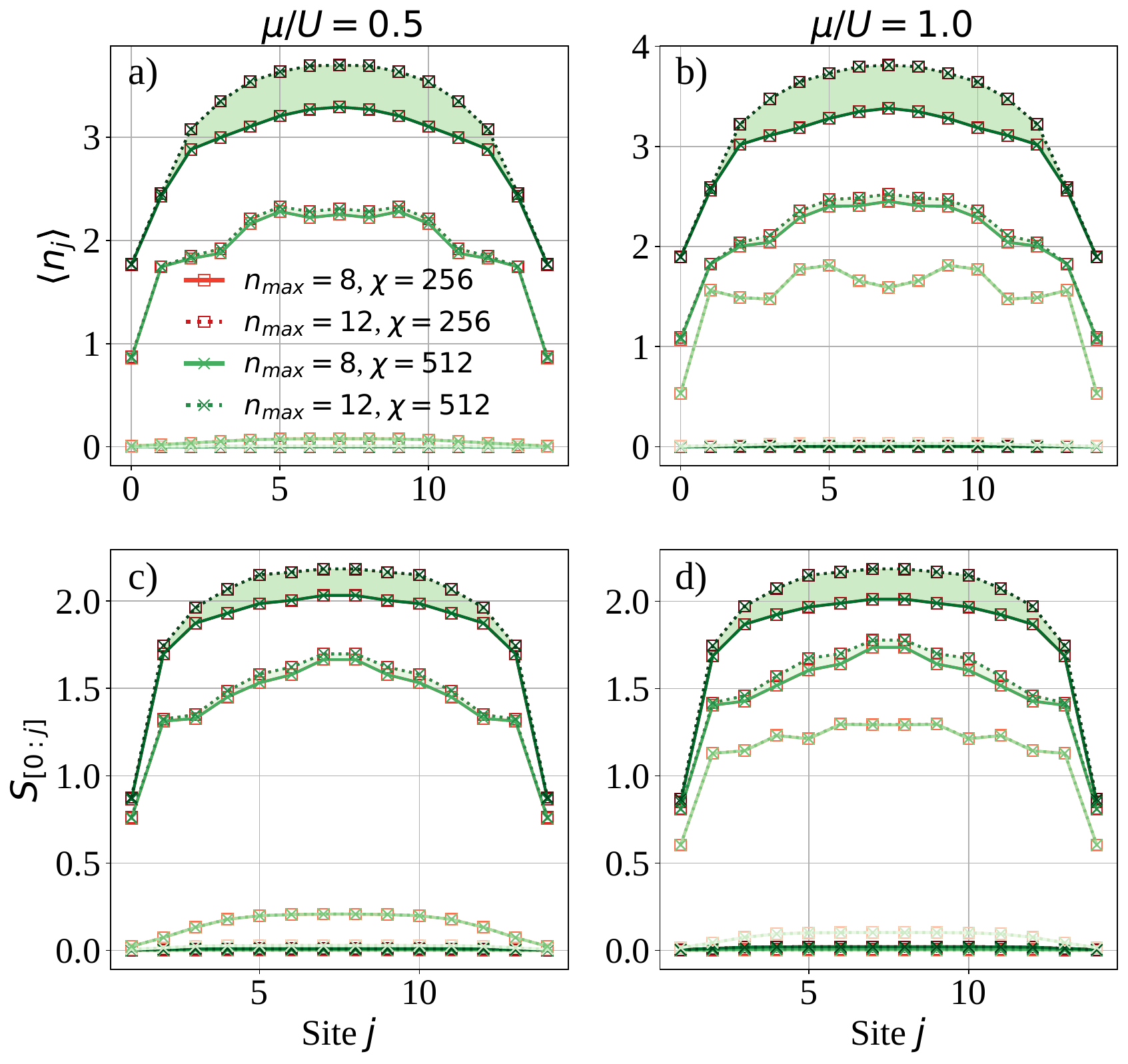}
    \caption{Convergence in bond dimension  $\chi$ and maximum number of bosons per site $n_{\textrm{max}}$, for incommensurate model with $q=2, p=1$ and finite system size with open boundary conditions. Panels on left and right columns show data for uniform chemical potential with $\mu/U=0.5, 1$ respectively. The ratio $J/U\in[0.25, 2]$ goes from lighter to stronger color intensity for increasing $J/U$ in steps of $0.25$. Panels (a-b) in the first row show the profile density $\langle \nh_j\rangle$, while panels (c-d) in the second show the bi-partite entanglement entropy $S_{[0:j]}$. The area between two nearby curves show the variation when increasing $n_{\text{max}}$ from $8$ to $12$. }
    \label{fig:incomm_detail}
\end{figure}

In this appendix we provide numerical results for the convergence of finite DMRG simulations of the QLRO phase for incommensurate $k^*$ with $q=2,p=1$ and uniform chemical potential $\mu/U=0.5,1$. These are shown in Fig.~\ref{fig:incomm_detail}. First and second rows show the convergence in the density profile $\langle \nh_j \rangle$ and entanglement profiles $S_{[0:j]}$ for bond dimensions $\chi=256, 512$ and a maximum number of boson per site $n_{\text{max}}=8,12$. Results are shown for different values of $J/U\in \{0.25,0.5,0.75,1.,1.25,1.5,1.75,2\}$ from lighter to darker green color. The colored area betwen two nearby data sets correspond to the variation of the numerical results when increasing $n_{\text{max}}$ for the same value of $J/U$ and bond dimension. 
\end{appendix}
\twocolumngrid
%\bibliography{biblio.bib}

\end{document}